%% file: mrssa.tex
\documentclass[article, shortnames, nojss]{jss}
\pdfoutput=1

\author{Nina Golyandina\\St.~Petersburg State University \And
        Anton Korobeynikov\\St.~Petersburg State University\AND
        Alex Shlemov\\St.~Petersburg State University\And
        Konstantin Usevich\\GIPSA-lab, CNRS, University of Grenoble}
\title{Multivariate and 2D Extensions of Singular Spectrum Analysis with the \pkg{Rssa} Package}

\Plainauthor{Nina Golyandina, Anton Korobeynikov, Alex Shlemov, Konstantin Usevich} 
\Plaintitle{Multivariate and 2D Extensions of Singular Spectrum Analysis with the Rssa Package} 
\Shorttitle{Multivariate and 2D Extensions of SSA with \pkg{Rssa}} 

\Abstract{
    Implementation of multivariate and 2D extensions of singular spectrum analysis (SSA)
    by means of the \proglang{R}-package \pkg{Rssa} is considered.
    The extensions include MSSA for simultaneous analysis and forecasting of several time series
    and 2D-SSA for analysis of digital images.
    A new extension of 2D-SSA analysis called Shaped 2D-SSA is introduced
    for analysis of images of arbitrary shape,
    not necessary rectangular.  It is shown that implementation of Shaped
    2D-SSA can serve as a base for implementation of MSSA and other generalizations.
    Efficient implementation of operations with
    Hankel and Hankel-block-Hankel matrices through the fast Fourier transform is suggested.  Examples
    with code fragments in \proglang{R}, which explain the methodology and
    demonstrate the proper use of \pkg{Rssa}, are presented.
}
\Plainkeywords{singular spectrum analysis, time series, image processing,
    analysis, forecasting, decomposition, R package} 

\Address{
  Nina Golyandina\\
  Department of Statistical Modelling\\
  Faculty of Mathematics and Mechanics\\
  St.~Petersburg State University\\
  Universitetskij pr 28, 198504, St.Petersburg, Russia\\
  E-mail: \email{nina@gistatgroup.com}\\
  \\
  Anton Korobeynikov\\
  Department of Statistical Modelling\\
  Faculty of Mathematics and Mechanics\\
  St.~Petersburg State University\\
  Universitetskij pr 28, 198504, St.Petersburg, Russia\\
  E-mail: \email{anton@korobeynikov.info}\\
  \\
  Alex Shlemov\\
  Department of Statistical Modelling\\
  Faculty of Mathematics and Mechanics\\
  St.~Petersburg State University\\
  Universitetskij pr 28, 198504, St.Petersburg, Russia\\
  E-mail: \email{shlemovalex@gmail.com}\\
  \\
  Konstantin Usevich\\
  GIPSA-lab, CNRS UMR5216\\
  11 rue des Math\'{e}matiques\\
  Grenoble Campus, BP 46\\
  F-38402 St. Martin d'H\`{e}res cedex, France\\
  E-mail: \email{konstantin.usevich@gipsa-lab.grenoble-inp.fr}
}

\graphicspath{{img/}}
\usepackage{latexsym,amssymb, amsthm}
\usepackage{amsmath}
\usepackage{thumbpdf}
\input{letters_series_mathbb}

\input{newcommands}

\input{newcommands2dssa}

\newtheorem{proposition}{Proposition}
\newtheorem{corollary}{Corollary}
\newtheorem{theorem}{Theorem}
\newtheorem{remark}{Remark}
\newtheorem{lemma}{Lemma}
\newtheorem{definition}{Definition}
\newtheorem{example}{Example}
\newtheorem{fragment}{Fragment}
\newtheorem{algorithm}{Algorithm}

\usepackage{tikz}
\begin{document}


\section{Introduction}
\input{intro.tex}

\section{Common structure of algorithms and general notions}
\label{sec:common}
\input{common.tex}

\section[MSSA]{Multivariate singular spectrum analysis}
\label{sec:MSSA}
\input{multi_algorithm.tex}

\input{multi_theory.tex}
\input{multi_package_typicalcode.tex}

\input{multi_package_comments.tex}
\input{multi_examples_decomp.tex}
\input{multi_examples_mssa_vs_ssa.tex}

\section[2D-SSA]{2D singular spectrum analysis}
\label{sec:2DSSA}
\input{2dssa_algorithm_2dssa.tex}
\input{2dssa_package.tex}
\input{2dssa_examples_smoothing.tex}
\input{2dssa_examples_esprit.tex}

\section[Shaped 2D-SSA]{Shaped 2D singular spectrum analysis}
\label{sec:Shaped2DSSA}
\input{shaped_algorithm.tex}

\input{shaped_package.tex}
\input{shaped_variants.tex}

\section{Implementation}
\label{sec:implementation}
\input{implement_hankel.tex}
\input{implement_shaped.tex}

\input{implement_forecast.tex}

\section{Conclusion}
The paper contains an extended guide to the use of the \pkg{Rssa} package for analysis 
of multivariate and multidimensional objects by Singular Spectrum Analysis (SSA). 
The following extensions of SSA are considered: MSSA for multidimensional time series,
2D-SSA for two-dimensional arrays (images), and a new method Shaped 2D-SSA (ShSSA) for arrays of arbitrary shape.

Numerous examples for each SSA extension are included in the paper. The examples cover
typical tasks that occur in SSA analysis and show how these tasks can be performed
with the \pkg{Rssa} package. 
The examples also demonstrate plotting capabilities of \pkg{Rssa}.
Together with practical and implementation issues, the paper contains a summary of theoretical and 
methodological aspects of SSA that assists in the proper use of the package.

In the paper, we stress on a common form of theory, algorithms, package interface and implementation.
The algorithms for SSA extensions are presented as particular cases of the general SSA scheme, 
which is specialized in each case with the help of an appropriate embedding operator.
The examples with typical code demonstrate the common structure of the algorithms,
which is reflected in the common structure of \pkg{Rssa} interface for all SSA extensions.
In implementation of \pkg{Rssa}, a unified approach is proposed, based on the Shaped 2D-SSA algorithm.

We hope that our general approach to the variety of SSA versions will help to users 
to apply \pkg{Rssa} properly and effectively.

\section*{Acknowledgement}
The authors would like to thank the anonymous referees for their patience and
thoughtful suggestions and comments that led to the significantly improved
presentation of the paper.

The paper was partly supported by the project NG13-083 of the Dynasty Foundation
in the area of computer science. Konstantin Usevich was supported by
the European Research Council under the European Union's Seventh Framework
Programme (FP7/2007-2013) / ERC Grant agreements no. 258581 and 320594.

\bibliography{mrssa}

\appendix
\section{Appendix}
\label{sec:app}
\input{app_mssa_theory.tex}

\input{app_2dssa_theory.tex}

\end{document}

%% file: letters_series_mathbb.tex
\usepackage{euscript}

\newcommand{\cF}{\EuScript{F}}

\newcommand{\cL}{\EuScript{L}}

\newcommand{\cP}{\EuScript{P}}

\newcommand{\cR}{\EuScript{R}}

\newcommand{\cX}{\EuScript{X}}

\newcommand{\rmA}{\mathrm{A}}
\newcommand{\rmB}{\mathrm{B}}
\newcommand{\rmC}{\mathrm{C}}

\newcommand{\rmH}{\mathrm{H}}

\newcommand{\rmT}{\mathrm{T}}
\newcommand{\rmU}{\mathrm{U}}
\newcommand{\rmV}{\mathrm{V}}

\newcommand{\rmX}{\mathrm{X}}

\newcommand{\tF}{\mathbb{F}}
\newcommand{\tG}{\mathbb{G}}
\newcommand{\tH}{\mathbb{H}}

\newcommand{\tN}{\mathbb{N}}

\newcommand{\tR}{\mathbb{R}}
\newcommand{\tS}{\mathbb{S}}

\newcommand{\tX}{\mathbb{X}}
\newcommand{\tY}{\mathbb{Y}}
\newcommand{\tZ}{\mathbb{Z}}

\newcommand{\bfA}{\mathbf{A}}
\newcommand{\bfB}{\mathbf{B}}
\newcommand{\bfC}{\mathbf{C}}
\newcommand{\bfD}{\mathbf{D}}
\newcommand{\bfE}{\mathbf{E}}
\newcommand{\bfF}{\mathbf{F}}
\newcommand{\bfG}{\mathbf{G}}
\newcommand{\bfH}{\mathbf{H}}
\newcommand{\bfI}{\mathbf{I}}

\newcommand{\bfP}{\mathbf{P}}
\newcommand{\bfQ}{\mathbf{Q}}

\newcommand{\bfS}{\mathbf{S}}

\newcommand{\bfU}{\mathbf{U}}
\newcommand{\bfV}{\mathbf{V}}
\newcommand{\bfW}{\mathbf{W}}
\newcommand{\bfX}{\mathbf{X}}
\newcommand{\bfY}{\mathbf{Y}}
\newcommand{\bfZ}{\mathbf{Z}}

\newcommand{\bbR}{\mathbb{R}}

\newcommand{\calA}{\mathcal{A}}

\newcommand{\calF}{\mathcal{F}}

\newcommand{\calM}{\mathcal{M}}

\newcommand{\calT}{\mathcal{T}}

%% file: newcommands.tex
\newcommand{\bt}{\begin{theorem}}
\newcommand{\et}{\end{theorem}}
\newcommand{\bl}{\begin{lemma}}
\newcommand{\el}{\end{lemma}}
\newcommand{\bp}{\begin{proposition}}
\newcommand{\ep}{\end{proposition}}
\newcommand{\bc}{\begin{corollary}}
\newcommand{\ec}{\end{corollary}}

\newcommand{\bd}{\begin{definition}\rm}
\newcommand{\ed}{\end{definition}}
\newcommand{\bex}{\begin{example}\rm}
\newcommand{\eex}{\end{example}}
\newcommand{\br}{\begin{remark}\rm}
\newcommand{\er}{\end{remark}}

\newcommand{\btbh}{\begin{table}[!ht]}
\newcommand{\etb}{\end{table}}
\newcommand{\bfgh}{\begin{figure}[!ht]}
\newcommand{\efg}{\end{figure}}

\newcommand{\bea}{\begin{eqnarray*}}
\newcommand{\eea}{\end{eqnarray*}}
\newcommand{\be}{\begin{eqnarray}}
\newcommand{\ee}{\end{eqnarray}}

\newcommand{\suml}{\sum\limits}

\newcommand{\vphi}{\varphi}

\newcommand{\lm}{\lambda}
\def\wtilde{\widetilde}
\def\what{\widehat}

\newcommand{\ra}{\rightarrow}

\def\bproof{\textbf{Proof.\ }}
\def\eproof{\hfill$\Box$\smallskip}

\def\spaceN{\mathsf{N}}
\def\spaceR{\mathsf{R}}
\def\spaceC{\mathsf{C}} 

\newcommand{\bfw}{\mathbf{w}}

\def\last#1{{\underline{#1}}}
\def\llast#1{\underline{\underline{#1}}}
\def\first#1{{\mathstrut\overline{#1}}}
\def\ffirst#1{\mathstrut\overline{\mathstrut\overline{#1}}}

\def\bfpi{\mbox{\boldmath{$\pi$}}}
\def\bfmu{\mbox{\boldmath{$\mu$}}}

\def\bfcR{\mbox{\boldmath{$\cR$}}}

\def\mmod{\mathop{\mathrm{mod}}}
\def\sspan{\mathop{\mathrm{span}}}

\makeatletter
\def\adots{\mathinner{\mkern2mu\raise\p@\hbox{.}
\mkern2mu\raise4\p@\hbox{.}\mkern1mu
\raise7\p@\vbox{\kern7\p@\hbox{.}}\mkern1mu}}
\newcommand{\l@abcd}[2]{\hbox to\textwidth{#1\dotfill #2}}
\makeatother

\newcommand{\frob}{\calF}
\def\trajmat#1{\calT_{\mathrm{#1}}}

\def\unit{\mathfrak{i}}

%% file: newcommands2dssa.tex
\newcommand{\Nx}{N_x}
\newcommand{\Ny}{N_y}
\newcommand{\Kx}{K_x}
\newcommand{\Ky}{K_y}
\newcommand{\Lx}{L_x}
\newcommand{\Ly}{L_y}
\newcommand{\mvec}{\mathop{\mathrm{vec}}}
\newcommand{\revv}{\mathop{\mathrm{rev}}}

\newcommand{\Ns}{\mathfrak{N}}
\newcommand{\Ls}{\mathfrak{L}}
\newcommand{\Ks}{\mathfrak{K}}
\newcommand{\LsE}{\ell}
\newcommand{\KsE}{\kappa}
\newcommand{\shs}{+_{\!\mbox{-}}}

\def\tcirc#1{\bfC_{\mathrm{T}}(#1)}
\def\tbtcirc#1{\bfC_{\mathrm{TbT}}(#1)}
\def\hbhcirc#1{\bfC_{\mathrm{HbH}}(#1)}
\def\hcirc#1{\bfC_{\mathrm{H}}(#1)}
\def\fftw#1{\cF_{#1}}
\def\fftwone#1{\bfF_{#1}}
\newcommand{\elmat}{\bfE}
\newcommand{\Pmat}{P}
\newcommand{\trace}{\mathop{\mathrm{trace}}}
\newcommand{\diagsums}{\mathop{\mathrm{diagsums}}}
\def\unitvec#1#2{E^{(#1)}_{#2}}
\def\conjug#1{\mathop{\mathrm{conj}}(#1)}

\newcommand{\TWODSSA}{2D\mbox{\small-}SSA}

%% file: intro.tex
Singular spectrum analysis as a method of time series analysis has
well-elaborated theory and solves various problems: time series decomposition,
trend extraction, periodicity detection and extraction, signal extraction,
denoising, filtering, forecasting, missing data imputation, change point
detection, spectral analysis among them (see examples and references in
\citet{Vautard.Ghil1989,Golyandina.etal2001,Ghil.etal2002,Golyandina.Zhigljavsky2012}).
Since the method does not need a model given a priori, it is called
nonparametric and is well suited for exploratory analysis of time series.

Additionally, SSA allows the construction of a model during or after
exploratory analysis.  The underlying parametric model of the signal is the sum
of products of polynomial, exponential and sine-wave functions.  This is a
linear model in the following sense: such series constitute the class of
solutions of linear differential equations. In the case of discrete time, such
time series satisfy linear recurrent relations (LRRs).  There is a class of so
called subspace-based methods \citep{VanDerVeen.etal1993}, which are related to
estimation of parameters in the mentioned parametric model, in particular, to
estimation of frequencies.

Although some problems like frequency estimation do need the model, some
problems like smoothing do not need a model at all.  For forecasting in SSA, the
time series may satisfy the model approximately and locally, since the
forecasting methods in SSA are based on estimation of the signal subspace and not on estimation
of the parameters of the model.  Depending on the quality of approximation of the series by
the model, long or short horizon forecasts can be constructed.

\paragraph{SSA software} 
The aforementioned possibilities are implemented in different software based on different methodologies of SSA,
see \citet{Golyandina.Korobeynikov2013} for references. The sources of the methodology used in this paper
are the books \citet{Golyandina.etal2001,Golyandina.Zhigljavsky2012} and the
software from \citet{CaterpillarSSA}, where SSA analysis and forecasting of
one-dimensional and multivariate time series are implemented in an interactive
manner.  The same methodology is used and developed in the \pkg{Rssa} package for \proglang{R}
system for statistical computing \citep{r2013}; see
\citet{Golyandina.Korobeynikov2013}, where analysis, forecasting and parameter
estimation by means of \pkg{Rssa} for one-dimensional series are described.

At the present time, \pkg{Rssa} is extensively developed and includes
SSA-processing of one-dimensional time series (Basic SSA, or simply SSA),
systems of series (Multivariate or Multichannel SSA, shortened to MSSA) and of
digital images (2D-SSA).  The aim of this paper is to describe this multidimensional
part of the \pkg{Rssa} package as of version 0.10.  Note that the effective implementation of the
algorithms of multidimensional SSA extensions is very important, since the
algorithms are much more time-consuming than in the one-dimensional
case. Therefore, we pay attention to both the guidelines for the proper use of the
package and the efficient implementation techniques, which are based on the
methods described in \citet{Korobeynikov2010}.  In addition, a new method called
Shaped 2D-SSA (ShSSA for short) is introduced. This method can be applied to the images of
non-rectangular form, e.g., circular images, images with gaps and so on. Also, it is shown
that the implementation of Shaped 2D-SSA can serve as a common base for implementation
of many SSA extensions.

\paragraph{General scheme of SSA}
Let us introduce a general scheme of SSA-like algorithms, including
MSSA and 2D-SSA.  The SSA-like algorithms decompose the given data $\tX$ into a
sum of different components:
\be
\label{eq:SSA_result}
\tX = \tX_1 + \cdots + \tX_m.
\ee
A typical SSA decomposition of a time series is the decomposition into
slowly-varying trend, seasonal components and residual or the decomposition into some
pattern, regular oscillations and noise.  The input data can be a time series, a
multivariate time series, or a digital image.

The algorithm is divided into four steps. {The first step} is 
generation of a multivariate object on the base of the initial object (time
series or image) by moving a \emph{window} of some form and taking the elements from
the window. For one-dimensional time series, this window is an interval that
produces subseries of length $L$ of the time series, where $L$ is called window
length (in SSA).  For multivariate time series (a system of $s$ one-dimensional
series), the window also produces subseries of length $L$, but we apply this
window to all time series of the system (in MSSA).  For images, the window can be a
2D rectangle (in 2D-SSA) or some other 2D shape (in ShSSA).
Then all the subobjects (subseries or vectorized 2D
shapes) obtained by applying the window are stacked as columns into the \emph{trajectory} matrix.
The trajectory matrix has a specific
structure: Hankel, stacked Hankel, Hankel-block-Hankel or quasi-Hankel.

{The second step} consists in decomposition of the trajectory matrix into
a sum of elementary matrices of rank 1.  The most frequently used decomposition,
which has a lot of optimal approximation properties, is
the singular value decomposition (SVD).

{The third step} is grouping of the decomposition components. At the
grouping step, the elementary rank-one matrices are grouped and summed within
groups.

{The last forth step} converts the grouped matrix decomposition back to the
decomposition of the initial time series or image decomposition.  The elements
of each component of the grouped decomposition are obtained by averaging the
entries of the grouped matrices that correspond to the same element in the
initial object.

Thus, the result of the algorithm is the decomposition (\ref{eq:SSA_result}) of
the initial object into the sum of objects.  We assume that the initial object
is a sum of some identifiable components, for example, trend and seasonality or
signal and noise and that we observe only the sum.  Then the aim of the SSA-like
methods is to reconstruct these components. The possibility to reconstruct the
object components is called \emph{separability} of the components.

Complex SSA is the same as the Basic SSA but it is applied to complex-valued
one-dimensional time series with complex-valued SVD and therefore fits well into
the described scheme.  Certainly, Multivariate Complex SSA and 2D Complex SSA
can be considered in similar manner, but it is out the scope of this paper.
Note that Complex SSA can be applied to a system of two real-valued time series
considered as real and imaginary parts of a complex series.

We mentioned that a model of the series can be constructed during the SSA
processing.  The simplified form of the signal model
is $s_n=\sum_{k=1}^r A_k \mu_k^n$, where $\mu_k=\rho_k e^{\unit
2\pi \omega_k}$ (possible polynomial modulation is omitted).  The general approach for
estimation of the modulations $\rho_k$ and frequencies $\omega_k$ in this model
is to use the so-called signal subspace,
which can be estimated at the third step of the SSA algorithm. One of the subspace-based methods
for parameter estimation
is ESPRIT \citep{Roy.Kailath1989}. This approach has a 2D extension named 2D-ESPRIT
for estimation of parameters in the model $s_{ln}=\sum_{k=1}^r A_k \mu_k^l \nu_k^n$.

\paragraph{\pkg{Rssa} and related packages}
The range of problems solved by SSA and its multidimensional extensions is very wide,
as it can be seen from a brief review of SSA capabilities given above. That
is why we do not consider in the scope of the paper the comparison of \pkg{Rssa}
with other packages implementing decomposition, filtering, regression,
frequency estimation, etc. We refer the reader to
\url{http://cran.r-project.org/web/views/TimeSeries.html} for a review of
packages for processing of time series and to
\url{http://cran.r-project.org/web/views/Spatial.html} for a review of packages
for processing of spatial data. As an implementation of the SSA method,
\pkg{Rssa} is the only package on CRAN.

Let us put attention to several issues related to difference of SSA with many
other methods and algorithms implemented in various \proglang{R} packages.
SSA-like methods can be applied as model-free approaches that do not need to
know parametric models and the periods of possible periodic components in
advance.  This is their difference from methods like seasonal decomposition and
parametric regression. Also, an important point is that in SSA-like methods it is not essential for
data to have additive or multiplicative structure. In
addition, note that although the model of series governed by
linear recurrence relations may seem similar to the autoregressive models, they have
nothing in common.

The present version of the \pkg{Rssa} package deals with equidistant series and
spatial data given on a rectangular grid. For processing of the data given on an
irregular grid, two approaches can be used. Certainly, interpolation to a
rectangular grid can be performed (it is not implemented in the current version
of the package).  For univariate and multivariate series, one can formally deal
with non-equidistant data as with equidistant measurements, ignoring the
irregularity.

For processing of time series or of systems of time series, \pkg{Rssa} can deal
with vectors, lists of vectors, matrices and also with input data of classes
\code{ts}, \code{mts}, \code{zooreg} \citep{zeileis2005} (also, \code{zoo} data
can be used, but the contents of the index attribute of a \code{zoo} object is
simply ignored).  The package tries to preserve attributes of input data, making
operations with the decomposition results convenient: no conversion to the
original class is required.

For processing of multidimensional data like digital images, \pkg{Rssa} takes only
matrices as input data.  In theory, 2D-SSA can be used with data of different
nature: spatial, spatio-temporal or arbitrary digital images. However,
the \proglang{R} implementation of objects of these types usually carries with itself a lot of
additional information. It seems to be error-prone to try to preserve
such information in a generic way, and it also would introduce
many spurious dependencies of \pkg{Rssa} on other packages. Thus, it is
considered as a job of the user to convert the results of the decomposition back
to the appropriate form (see Fragment~\ref{frag:2dssa_brecon_rec} for an
example with the objects from the \pkg{raster} package \citep{Hijmans2013}).

Most of the plotting capabilities of the package are implemented via
\pkg{lattice} \citep{Sarkar2008}.  Therefore, the user can either plot
\pkg{lattice} objects returned by \code{plot} functions directly, or use the
capabilities of \pkg{lattice} to manipulate these plot objects in a convenient
way. Also, since the input time series classes may vary, \pkg{Rssa}, where
possible, allows one to fall back to native plotting functions of such
classes.

Finally we dwell on the issues related to efficient implementation of SSA-related methods
in the \pkg{Rssa} package.
The two main points that underlie the implementation are (1) the use of fast methods for
the singular value decomposition and (2) the use of fast multiplication of
Hankel-related matrices and matrix hankelization by means of the fast Fourier
transform (FFT), see \citet{Korobeynikov2010}. However, the usage of the
algorithms possessing good theoretical complexity will not automatically yield the
efficient implementation per se. The package also implements many
practical considerations which make the core SSA steps really fast. Besides
this, many of the functions in the package employ on-demand calculation of the
necessary objects and memoization. These points are very important for large
window sizes and long time series, and are crucial for 2D-SSA and ShSSA.


\paragraph{Contribution of the paper}
On the theoretical/algorithmic side, this paper has several new
contributions.  First, we describe a fast algorithm of the vector SSA
forecasting and provide its implementation.  Second, we introduce a new
extension of 2D-SSA, Shaped 2D-SSA, which should extend the application area of
the method in digital image processing. Then, fast FFT-based algorithms are
described and implemented for each considered SSA-like method.  Finally, we show
that all the considered extensions of SSA (and several related extensions) are
special cases of Shaped 2D-SSA.

\paragraph{Structure of the paper}
The structure of the paper is as follows. Section~\ref{sec:common}
contains an overview of the SSA approach, including common scheme
of the algorithms, general notions, and a summary of implementation details.
Multivariate version of SSA for analysis and forecasting of systems of series is described in
Section~\ref{sec:MSSA}.  2D-SSA for image processing together with 2D-ESPRIT for
parameter estimation are considered in Section~\ref{sec:2DSSA}.
Section~\ref{sec:Shaped2DSSA} devoted to ShSSA for
analysis of images of non-rectangular form.
Section~\ref{sec:implementation} contains details on the fast algorithms, including
FFT-based matrix multiplication, hankelization and the fast algorithm of the vector forecasting.
Theoretical details of MSSA and 2D-SSA are put in Appendix~\ref{sec:app}.

Each of Sections~\ref{sec:MSSA}--\ref{sec:Shaped2DSSA} contains the implemented
algorithms, description of the \pkg{Rssa} functionality with a typical \proglang{R} code,
simulated and real-life examples with the corresponding fragments of \proglang{R} code
accompanied by guidelines for the proper use. In Section~\ref{sec:shaped_special},
it is shown that all the considered SSA extensions are special cases of ShSSA.
Thus the fast FFT-based algorithms in Section~\ref{sec:implementation} are
provided only for Shaped 2D-SSA.

This paper contains a comprehensive description of multivariate extensions of SSA.
Although it is impossible to consider all applications of multivariate extensions,
we provide examples for the most common ones, with code fragments and references to the literature.
We also demonstrate plotting capabilities of the package which are necessary for
the proper use of the SSA methodology and representation of the results.


%% file: common.tex
Before introducing the details of the SSA-like algorithms and the corresponding notions, we present
a general scheme of the  algorithms in Figure~\ref{fig:Scheme}.

\bfgh
\begin{center}
    \vspace*{0.5cm}
    \input{common_scheme}
\end{center}
\caption{Scheme of the SSA-like algorithms.}
\label{fig:Scheme}
\efg

The algorithms consist of four steps. The input data are the object $\tX$ and the mapping
$\calT$.  The result of the algorithms is a decomposition of $\tX$ into sum of
additive components.  The intermediate outcomes of the steps can be used for
solving additional problems like forecasting and parameter estimation.

Table~\ref{tab:kinds} contains a summary of SSA extensions considered in the paper.

\begin{table}[h]
\begin{center}
\begin{tabular}{|l|c|c|c|c|}
\hline
Method & Data & Notation & Trajectory matrix  & Section \\
\hline
SSA     &  time series   & $\tX=(x_1,\ldots,x_N)$                               & Hankel          & \ref{sec:common_alg} \\
MSSA    &  system of time series & $\tX^{(p)}$, $p=1,\ldots,s$        & Stacked Hankel  & \ref{sec:MSSA} \\
CSSA    &  complex time series   & $\tX^{(1)} + \unit\, \tX^{(2)}$ & Complex Hankel& \ref{sec:MSSA}   \\
2D-SSA  &  rectangular image   & $\tX = (x_{ij})_{i,j=1}^{\Nx,\Ny}$ & Hankel-block-Hankel            & \ref{sec:2DSSA}   \\
ShSSA   &  shaped image  & $\tX = (x_{(i,j)})_{(i,j) \in \mathfrak{N}}$            & Quasi-Hankel    & \ref{sec:Shaped2DSSA}   \\
\hline
\end{tabular}
\caption{Kinds of SSA-like multivariate extensions.}
\label{tab:kinds}
\end{center}
\end{table}

Further we discuss the steps of the algorithms of the SSA-like methods in detail.

\subsection{Details of the algorithms}
\label{sec:common_alg}
\textbf{Step 1. Embedding}. The first step consists in the construction of the 
so-called \emph{trajectory matrix} $\bfX=\calT(\tX)$ by means of a map $\calT$.

In the Basic SSA algorithm applied to a one-dimensional time
series $\tX=(x_1,\ldots,x_N)$ of length $N$, $\calT$ maps $\spaceR^N$ to the space of Hankel
matrices $L\times K$ with equal values on anti-diagonals:
\begin{equation}
\label{eq:ssa_embedding}
\calT_\mathrm{SSA}(\tX)=\left(
    \begin{array}{lllll}
        x_1&x_2&x_3&\ldots&x_{K}\cr
        x_2&x_3&x_4&\ldots&x_{K+1}\cr
        x_3&x_4&x_5&\ldots&x_{K+2}\cr
        \vdots&\vdots&\vdots&\ddots&\vdots\cr
        x_{L}&x_{L+1}&x_{L+2}&\ldots&x_{N}\cr
    \end{array}
\right),
\end{equation}
where $L$ is the window length and $K=N-L+1$.

Let us describe the scheme of this step in a more formal manner for the general case of
SSA-like methods.  Denote $\calM_{p,q}$ the set of $p\times q$ real matrices.  
Let $\calM$ be the linear space of all possible $\tX$,
where $\tX$ is an ordered set of real values representing initial data
(i.e., time series or image).  Then the \emph{embedding} $\calT$ is a one-to-one mapping
from $\calM$ to $\calM_{L,K}^{(H)}\subset \calM_{L,K}$, where
$\calM_{L,K}^{(H)}$ is the set of matrices with a Hankel-like structure, 
which is determined together with $L$ and $K$ by the method parameters. 

The mapping $\calT$ puts elements of $\tX$ in 
some places of the trajectory matrix $\bfX = \calT(\tX)$.  
Let $x_k$ denote the $k$th element of $\tX$. (The index $k$ may be two-dimensional,
see e.g. Table~\ref{tab:kinds}.)
Then the mapping $\calT$ determines the set of indices $\calA_k$ such
that $(\bfX)_{ij}=x_k$ for any $(i,j)\in \calA_k$. Formally, let $E_k\in\calM$ be
the object with the $k$th element equal to 1 and all the other elements equal to zero.
Then the set $\calA_k$ corresponds to the places of `1' in the matrix $\calT(E_k)$.

For example, in the Basic SSA algorithm applied to a one-dimensional time
series $\tX$, $\calM = \spaceR^N$, $\calM_{L,K}^{(H)}$ is the set of Hankel
matrices and the mapping $\calT$ is given in (\ref{eq:ssa_embedding}). The objects $E_k$, 
$k=1,\ldots, N$, are the standard unit vectors in $\spaceR^N$, and $\calA_k$
consists of the positions of the elements on the $k$th anti-diagonal of $\bfX$.

\medskip
\textbf{Step 2. Singular value decomposition}.
Let $\bfS=\bfX \bfX^\top$, $\lambda_1\geq \ldots \geq \lambda_L\geq 0$ be {\em
    eigenvalues} of the matrix $\bfS$, $d=\max\{j:\,\lm_j >0\}$,
$U_1,\ldots,U_d$ be the corresponding {\em eigenvectors}, and $V_j=\bfX^\top
U_j/\sqrt{\lm_j}$, $j=1,\ldots,d$, be the {\em factor vectors}. 
Denote $\bfX_j=\sqrt{\lm_j}U_j V_j^\top$. Then the SVD of the trajectory matrix $\bfX$
can be written as
\be
\label{eq:mmexp}
    \bfX = \bfX_1 + \ldots + \bfX_d.
\ee
The values  $\sqrt{\lambda_j}$ are exactly the singular values of $\bfX$. The vectors
$U_j$ (respectively, $V_j$) are exactly the left (respectively, right) singular vectors
of $\bfX$. The triple $(\sqrt{\lambda_j}, U_j, V_j)$ is called the \emph{$j$th eigentriple} (or ET$j$ for short).

\medskip
\textbf{Step 3. Grouping}.
Once the expansion (\ref{eq:mmexp}) has been obtained,
the grouping procedure partitions the set of indices
$\{1,\ldots,d\}$ into $m$ disjoint subsets $I_1,\ldots,I_m$.
For a subset $I=\{i_1,\ldots,i_p\}$, the matrix $\bfX_I$ corresponding to
the group $I$ is defined as
$\bfX_I=\bfX_{i_1}+\ldots+\bfX_{i_p}$.
Thus, we have the \emph{grouped matrix decomposition}
\be
\label{eq:mmexp_g}
\bfX=\bfX_{I_1}+\ldots+\bfX_{I_m}.
\ee
For example, grouping can be performed on the base of the form of eigenvectors,
which reflect the properties of initial data components. The grouping with $I_j=\{j\}$ is called \emph{elementary}.

\medskip
\textbf{Step 4. Reconstruction}.
At this final step, each matrix of the decomposition (\ref{eq:mmexp_g}) is
transferred back to the form of the input object $\tX$.  It is performed
optimally in the following sense: for a matrix $\bfY\in \calM_{L,K}$ we seek for the object
$\widetilde\tY \in \calM$ that provides the minimum to $\|\bfY -
\calT(\widetilde\tY)\|_\calF$, where $\|\bfZ\|_\calF=\sqrt{\sum^{L,K}_{i,j=1} (\bfZ)_{ij}^2}$
is the Frobenius norm.

Denote $\Pi^{(H)}: \calM_{L,K} \to \calM_{L,K}^{(H)}$ the orthogonal projection
on $\calM_{L,K}^{(H)}$ in Frobenius norm.  Then $\widetilde\tY =
\calT^{-1}(\Pi^{(H)} \bfY)$. 
By the embedding nature of $\calT$, the
projection $\Pi^{(H)}$ is the averaging of the entries along the sets $\calA_k$,
that is, the elements of $\widetilde\tY$ are equal to
$\widetilde{y}_k=\sum_{(i,j)\in \calA_k} (\bfY)_{ij}/|\calA_k|$, 
where $|\calA_k|$ is the number of elements in $\calA_k$.
The same averaging can be rewritten as
$\widetilde{y}_k = {\langle \bfY, \calT(E_k)\rangle_\calF}/{\|\calT(E_k)\|^2_\calF}$,
where $\langle\bfY,\bfZ\rangle_\calF = \sum_{i,j=1}^{L,K} (\bfY)_{ij} (\bfZ)_{ij}$ is the Frobenius inner product.

In Basic SSA, the set  $\calA_k=\{(i,j):\,1\le i\le L, 1\le j\le K, i+j=k+1\}$ corresponds to the $k$th anti-diagonal, and the composite mapping $\calT^{-1}\circ \Pi^{(H)}$ is the averaging along
anti-diagonals.  Hence, in the SSA literature the reconstruction step is often
called \emph{diagonal averaging}.

Thus, denote $\what\bfX_k = \bfX_{I_k}$ the reconstructed matrices,
$\wtilde\bfX_k=\Pi^{(H)} \what\bfX_k$ the trajectory matrices of the
reconstructed data and $\wtilde\tX_k=\calT^{-1}(\wtilde\bfX_k)$ the
reconstructed data themselves.  Then the resultant decomposition of the initial
data has the form
\be
\label{eq:res_decomp}
\tX=\widetilde\tX_1+\ldots+\widetilde\tX_m.
\ee
If the grouping is elementary, then $m=d$, $\bfX_{I_k} = \bfX_{k}$, and the reconstructed objects $\wtilde\tX_k=\calT^{-1}\circ \Pi^{(H)} \bfX_k$ are called
\emph{elementary components}.

\subsection{General notions}
\label{sec:general}
\paragraph{Separability} A very important notion in SSA is separability.
Let $\tX=\tX_1+\tX_2$. (Approximate) separability means that there exist such
grouping that the reconstructed series $\widetilde\tX_1$ is
(approximately) equal to $\tX_1$. Properties of the SVD yield (approximate)
orthogonality of columns and orthogonality of rows of trajectory matrices $\bfX_1$ and
$\bfX_2$ of $\tX_1$ and $\tX_2$ as the separability condition.  There is a
well-elaborated theory of separability of one-dimensional time series
\citep[Sections 1.5 and 6.1]{Golyandina.etal2001}.  It appears that many
important decomposition problems, from noise reduction and smoothing to trend,
periodicity or signal extraction, can be solved by SSA.  Certainly, the
embedding operator $\calT$ determines separability conditions.  It could be said
that the success of SSA in separability is determined by Hankel structure of the
trajectory matrix and optimality features of the SVD.

\paragraph{Information for grouping} The theory of SSA provides various ways to
detect the SVD components related to the series component in order to perform proper
grouping in conditions of separability.  One of the rules is that the eigenvector
produced by a data component repeats the properties of this component. For
example, in SSA the eigenvectors produced by slowly-varying series components
are slowly-varying, the eigenvectors produced by a sine wave are sine waves with
the same frequencies, and so on. These properties help to perform the
grouping by visual inspection of eigenvectors and also by some automatic
procedures (see \citet{Alexandrov2009} and \citet[Section
2.4.5]{Golyandina.Zhigljavsky2012}).

To check separability of the reconstructed components $\wtilde\tX_1$ and
$\wtilde\tX_2$, we should check the orthogonality of their reconstructed
trajectory matrices $\wtilde\bfX_1$ and $\wtilde\bfX_2$.  A convenient measure
of their orthogonality is the Frobenius inner product
$\langle\wtilde\bfX_1,\wtilde\bfX_2\rangle_\calF$.  The
normalized measure of orthogonality is $\rho(\wtilde\bfX_1,\wtilde\bfX_2)=\langle
\wtilde\bfX_1,\wtilde\bfX_2\rangle_\calF/ (\|\wtilde\bfX_1\|_\calF
\|\wtilde\bfX_2\|_\calF)$.

Since the trajectory matrix consists of $w_k=|\calA_k|=\|\calT(E_k)\|_\calF^2$
entries corresponding to the $k$th element $x_k$ of the initial object, we can introduce
the \emph{weighted inner product} in the space $\calM$: $(\tY,\tZ)_\bfw=\sum_k w_k y_k
z_k$, which induces the \emph{weighted norm} $\|\tY\|_{\bfw} = \sqrt{(\tY,\tY)_\bfw}$.  Then
\be
\rho_\bfw(\wtilde\tX_1,\wtilde\tX_2) = \rho(\wtilde\bfX_1,\wtilde\bfX_2)=\frac{(\wtilde\tX_1,\wtilde\tX_2)_\bfw}
{\|\wtilde\tX_1\|_\bfw \|\wtilde\tX_2\|_\bfw}
\ee
is called \emph{$\bfw$~correlation} by statistical analogy. Note that in this definition means are not subtracted.

Let $\wtilde\tX_j$ be the elementary reconstructed components produced by the
elementary grouping $I_j=\{j\}$. Then the matrix of $\rho^{(\bfw)}_{ij}=
\rho_\bfw(\widetilde\tX_i,\widetilde\tX_j)$ is called \emph{$\bfw$~correlation
matrix}.

The weighted norm $\|\cdot\|_\bfw$  serves as a
measure of contribution of components to the decomposition
(\ref{eq:res_decomp}): the contribution of $\wtilde\tX_j$ is defined as
$\|\wtilde\tX_j\|_\bfw^2\big/\|\tX\|_\bfw^2$.

\paragraph{Trajectory spaces and signal subspace}
Let us introduce several notions related to subspaces generated by the data.
For the data $\tX$ the column (row) subspace of its trajectory matrix $\bfX$ is
called \emph{column (row) trajectory space}.  The term `trajectory space'
usually means `column trajectory space'.  The column trajectory space is a
subspace of $\spaceR^L$, while the row trajectory space is a subspace of
$\spaceR^K$.  In general, for real-world data the trajectory spaces coincide with
the corresponding Euclidean spaces, since they are produced by a signal
corrupted by noise. However, if the signal has rank-deficient trajectory matrix,
then the signal trajectory space can be called `signal subspace'.
Column and row signal subspaces can be considered.
Note that the dimensions of row and column subspaces coincide.

\paragraph{Objects of finite rank} The class of objects that suit the SSA method
are the so-called objects of finite rank.  We say that the object (time series
or image) has rank $r$ if the rank of its trajectory matrix is equal to
$r<\min(L,K)$, that is, the trajectory matrix is rank-deficient.  If the rank
$r$ does not depend on the choice of $L$ for any sufficiently large object and
trajectory matrix sizes, then we say that the object has finite rank (rank does
not tend to infinity as the size of the object tends to infinity), see
\citet[Chapter 5]{Golyandina.etal2001} and \citet{Golyandina.Usevich2010} for
rigorous definitions.

Since the trajectory matrices considered in SSA methods are Hankel or consist
of Hankel blocks, the rank-deficient Hankel matrices
are closely related to objects satisfying to some linear relations. These linear
relations can be taken as a basis for forecasting algorithms.
In the one-dimensional case, rank-deficient Hankel matrices are closely related to
linear recurrent relations $x_n=\sum_{i=1}^r a_i x_{n-i}$ and therefore
to time series which can be expressed as a sum of products of exponentials,
polynomials and sinusoids.

Each specific SSA extension produces a class of specific objects of finite rank.
The knowledge of ranks of objects of finite rank can help to group the corresponding SVD
components (whose number is equal to the rank value).
For example, to reconstruct the exponential trend in the
one-dimensional case, we need to group only one SVD component (the exponential
has rank 1), while to reconstruct a sine wave we generally need to group two SVD
components (the rank equals 2).

Surely, most of real-life time series or images are not of finite rank. For
example, a time series can be a sum of a signal of rank $r$ and noise. Then, due
to approximate separability, we can use SSA  to extract the signal and then apply  methods
 designed for series of finite rank.

\subsection{Typical code}
\label{sec:ssa_typical}
The general structure of the \pkg{Rssa} package is described in
\citet{Golyandina.Korobeynikov2013} and holds for the multivariate
extensions. Therefore, let us briefly discuss the base \pkg{Rssa} functions
for analysis and forecasting of one-dimensional time series.  Implementation of
SSA analysis and forecasting is mostly contained in the functions \code{ssa}
(decomposition), \code{reconstruct} (reconstruction), \code{predict}, \code{rforecast} and
\code{vforecast} (forecasting), \code{plot} (plotting), \code{wcor} (weighted
correlations for grouping).

For demonstration, we consider the series of sales of fortified wines (shortly
FORT) taken from the dataset ``Monthly Australian wine sales: thousands of
litres. By wine makers in bottles <= 1 litre'' \citep{HyndmanTSDL}. The full dataset
contain sales from January, 1980, to July, 1995 (187 points). However, the data
after June, 1994 have missing values. Therefore, we begin with the first 174
points.

Fragment \ref{frag:wines_input} contains the standard code for loading the
package \pkg{Rssa} and input of the data included into the package.

\begin{fragment}[Australian Wines: Input]
\label{frag:wines_input}
\input{fragments/wines_input.tex}
\end{fragment}

Fragment~\ref{frag:fort_rec} contains a typical code for extraction of the
trend and seasonality.  The resultant decomposition is depicted in
Figure~\ref{fig:fort_rec}.

\bfgh
    \begin{center}
        \includegraphics[width=12 cm]{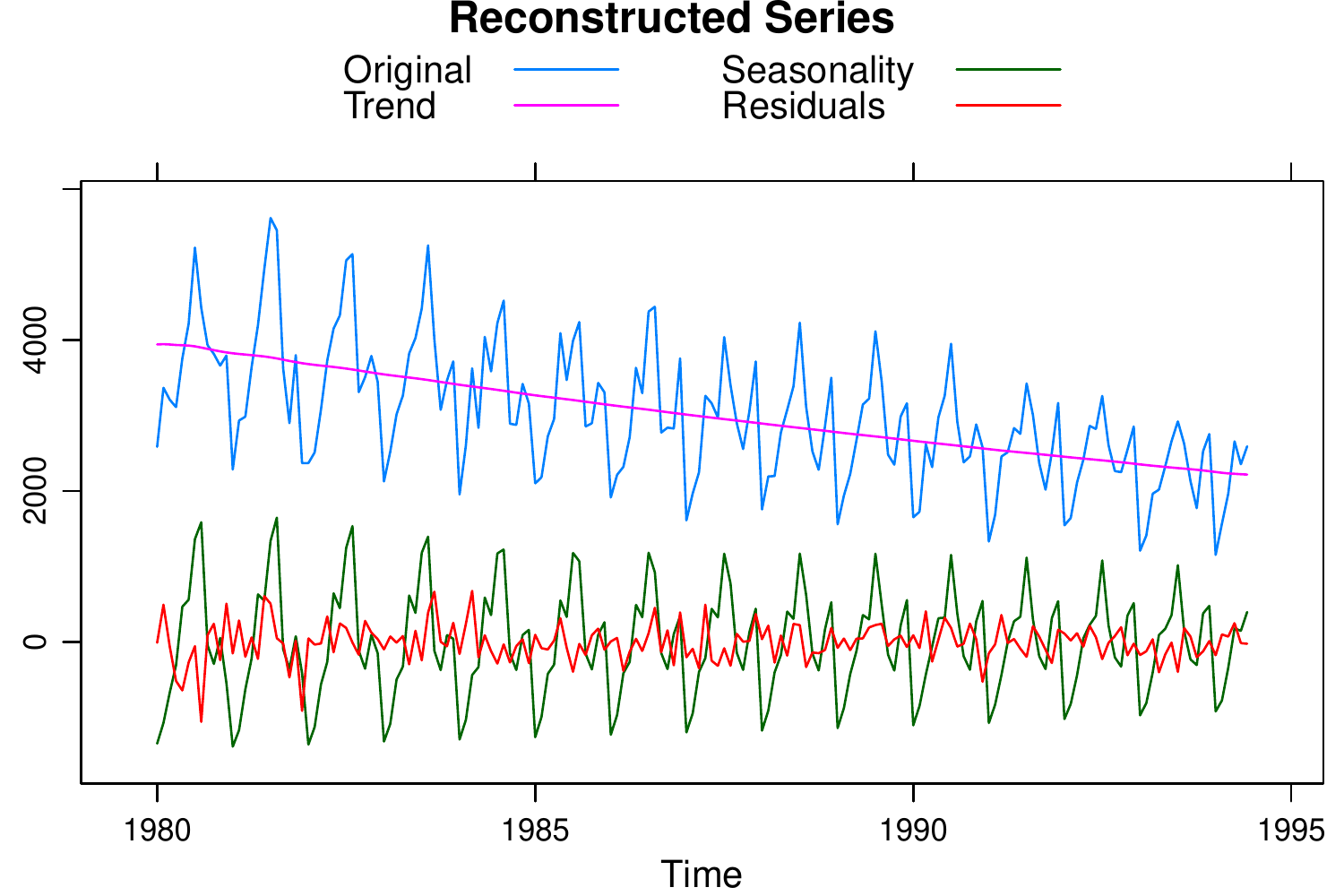}
    \end{center}
    \caption{FORT: Decomposition.}
    \label{fig:fort_rec}
\efg

\begin{fragment}[FORT: Reconstruction]
\label{frag:fort_rec}
\input{fragments/fort_rec.tex}
\end{fragment}

Roughly speaking (see details in \citet{Golyandina.Korobeynikov2013}), \code{ssa}
performs steps 1 and 2 of the algorithm described in Section~\ref{sec:common_alg}, 
while \code{reconstruct} performs steps 3 and 4 of the algorithm.  The argument values
\code{kind = "1d-ssa"}  and \code{svd.method = "auto"} are default and can be omitted.  Note
that the function \code{plot} for the reconstruction object implements different
special kinds of its plotting. In Fragment~\ref{frag:fort_rec}, the last two
parameters of \code{plot} are the parameters of the function \code{xyplot} from
the standard package \pkg{lattice}.

The grouping for reconstruction was made on the base of the following
information obtained
from the \code{ssa} object:
\begin{enumerate}
\item
one-dimensional (1D) figures of eigenvectors $U_i$ (Figure~\ref{fig:fort_1d}),
\item
two-dimensional (2D) figures of eigenvectors $(U_i, U_{i+1})$ (Figure~\ref{fig:fort_2d}), and
\item
matrix of $\bfw$~correlations $\rho_\bfw$ between elementary reconstructed
series (functions \code{wcor} and \code{plot}, Figure~\ref{fig:fort_wcor}).
\end{enumerate}

The following fragment shows the code that reproduces Figures~\ref{fig:fort_1d}--\ref{fig:fort_wcor}.
\begin{fragment}[FORT: Identification]
\label{frag:fort_identific}
\input{fragments/fort_identific.tex}
\end{fragment}

Let us explain how the figures obtained by means of
Fragment~\ref{frag:fort_identific} can help to perform the grouping.
Figure~\ref{fig:fort_1d} shows that the first eigenvector is slowly-varying and
therefore the eigentriple (abbreviated as ET) ET1 should be included to the
trend group.  Figure~\ref{fig:fort_2d} shows that the pairs 2--3, 4--5, 6--7,
8--9, 10--11 are produced by modulated sine-waves, since the corresponding 2D-scatterplots
of eigenvectors are similar to regular polygons. This way of identification is
based on the following properties: a sine wave has rank 2 and produces two
eigentriples, which are sine waves with the same frequency and have a phase
shift exactly or approximately equal to $\pi/2$, due to orthogonality of
eigenvectors. 

By counting the numbers of polygon vertices in Figure~\ref{fig:fort_2d}, the periods of the sine-waves can be determined as 12, 4, 6, 2.4, 3. Alternatively, automatic methods of frequency calculation can be employed, such as  
LS-ESPRIT and TLS-ESPRIT methods \citep{Roy.Kailath1989}.
These methods are implemented in \pkg{Rssa} in the function
\code{parestimate} and are described in \citet[Sections 2.4.2.4. and
3.8.2]{Golyandina.etal2001} and \citet{Golyandina.Korobeynikov2013} for
one-dimensional time series.
The periods calculated by the automatic \code{parestimate} method in Fragment~\ref{frag:fort_identific} agree with the numbers of
vertices in Figure~\ref{fig:fort_2d} for the five pairs listed.

The matrix of absolute values of $\bfw$~correlations in Figure~\ref{fig:fort_wcor} is depicted in
grayscale (white color corresponds to zero values, while black color
corresponds to the absolute values equal to 1).
Figure~\ref{fig:fort_wcor} confirms that the indicated pairs are separated
between themselves and also from the trend component, since the correlations between the pairs are small,
while correlations between the components from one
pair are very large. The block of 12--84 components is ``gray'', therefore we
can expect that these components are mixed and are produced by noise.

Figure~\ref{fig:fort_recsine} contains four reconstructed modulated sine waves and shows
that several sine waves have increasing amplitudes, while others are decreasing;
the same can be seen in Figure~\ref{fig:fort_1d}.  In Figure~\ref{fig:fort_rec}, we
grouped the modulated sine waves and obtained the seasonal component with
varying form.

 \bfgh
        \begin{center}
        \includegraphics[width=120 mm]{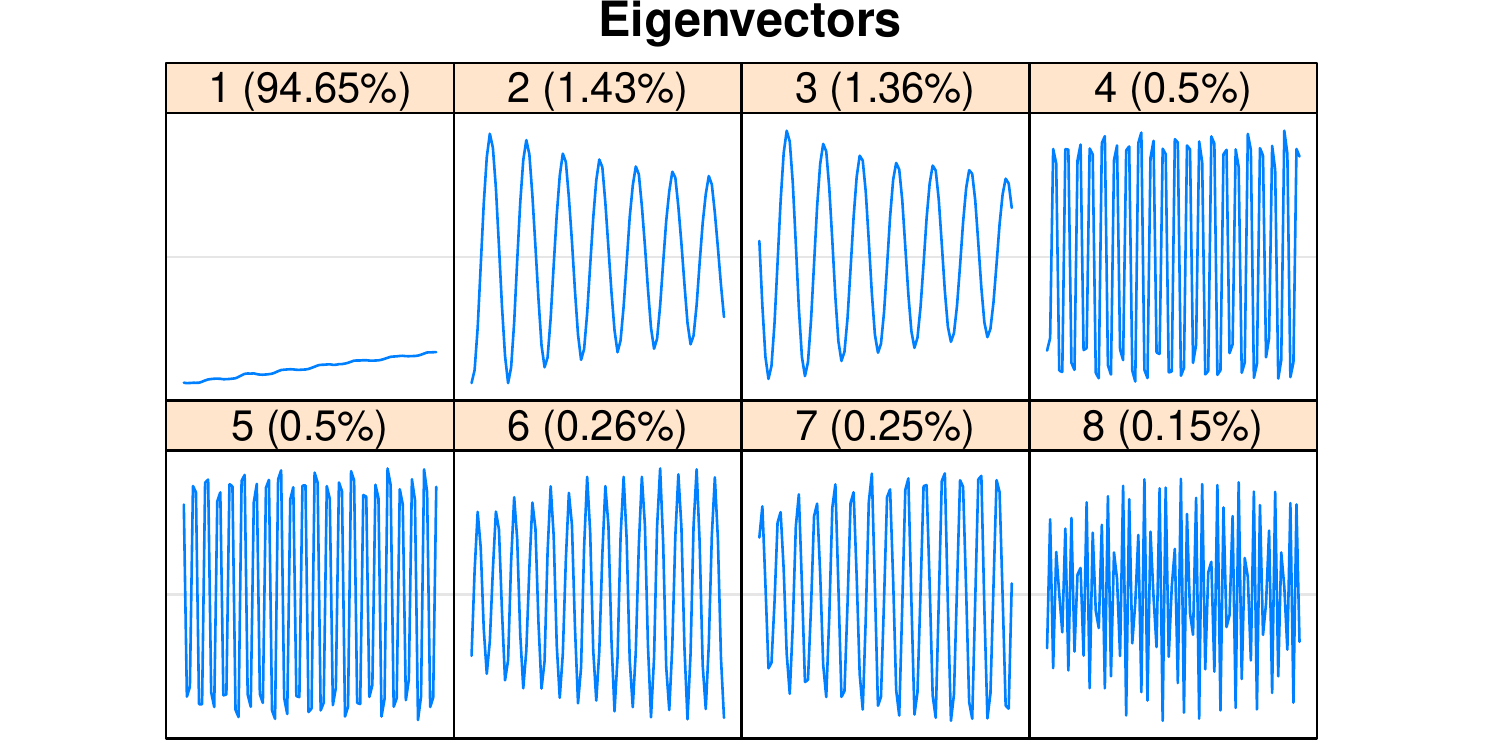}
        \end{center}
        \caption{FORT: 1D graphs of eigenvectors.}
        \label{fig:fort_1d}
 \efg
 \bfgh
        \begin{center}
        \includegraphics[width=120 mm]{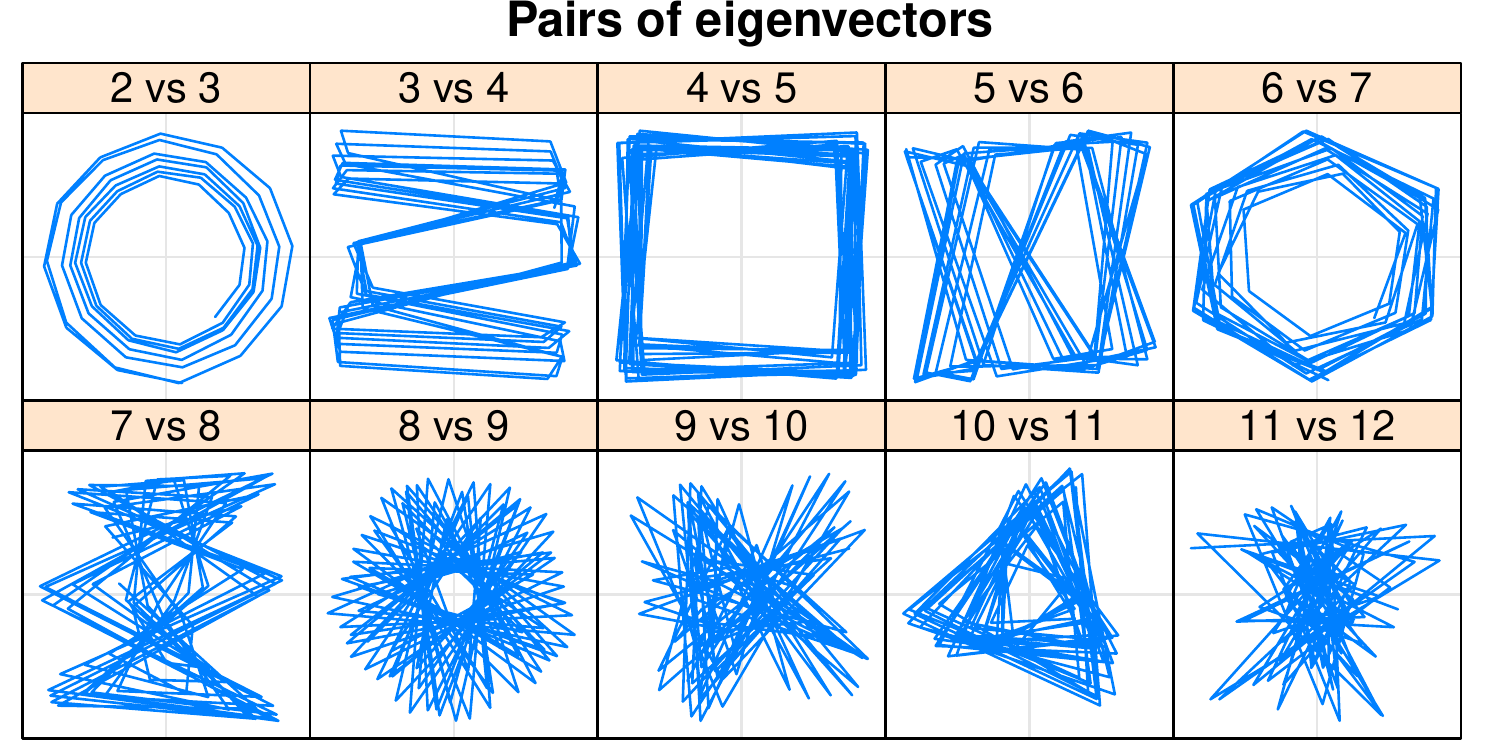}
        \end{center}
        \caption{FORT: 2D scatterplots of eigenvectors.}
        \label{fig:fort_2d}
 \efg
 \bfgh
        \begin{center}
        \includegraphics[width=80 mm]{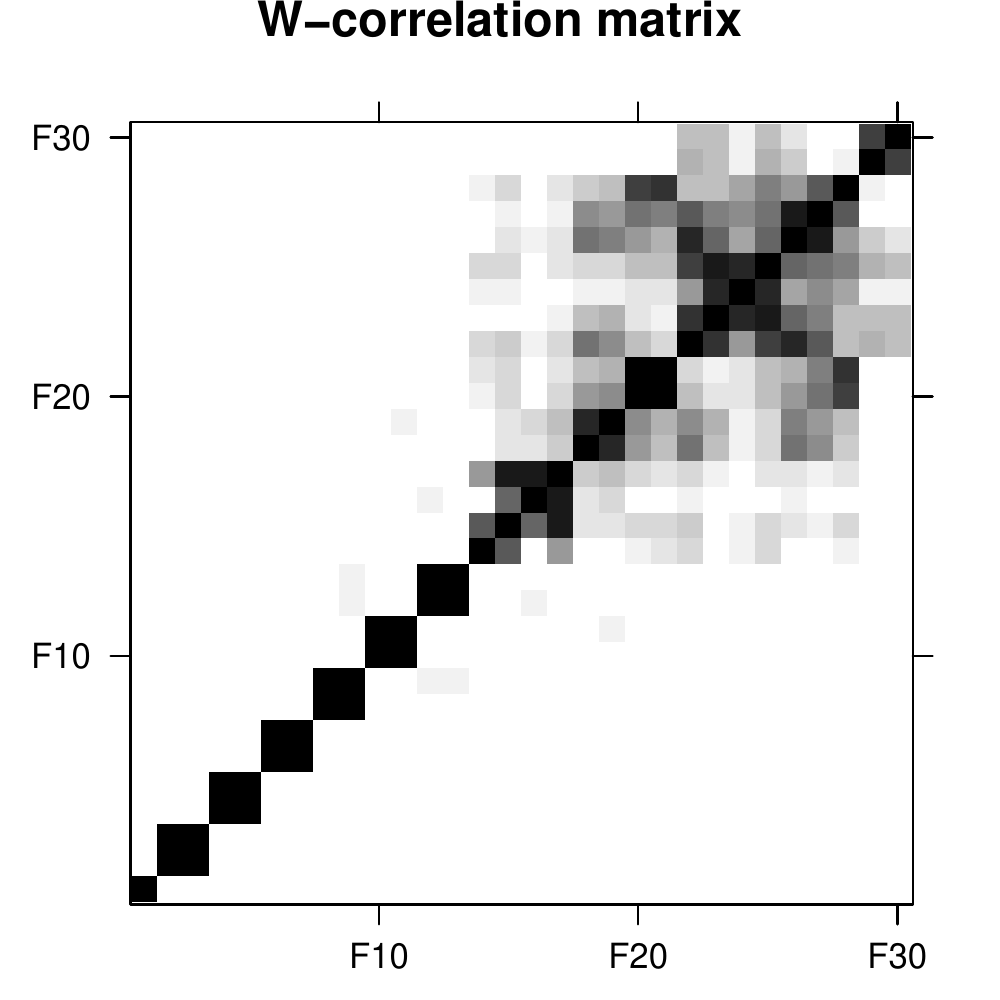}
        \end{center}
        \caption{FORT: Weighted correlations.}
        \label{fig:fort_wcor}
 \efg

 \bfgh
        \begin{center}
        \includegraphics[width=120 mm]{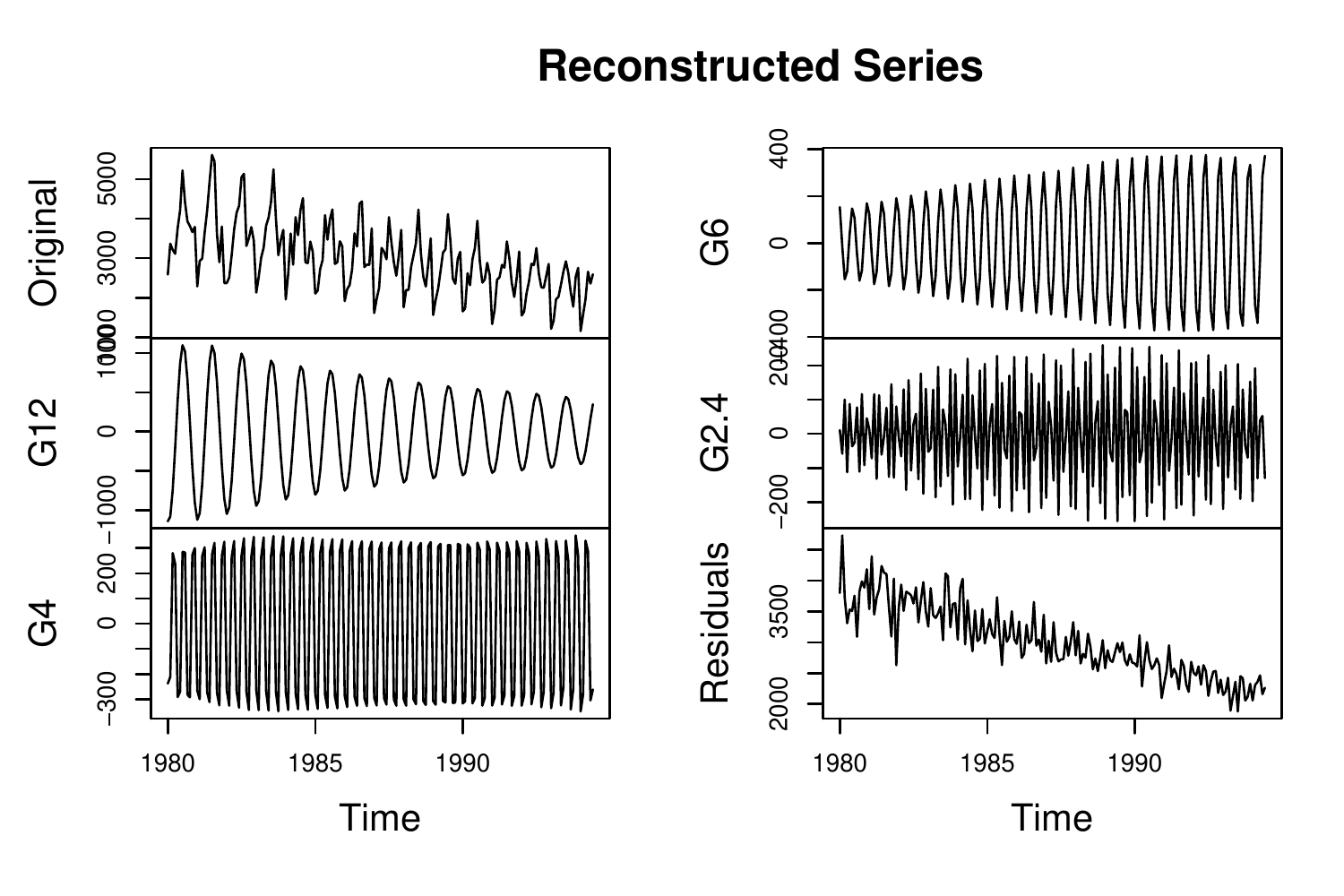}
        \end{center}
        \caption{FORT: Reconstructed sine waves.}
        \label{fig:fort_recsine}
 \efg

Finally, Fragment~\ref{frag:fort_forecast} contains an example of forecasting the series components (trend and signal), and the result is depicted in Figure~\ref{fig:fort_forecast}. Two forecasting methods are implemented for one-dimensional time
series: recurrent (function \code{rforecast}) and vector forecasting (function
\code{vforecast}). Both forecasting methods are based on estimating linear recurrent relations that govern time series, and are described in detail in Section~\ref{SEC:continuation}.

In Fragment~\ref{frag:fort_forecast}, the function \code{vforecast} is used. Alternatively, one can use all-in-one \code{predict} wrapper.

\begin{fragment}[FORT: Forecast]
\label{frag:fort_forecast}
\input{fragments/fort_forecast.tex}
\end{fragment}

 \bfgh
        \begin{center}
        \includegraphics[width=120 mm]{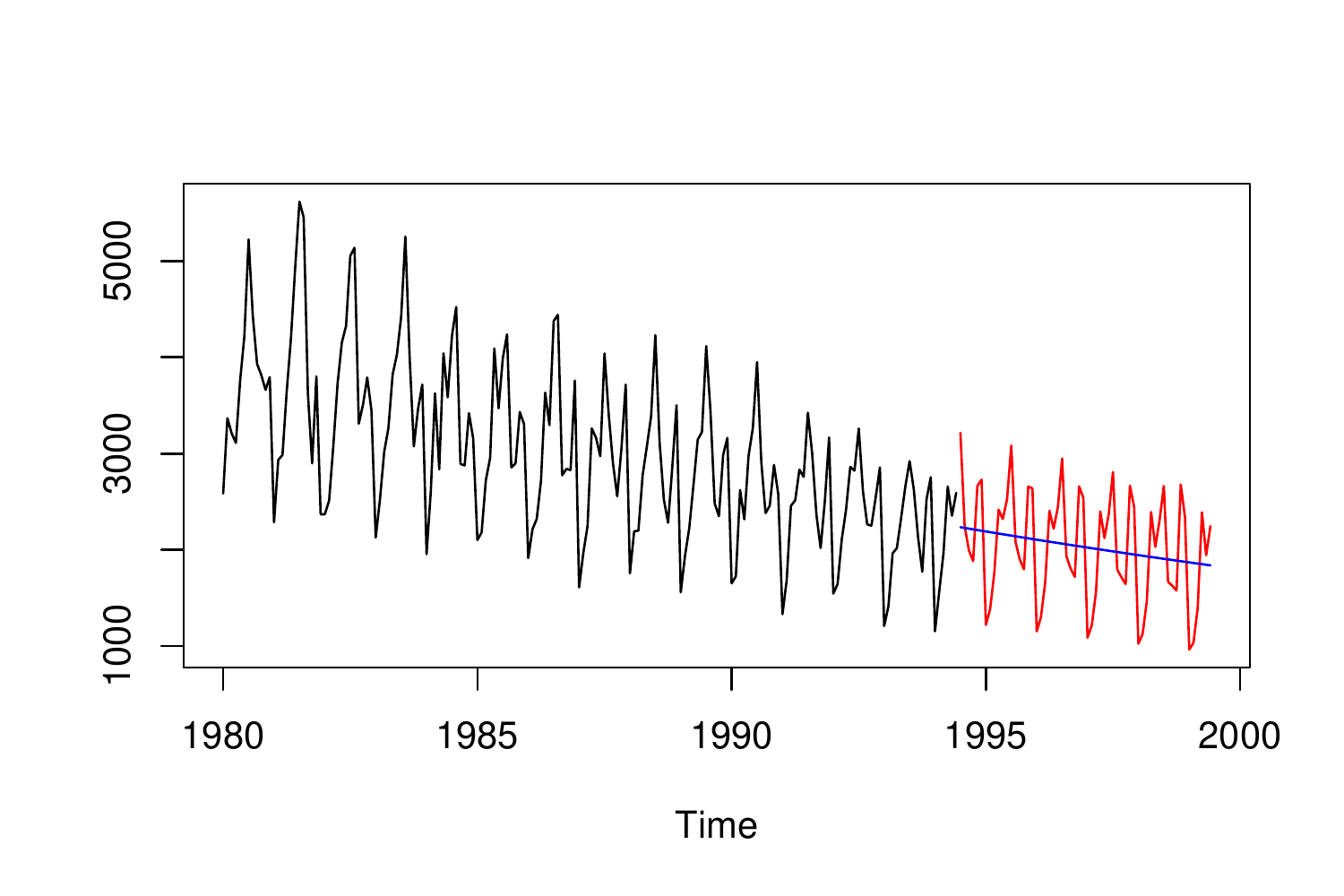}
        \end{center}
        \caption{FORT: Forecasts of trend and signal.}
        \label{fig:fort_forecast}
 \efg

\subsection[Comments on Rssa]{Comments on \pkg{Rssa}}
The detailed description of the \pkg{Rssa} package structure and the principles
of implementation is contained in \citet{Golyandina.Korobeynikov2013}. It is
related to the use of \pkg{Rssa} for the SSA processing of one-dimensional time
series, but most of the given information is also valid for other considered
types of data.  The typical code demonstrates the main logic of the SSA
processing by \pkg{Rssa} and thereby the structure of functions from the
package.  Below we briefly comment on the most important issues and main
differences with the one-dimensional case. More details are
contained in the corresponding sections as comments to the typical code or the
package.

\subsubsection{Formats of input and output data}
The inputs of SSA can be quite different depending on the kind of SSA used. For
example, the inputs can be vectors, \code{ts} objects, matrices, data frames, lists of
vector-like objects of different lengths, images in the form of matrices.
For Shaped 2D-SSA, the images can contain missing \code{NA}
values. For MSSA the missing values can be used to construct the series of
different lengths.

The routines in the packages are designed in a such way that they preserve all
the attributes (shape, time scale, etc.) of the input object. So, the result of
the Reconstruction step is exactly of the same type  as the
input object was. This can be seen in Figure~\ref{fig:fort_recsine}, where \code{plot.ts}
(corresponding to the default \code{plot.method = 'native'}) was used to draw
the result of the Reconstruction step.

All the forecasting routines try to use the attributes of the input object
for the resulting object (in particular, they try to add the time scale to the
result). Unfortunately, this cannot be done in a class-neutral way, as it is done in
the reconstruction case, and needs to be handled separately for each possible
type of the time series classes. The forecasting routines know how to impute the
time indices for some standard time series classes like \code{ts}, \code{mts}.

\subsubsection{Plotting specifics}
The package implements most of its own plotting capabilities with the help of
the \pkg{lattice} package \citep{Sarkar2008}; thus, the majority of the plotting
specifics comes from \pkg{lattice}. In particular, the results of plotting
functions of \pkg{Rssa} are proper \code{trellis} objects and can be modified,
combined and otherwise altered using the standard \pkg{lattice} functionality.

A very convenient way of plotting the reconstructed series is via
\code{'xyplot'} method of \code{plot} for SSA reconstruction objects. There are
powerful specializations of \code{xyplot} for \code{ts} and \code{zoo}
objects and \pkg{Rssa} provides stub implementation of
\code{xyplot} for bare matrices.

\subsubsection{SVD method}
\pkg{Rssa} allows one to use several SVD implementations: either full
decompositions implemented via \proglang{R} functions \code{eigen} and
\code{svd}, or truncated Lanczos SVDs from the package \pkg{svd}
\citep{svd2013}. These implementations differ in terms of speed, memory
consumption and numerical stability, see \citep{Golyandina.Korobeynikov2013} for
discussion concerning these properties. Here we note that the use of fast SVD
methods is the key point in the SSA processing of images.

By default, the \code{ssa} routine tries to select the best SVD implementation
given the series length, window length and the desired number of
eigentriples. This corresponds to the selection of \code{'auto'} for
\code{svd.method} argument. However, one can override this default setting and
select the desired SVD implementation, if necessary. This differs from the
behavior of previous versions of \pkg{Rssa}, when the package tried to 'fix' the
SVD method settings, when it thought it would be bad idea to proceed. The
present and next package versions (as of \pkg{Rssa} 0.10 and later on) will
always tolerate explicit user choice.

\subsubsection{Efficient implementation}
Complementary to the use of efficient SVD methods, the considerable speed-up is
achieved by means of special methods of actions with Hankel-related matrices,
see Section~\ref{sec:implementation} for the algorithms description.  This is
important, since multiplication by a Hankel matrix and and hankelization of
matrices, which can be very large, take a substantial part of the SSA
algorithms. The efficient implementation of these routines relies on the
possibility to compute the fast Fourier transform of a vector of arbitrary
length with the optimal $O(N \log{N})$ complexity. In order to achieve this,
\pkg{Rssa} uses \texttt{FFTW} library \citep{Frigo.Johnson2005}. While we
strongly encourage to always complement \pkg{Rssa} installation with
\texttt{FFTW}, the package will fallback to \proglang{R}'s
FFT implementation if the package was compiled without \texttt{FFTW}.
Pre-built Windows and Mac packages on CRAN are statically linked
with \texttt{FFTW}; for other platforms it is usually possible to install the
library using the standard package manager. (Note that a \texttt{-dev} version of
the \texttt{FFTW} package is usually required.) If \texttt{FFTW} is not installed,
only inherently one-dimensional analysis (1D-SSA, Toeplitz SSA and Complex
SSA) will be available. The computational speed will be slower in this case too.

Following  \citet{Korobeynikov2010},
let us briefly describe the algorithm complexity for the one-dimensional case to
show the order of speed-up.  The direct
implementation of the SSA algorithms has $O(N^{3})$ computational and space
complexity in the worst case $L \sim K \sim N/2$ (this case is standard for
SSA), where $N$ is the series length.  Therefore, it is important to provide
efficient implementations which makes non-trivial cases feasible.

\begin{itemize}
    \item The methods of an efficient Hankel
    matrix-vector multiplication by the means of the fast Fourier transform (see
    Section~\ref{ssec:hmatmul}) and the usage of fast Lanczos-based
    truncated SVD implementations drop the complexity from $O(N^{3})$
    down to $O(k N \log{N} + k^{2}N)$, where $k$ is the number of calculated eigentriples.
    \item Second, one can represent the computation of the elementary series,
    which is \emph{rank 1 hankelization}, as a special form of convolution. This
    way, the efficient implementation of the hankelization (diagonal averaging)
    procedure is again possible via the fast Fourier transform and has similar
    improvement of complexity.
    \item Third, it is possible to implement the vector forecast much more
    efficiently than using the direct implementation. The details can be found in
    Section~\ref{sec:implement_forecast}.
\end{itemize}

Note that the principle of automatic calculation of necessary objects is used in
the implementation of the package. For example, if 10 eigentriples were
calculated while decomposing, then the user could still perform reconstruction
by the first 15 components, since the decomposition will be automatically
continued to calculate 11--15 eigentriples. Also, the routines reuse the results
of the previous calculations as much as possible in order to save time (hence
the argument \code{cache} of many routines). For example, the elementary series
once calculated are stored inside the SSA object, so next time the function
\code{reconstruct} might not need to calculate the resulting series from
scratch.

All these speed-up improvements make it possible to perform in reasonable time
the tasks of analysis of long series, image processing,
tracking procedures and batch processing.


%% file: common_scheme.tex
\begin{tikzpicture}[scale=1, auto]
    \small
     \quad 
\tikzstyle{block} = [draw,rectangle,thick,minimum width = 3cm, minimum height=2cm, anchor=west]
\tikzstyle{sum} = [draw,circle,inner sep=0mm,minimum size=2mm]
\tikzstyle{connector} = [->,thick]
\tikzstyle{line} = [thick]     

    \draw (0,0) node[block] (Data)	{$\begin{matrix}
                                      \tX\quad \mbox{---} \\
                                      \mbox{time series,} \\
                                      \mbox{system of t.s.,} \\
                                      \mbox{array (image), ...} 
                                      \end{matrix}$ };
    \draw (Data.north) node[anchor=south] {Data};

    \draw (6,0) node[block] (Traj) {$\begin{matrix}
                                     \bfX \in \calM_{L,K}^{(H)}: \\
                                     \bfX = \calT(\tX) 
                                     \end{matrix}$ };
    \draw (Traj.north) node[anchor=south] {Trajectory matrix};

    \draw (12,0) node[block] (Decomp)	{$\begin{matrix}
                                        \bfX = \suml_{j=1}^d \bfX_j  \\
                                        \bfX_j = \sqrt{\lambda_j} U_j V_j^{\top}
                                        \end{matrix}$};
    \draw (Decomp.north) node[anchor=south] {Sum of rank-$1$ matrices};

    \draw [connector] ([xshift=0.3cm] Data.east) -- node {1. Embedding} 
                      ([xshift=-0.3cm] Traj.west);
    \draw [connector] ([xshift=0.3cm] Traj.east) -- node {2. Decomposition} 
                      ([xshift=-0.3cm] Decomp.west);

    \draw (6.8,-4) node[block,minimum width = 3.5cm] (Grouped)	
                                   {$\begin{matrix}
                                     \bfX = \bfX_{I_1}+\cdots+\bfX_{I_m} \\
                                     \bfX_{I} = \suml_{j\in I} \bfX_j
                                     \end{matrix}$ };
    \draw (Grouped.north) node[anchor=south] {Grouped matrices};
    \draw [connector] ([yshift=-0.3cm] Decomp.south) -- node[above,sloped] {3. Grouping} ([xshift=0.3cm] Grouped.east);

    \draw (0,-4) node[block, minimum width = 3.5cm] (Reconstr)	{$\begin{matrix}
                          \tX = \widetilde\tX_1+\ldots+\widetilde\tX_m, \\
                          \widetilde\tX_k = \calT^{-1}\circ \Pi^{(H)} (\bfX_{I_k}) \end{matrix}$ };
    \draw (Reconstr.north) node[anchor=south] {SSA decomposition};
    
    \draw [connector] ([xshift=-0.3cm] Grouped.west) -- node[above,sloped] {4. Reconstruction} 
([xshift=0.3cm] Reconstr.east);
 
\end{tikzpicture}        

%% file: fragments/wines_input.tex
\begin{CodeChunk}
\begin{CodeInput}

> library("Rssa")
> data("AustralianWine")
> wine <- window(AustralianWine, end = time(AustralianWine)[174])
\end{CodeInput}

\end{CodeChunk}

%% file: fragments/fort_rec.tex
\begin{CodeChunk}
\begin{CodeInput}

> fort <- wine[, "Fortified"]
> s.fort <- ssa(fort, L = 84, kind = "1d-ssa")
> r.fort <- reconstruct(s.fort, groups = list(Trend = 1,
+                                             Seasonality = 2:11))
> plot(r.fort, add.residuals = TRUE, add.original = TRUE,
+      plot.method = "xyplot",
+      superpose = TRUE, auto.key = list(columns = 2))
\end{CodeInput}

\end{CodeChunk}

%% file: fragments/fort_identific.tex
\begin{CodeChunk}
\begin{CodeInput}

> plot(s.fort, type = "vectors", idx = 1:8)
> plot(s.fort, type = "paired", idx = 2:11, plot.contrib = FALSE)
> parestimate(s.fort, groups = list(2:3, 4:5), method = "esprit-ls")
$F1
   period     rate   |    Mod     Arg  |     Re        Im
   12.003  -0.006572 |  0.99345   0.52 |  0.86042   0.49661
  -12.003  -0.006572 |  0.99345  -0.52 |  0.86042  -0.49661
$F2
   period     rate   |    Mod     Arg  |     Re        Im
    4.005   0.000037 |  1.00004   1.57 |  0.00189   1.00003
   -4.005   0.000037 |  1.00004  -1.57 |  0.00189  -1.00003
> plot(wcor(s.fort, groups = 1:30),
+           scales = list(at = c(10, 20, 30)))
> plot(reconstruct(s.fort, add.residuals = FALSE, add.original = FALSE,
+                  groups = list(G12 = 2:3, G4 = 4:5, G6 = 6:7, G2.4 = 8:9)))
\end{CodeInput}

\end{CodeChunk}

%% file: fragments/fort_forecast.tex
\begin{CodeChunk}
\begin{CodeInput}

> f.fort <- vforecast(s.fort,
+                     groups = list(Trend = 1, Signal = 1:11),
+                     len = 60, only.new = TRUE)
> plot(cbind(fort, f.fort$Signal, f.fort$Trend),
+      plot.type = "single", col = c("black", "red", "blue"), ylab = NULL)
\end{CodeInput}

\end{CodeChunk}

%% file: multi_algorithm.tex
Let us consider the problem of simultaneous decomposition, reconstruction and
forecasting for a collection of time series from the viewpoint of SSA.
The method is called Multichannel SSA or Multivariate SSA, shortened to MSSA.  The
main idea of the algorithm is the same as in the Basic SSA, the difference
consists in the way of construction of the trajectory matrix.

In a sense, MSSA is a straightforward extension of SSA.  However, the algorithm
of MSSA was published even earlier than the algorithm of SSA; see
\citet{Weare.Nasstrom1982}, where the MSSA algorithm was named Extended
Empirical Orthogonal Function (EEOF) analysis. Formally, the algorithm of MSSA
in the framework of SSA was formulated in \citet{Broomhead.King1986b}.

Here we consider the algorithm of MSSA for analysis and forecasting of
multivariate time series following the approach described in \citet[Chapter
2]{Golyandina.etal2001} for one-dimensional series and in
\citet{Golyandina.Stepanov2005} for multidimensional ones.
In this section, we also describe
the complex-valued version of SSA (called CSSA), since it can be considered as a
multidimensional version of SSA for analysis and forecasting of a system of two time
series.

The theory studying the ranks of the multivariate time series and the
separability of their components for MSSA and CSSA is very similar to that of SSA and is briefly
described in Section~\ref{sec:mssa_theory}.  Let us start with the algorithm
description.

\subsection{MSSA and CSSA algorithms}
\label{SEC:AlgoMandC}

\subsubsection{CSSA analysis}
Let the system of series consist of two series, that is, $s=2$. Then we can consider
one-dimensional complex-valued series $\tX=\tX^{(1)}+ \unit
\tX^{(2)}$ and apply the complex version of SSA to this
one-dimensional series.  Since the general algorithm in Section~\ref{sec:common}
is written down in real-valued form, there is the difference in the form of the
SVD performed in the complex-valued space, where the transposition should be
Hermitian.

Also, there is a difference regarding the uniqueness of the SVD expansion.  If
the singular values are different, then the SVD is unique up to multiplication of
left and right singular vectors by $c$, where $|c|=1$. In the real-valued case,
$c=\pm 1$, while in the complex-valued case there are a lot of suitable
constants $c$.

\subsubsection{MSSA analysis}
Consider a multivariate time series, that is, a collection
$\{\tX^{(p)}=\big(x_j^{(p)}\big)_{j=1}^{N_p},\;\; p=1, \ldots, s\}$ of $s$ time
series of length $N_p$, $p=1,\ldots,s$.

Denote $\tX=(\tX^{(1)},\ldots, \tX^{(s)})$ the initial data for the MSSA
algorithm.  Since the general scheme of the algorithm described in
Section~\ref{sec:common} holds for MSSA, we need to define the embedding
operator $\calT_{\mathrm{MSSA}}(\tX)=\bfX$ only.

Let $L$ be an integer called window length, $1<L<\min(N_p, p=1,\ldots,s)$.  For
each time series $\tX^{(p)}$, the embedding procedure forms $K_p =N_p-L+1\,\,$
$L$-lagged vectors $X_j^{(p)}=(x_{j}^{(p)},\ldots,x_{j+L-1}^{(p)})^\top$, $1\leq
j\leq K_p$. Denote $K=\sum_{p=1}^s K_p$.  The {\em trajectory matrix} of the
multidimensional series $\tX$ is the $L\times K$ matrix of the form
\be
\label{eq:mssa_embedding}
\calT_{\mathrm{MSSA}}(\tX)={\bf X}
=[X_1^{(1)}:\ldots:X_{K_1}^{(1)}:\ldots:X_1^{(s)}:\ldots:X_{K_s}^{(s)}]
=[{\bf X}^{(1)}:\ldots:{\bf X}^{(s)}],
\ee
where $\bfX^{(p)}=\calT_{\mathrm{SSA}}(\tX^{(p)})$ is the trajectory matrix of
the one-dimensional series $\tX^{(p)}$ defined in
(\ref{eq:ssa_embedding}). Thus, the trajectory matrix of a system of time
series has \emph{stacked Hankel} structure.

The eigenvectors $\{U_i\}$ in the SVD (\ref{eq:mmexp}) of $\bfX$ form the common
basis of the column trajectory spaces of all time series from the system.
Factor vectors $V_i$ (named EEOF in climatology applications) consist of parts
related to each time series separately, that is,
\begin{equation}
    V_i=\left(
        \begin{array}{c}
            V_i^{(1)}\\
            \vdots\\
            V_i^{(s)}
        \end{array}
    \right),
\end{equation}
where $V_i^{(p)} \in \spaceR^{K_p}$ and belong to the row trajectory spaces of the $p$th series.

The eigenvectors $U_i$ reflect the common features of time series, while the factor
subvectors $V_i^{(1)}$ show how these common features appear in each series. It
is natural to transform a factor vector to a \emph{factor system} of factor
subvectors $V_i^{(p)}$. Then the form of transformed factor vectors will be
similar to the initial system of series.

Analogously to one-dimensional case, the main result of application of MSSA is
the decomposition \eqref{eq:res_decomp} of the multivariate time series $\tX$
into a sum of $m$ multivariate series. 

\subsubsection{Remarks}

\begin{enumerate}
    \item Note that the indexing of time points $1,\ldots,N_p$
    ($p=1,\ldots,s$) starting from 1 does not mean that the series start at the
    same time and can finish at different times if lengths of time series are
    different.  The resultant decomposition obtained by the MSSA algorithm does
    not depend on the shift between series and therefore this numeration is just
    formal. Even more, decompositions of two series measured at the same time
    and in disjoint time intervals do not differ.

    \item The original time ranges of series $\tX^{(p)}$ can be useful
    for depicting and interpreting them. Certainly, the reconstructed series
    have the same time ranges as the original ones.  Factor subvectors from the
    factor system can also be synchronized for plotting based on the ranges of the
    initial series. Although factor vectors are shorter than the initial series, their
    time shifts are the same.

    \item For the SSA analysis of one time series, it makes sense to consider
    window lengths $2\le L\le \lfloor (N+1)/2 \rfloor$, since the SVD expansions for window
    lengths $L$ and $N-L+1$ coincide. For the MSSA-analysis of more than one
    time series the expansions for all possible window lengths $2\le L\le \min_p N_p-1$
    are generally different.

    \item For simultaneous analysis of several time series, it is recommended
    to transfer them into the same scale. Otherwise, the structure of one time
    series will overweigh the results.  To balance the time series, they can be
    either standardized (centered and normalized; in additive models) or only
    normalized (in multiplicative models).  On the other hand, the scale of
    series can be used instead of their weights in the common decomposition if,
    e.g., one of the series is more important or has smaller noise level.

    \item In a sense, the most detailed decomposition can be obtained if the
    trajectory matrix $\bfX$ has maximal rank. In the general case of arbitrary
    time series, this corresponds to the case of square matrix. Thus, for a
    system of $s$ time series of length $N$ the window length providing the
    square (or the closest to square) trajectory matrix $\bfX$ is approximately 
    $s(N+1)/(s+1)$ for MSSA.  In the case of two time series,
    this corresponds to $2(N+1)/3$ for MSSA, while CSSA gives $(N+1)/2$.

    \item The MSSA algorithm might be modified by the same ways as the SSA one;
    for example, Toeplitz MSSA, MSSA with centering can be considered. However, these options
    are not implemented in the described version of \pkg{Rssa}.
\end{enumerate}

%% file: multi_theory.tex
\subsection{Comments to the algorithms}
\label{sec:comments_mssa}
\subsubsection{Covariance structure}
Consider in more detail the case of two time series $\tX=(\tF,\tG)$ and let
$\bfF$ and $\bfG$ be the trajectory matrices of $\tF$ and $\tG$ correspondingly.
Then MSSA considers the eigendecomposition of ${\bf S}={\bf X}{\bf X}^\top={\bf
    F}{\bf F}^\top+{\bf G}{\bf G}^\top$, that is, MSSA analyzes the averaging
structure of two time series.

Since the SVD of a matrix coincides with the SVD of its transpose,
we can consider the decomposition of the time
series trajectory matrices on the basis of right singular vectors or, equivalently,
to transpose $\bfX$ and consider eigenvectors of
\bea
{\bf S}=\left(
\begin{matrix}
\bfF^\mathrm{T} \bfF & \bfF^\mathrm{T} \bfG\\
\bfG^\mathrm{T} \bfF & \bfG^\mathrm{T} \bfG
\end{matrix}
\right)
\eea
called EEOFs.
The last formula demonstrates more clearly that MSSA takes into consideration
cross-covariances of time series (more precisely, cross-covariances are
considered if centering is used).

Since the eigendecomposition of a complex-valued matrix $\bfA+\unit\bfB$
can be reduced to the eigendecomposition of the real-valued matrix
    \bea
    {\bf D}=
    \left(
    \begin{array}{cr}
     \bf A&\bf -B\cr
     \bf B&\bf A\cr
    \end{array}
    \right),
    \eea
in the case of CSSA we in fact analyze eigenvectors of the matrix
\bea
{\bf S}=
\left(
\begin{matrix}
\bfF^\mathrm{T} \bfF & \bfF^\mathrm{T} \bfG\\
\bfG^\mathrm{T} \bfF & \bfG^\mathrm{T} \bfG
\end{matrix}
\right)+
\left(
\begin{matrix}
\bfG^\mathrm{T} \bfG & -\bfG^\mathrm{T} \bfF\\
-\bfF^\mathrm{T} \bfG & \bfF^\mathrm{T} \bfF
\end{matrix}
\right),
\eea
that is, structures of the time series are mixed in greater degree in comparison with MSSA.

\subsubsection{Matching of series}
Simultaneous analysis of several time series is usually performed to identify
their relation and extract the common structure. Recall that the structure in
SSA means that the trajectory matrix is rank-deficient.  Certainly, for
real-world series, the trajectory matrix is of full rank, since at least noise
has no structure.  Therefore, in what follows we say about rank of signal or its components.

Consider the system of series with rank-deficient trajectory matrix.  The
structure of the series is reflected by its trajectory space.  Therefore, we can
say that two time series have the same structure if their trajectory spaces
coincide. For example, the trajectory spaces of two sine waves with equal periods 
coincide irrespective of the amplitudes and phases. This fact is evident,
since the trajectory space is the span of subseries of length $L$ of the initial
series.  To the contrary, sine waves with different frequencies have totally
different structure and their combined trajectory space is the direct sum of
the trajectory spaces of individual time series.

If two time series are fully matched, then the trajectory space of one time
series can be used for reconstruction or forecasting of the second series.  If
the series are totally different, then the first series is useless for the
analysis of the second one.

For MSSA, the shift between time series, for example,
the difference between phases of two matched sine waves, is of no consequence. Therefore, we cannot
say anything about direction of causality (if any; see \citet{Hassani.etal2013}, where the
attempt to detect causality is performed).  Moreover, asymmetry of influence
of one time series to the another series can be caused by different levels of
noise.

\subsubsection{Relation between SSA, MSSA and CSSA ranks}

For MSSA and CSSA, the notions of time series of finite rank and of time series
satisfying linear recurrence relations are analogous to these notions in
SSA. However ranks of the same time series can differ depending on the applied
method. Therefore, we can think about SSA-, MSSA- and CSSA-ranks.

If the individual time
series have the same structure (and therefore the same SSA-ranks), then the
MSSA-rank is equal to the SSA-rank of each time series. As for CSSA-rank, it can
be even less than all the individual SSA-ranks for some special cases.

Consider a collection $\tH^{(p)}=\big(h_j^{(p)}\big)_{j=1}^{N}$, $p=1, \ldots,
s$, of $s$ signals of length $N$.  Let $r_p$ denote the SSA rank of
$\tH^{(p)}$ (i.e., dimension of the trajectory spaces generated by
one-dimensional SSA applied to this time series) and $r$ denote the MSSA rank of
$(\tH^{(1)},\ldots,\tH^{(s)})$.
The relation between $r$ and $r_p$, $p=1, \ldots, s$, is considered in
Section~\ref{sec:mssa_theory}.  In particular, it is shown that $r_{\rm min}\le
r \le r_{\rm max}$, where $r_{\rm min}=\max\{r_p, p=1, \ldots, s\}$ and $r_{\rm
    max}=\sum_{p=1}^s r_p$.  The case $r = r_{\rm max}$ is the least favorable
for MSSA and means that different time series do not have matched components.
The case $r < r_{\rm max}$ indicates the presence of matched components and can
lead to advantages of simultaneous processing of the time series system.

In terms of matching, if all the series have the same characteristic roots (see
Section~\ref{sec:mssa_theory} for definition), then this means that the time
series $\tX^{(p)}$, $p=1,\ldots,s$, consist of additive components of the same
SSA-structure.  Such time series are {fully matched}. For fully matched time
series the MSSA-rank is much smaller than the sum of the SSA-ranks of the
separate time series from the system.  On the contrary, if the sets of
characteristic roots do not intersect, then the time series have no common
structure.  In this case, the MSSA-rank is exactly equal to the sum of the
SSA-ranks of the separate time series from the system.  For the real-world time
series, we are usually between these extreme cases.

\subsubsection{Separability}

The notion of separability for multidimensional time series is analogous to that for
one-dimensional series briefly commented in Section~\ref{sec:general} and
thoroughly described in \citet[Sections 1.5 and 6.1]{Golyandina.etal2001}.
Section~\ref{sec:mssa_theory} contains several results on separability of
multidimensional series together with definitions of weak and strong separability.  
Generally, conditions of separability of
multidimensional time series are more restrictive than that for one-dimensional
series.  In particular, the sufficient condition for separability of two series
systems is the separability of each series from one collection with each series
from the second one.
However, for matched signals, their (weak) separability from the noise can be
considerably improved by their simultaneous MSSA analysis.

Since weak separability is not enough for extraction of time series
components, we should pay attention to strong separability related to eigenvalues
produced by time series components. It appears (see Example~\ref{ex:mssa_ev} in
Section~\ref{sec:mssa_theory}) that the time series $(\tF^{(1)}, \tF^{(2)})$
can produce different eigenvalues in SSA, MSSA and CSSA.  Therefore, the
application of an appropriate multidimensional modification of SSA can improve
the strong separability. Again, matching of time series diminishes the number of
eigenvalues related to the signal and thereby weakens the chance of mixing of
the signal with the residual.

\subsubsection{Choice of window length}

Recommendations on the choice of window length for one-dimensional SSA analysis
can be found, e.g., in \citet[Section 1.6]{Golyandina.etal2001} and
\citet{Golyandina2010}.  However, the problem of the choice of window length in
MSSA is more complicated than that for SSA. To the best of the authors' knowledge, there is no
appropriate investigation of the choice of optimal window length for analysis
and, to a greater extent, forecasting of multidimensional time series.
Moreover, the choice of the best window length for MSSA forecasting differs for
different types of forecasting methods, see numerical comparison in Section~\ref{sec:examples_mssa}.
Section~\ref{sec:examples_mssa}, in particular, contains an extension of
the numerical investigation done in \citet{Golyandina.Stepanov2005}.

By analogy with the one-dimensional case, we can formulate some principles for the choice of $L$.
The main principle is the same as for SSA and states that the choice of $L$ should provide
(approximate) separability of series. However, the MSSA case has additional features.
For example, while in SSA analysis it makes no sense to take
$L>(N+1)/2$, in MSSA analysis large $L$ with small $K_p=N_p-L+1$ can be taken
instead of small $L$ for trend extraction and smoothing.

Various special techniques can be transferred from SSA to MSSA, such as Sequential SSA. 
Sequential SSA is based on successive application of SSA with different window lengths, see \citet[Section 2.5.5]{Golyandina.Zhigljavsky2012} for more details. For example, if trends are of complex or of different
structure, a smaller window length can be applied to achieve the similarity
of eigenvectors and to improve separability.  Then the residual can be decomposed
with a larger window length. For multidimensional time series, Sequential MSSA can be applied
by analogy with Sequential SSA. 

Two alternatives to Sequential SSA are described in
\citet{Golyandina.Shlemov2013} and are implemented for one-dimensional time
series in the version 0.11 of the \pkg{Rssa} package.  However, these approaches
can be extended to multivariate and multidimensional cases.

Since $L$ in SSA does not exceed the half of the time series length, the
divisibility of $L=\min(L,K)$ on possible periods of oscillations is recommended
in SSA. In MSSA, $\min(L,K_p)$ is not necessary equal to $L$ and therefore one
puts attention on values of $K_p$.

Numerical investigations show that the choice $L=\lfloor sN/(s+1) \rfloor$ is appropriate for
the decomposition of several (a few) time series (see simulation results in
Section~\ref{sec:examples_mssa}), but evidently cannot be applied for the
system of many short series ($K_p$ become too small for separability).
In general, the choice $L=\lfloor N/2\rfloor$ is still appropriate even for multivariate SSA.

\subsection{Multivariate SSA forecasting}
\label{SEC:continuation}
Forecasting in SSA is performed for a time series component which can be separated by
SSA and is described (maybe, approximately) by a linear recurrent relation (LRR).
For brevity, we will say
about forecasting of a signal. If the forecasted series component (signal) is
governed by an LRR, then it can be forecasted by this or extended LRR.  

Without loss of generality, we assume that the first component $\wtilde{\tX}_1$ of the decomposition
\eqref{eq:res_decomp} is forecasted. In order to simplify the notation,
we denote the reconstructed series by 
$\wtilde{\tX} = \wtilde{\tX}_1 = (\wtilde{\tX}^{(1)}, \ldots, \wtilde{\tX}^{(s)})$,
where $\wtilde{\tX}^{(p)}=\big(\tilde{x}_j^{(p)}\big)_{j=1}^{N_p}$, $p=1, \ldots, s$.
The methods of SSA forecasting aim at obtaining the forecasted points $\tilde{x}_j^{(p)}$
for $j > N_p$, based on the reconstructed time series $\wtilde{\tX}$ and the corresponding
reconstructed matrix $\what{\bfX} = \what{\bfX}_1$. The forecasting is called 
\emph{$M$-step ahead} if the points $(\tilde{x}_{N_p+1}^{(p)},\ldots,\tilde{x}_{N_p+M}^{(p)})$, $p=1,\ldots,s$ are being obtained.

There are two main methods of SSA forecasting: recurrent and vector. In \emph{recurrent}
forecasting, an estimated LRR is applied to the the reconstructed series $\wtilde{\tX}$
in order to obtain a $1$-step ahead forecast. The $M$-step ahead forecast
is obtained by recurrence (by applying $M$ times  the $1$-step ahead forecasting with
the same LRR), which explains the name of the method. 
In \emph{vector} forecasting, the reconstructed matrix $\what{\bfX}$ is first 
extended, by continuation of the reconstructed lagged vectors in a given subspace (row or column 
space of  $\what{\bfX}$). The forecasted points are then obtained by diagonal averaging of 
the extended reconstructed matrix. 

Recurrent and vector methods of one-dimensional SSA forecasting ($s=1$) are  
fully described in \citet[Chapter 2]{Golyandina.etal2001}.
For CSSA, the forecasting algorithms are straightforward extensions of SSA forecasting
algorithms to the complex-valued case, therefore we do not discuss them here.
The methods of MSSA forecasting, however, need special attention.

As in SSA, methods of MSSA forecasting can be subdivided into recurrent and vector forecasting. 
In contrast with SSA, rows and columns of the trajectory matrix in MSSA have different structure. Therefore, there exist two kinds of MSSA forecasting: row forecasting and column forecasting, depending on which space is used (row or column space respectively). In total, there are four variants of MSSA forecasting: recurrent column forecasting, recurrent row forecasting, vector column forecasting and vector row forecasting.

In the column forecasting methods, each time series in the system is forecasted separately, but in a given common subspace (i.e., using the common LRR). In the row forecasting methods, each series is forecasted with the help of its own LRR applied to the whole set of series from the system. Next, we describe all the variants of MSSA forecasting in detail.


\subsubsection{Common notation}
First, we introduce some common notation used for description of all the variants of MSSA forecasting.

For a vector $A\in \spaceR^Q$, we denote by $\first{A}\in \spaceR^{Q-1}$ the vector of
the last $Q-1$ coordinates and by $\last{A}\in\spaceR^{Q-1}$ the vectors of the first $Q-1$
coordinates. The line on the top (respectively, on the bottom) indicates that the the first
(respectively, the last) coordinate is removed from the vector $A$. Also, denote by $\pi(A)$
the last coordinate of the vector.  For a matrix $\bfA=[A_1:\ldots:A_r]$ denote
$\first\bfA=[\first{A}_1:\ldots:\first{A}_r]$ and
$\last\bfA=[\last{A}_1:\ldots:\last{A}_r]$ and let
$\bfpi(\bfA)=(\pi(A_1),\ldots,\pi(A_r))^\top$ be the last row of the matrix
$\bfA$.

For a vector $B\in \spaceR^K$, where $K=\sum_{p=1}^s K_p$, we use the following notation:
\be
    B=\left(
    \begin{array}{c}
    B^{(1)}\\
    B^{(2)}\\
    \vdots\\
    B^{(s)}
    \end{array}
    \right),
    \quad
    \llast{B}=
    \left(
    \begin{array}{c}
    \last{B}^{(1)}\\
    \last{B}^{(2)}\\
    \vdots\\
    \last{B}^{(s)}
    \end{array}
    \right),
    \quad
    \ffirst{B}=
    \left(
    \begin{array}{c}
    \first{B}^{(1)}\\
    \first{B}^{(2)}\\
    \vdots\\
    \first{B}^{(s)}
    \end{array}
    \right),
    \quad
    \bfmu(B)=
    \left(
    \begin{array}{c}
    \pi({B}^{(1)})\\
    \pi({B}^{(2)})\\
    \vdots\\
    \pi({B}^{(s)})
    \end{array}
    \right).
\ee
where $B^{(p)}\in \spaceR^{K_p}$. For $\bfB=[B_1:\ldots:B_r]$, let $\llast{\bfB}=[\llast{B}_1:\ldots:\llast{B}_r]$ and
$\bfB^{(p)}=[B_1^{(p)}:\ldots:B_r^{(p)}]$.

For simplicity we assume that the set $I =I_1$ corresponding to the forecasted component
is given by the set of the leading components; that is, $I=I_1=\{1,\ldots,r\}$.
(This is made just for simplification of formulas.)
Thus, let $r$ leading eigentriples $(\sqrt{\lm_j},U_j,V_j)$ be identified and chosen
as related to the signal of rank $r$, and denote
$\bfU=[U_1:\ldots:U_r]$, $\bfV=[V_1:\ldots:V_r]$.

The reconstructed series
$\wtilde{\tX}$, its trajectory matrix $\wtilde{\bfX}$ and the reconstructed
matrix $\what{\bfX}$ are defined in Section~\ref{sec:common_alg}.  Define
$\cL^{\mathrm{col}} =\sspan({U_i}, i\in I)$, ${\cL^{\mathrm{row}}}
=\sspan({V_i}, i\in I)$.  The reconstructed matrix
$\widehat{\bfX}=[\widehat{X}_1^{(1)}:\ldots:\widehat{X}_{K_1}^{(1)}:\ldots:\widehat{X}_1^{(s)}:\ldots:\widehat{X}_{K_s}^{(s)}]$ consists of column
vectors that are projections of column vectors of the trajectory matrix \eqref{eq:mssa_embedding}
on the chosen subspace $\cL^{\mathrm{col}}$. For convenience, we also denote by $\widehat{Y}_i$ the 
$i$th row of the matrix $\widehat{\bfX}$, such that $\widehat{\bfX}^\top=[\widehat{Y}_1:\ldots:\widehat{Y}_L]$.


\subsubsection{Recurrent MSSA forecast}

Denote by $R_N=\big(\tilde{x}^{(1)}_{N_1+1}, \tilde{x}^{(2)}_{N_2+1},
\ldots,\tilde{x}^{(s)}_{N_s+1}\big)^\top$ the vector of forecasted signal values
for each time series ($1$-step ahead forecast).  Recurrent forecasting is closely related to missing data
imputation for components of vectors from the given subspace and in fact uses
the formula (1) from \citet{Golyandina.Osipov2007}.  Following
\citet{Golyandina.Stepanov2005}, we can write out forecasting formulas for two
versions of the recurrent MSSA forecast: row (generated by $\{U_j\}_{j=1}^r$)
and column (generated by $\{V_j\}_{j=1}^r$). These one-term ahead forecasting
formulas can be applied for $M$-term ahead forecast by recurrence.

The column recurrent forecasting performs forecast by an LRR of order $L-1$,
applied to the last $L-1$ points of the reconstructed signal, that is, one LRR
and different initial data.  The row recurrent forecasting constructs $s$
different linear relations, each is applied to the set of $K_i-1$ last points of
series, that is, LRRs are different, but initial data for them are the same.

\paragraph{Column forecast}
Denote by $\bfZ$ the matrix consisting of the last $L-1$ values of the reconstructed signals:
\bea
    \bfZ=\left(
    \begin{array}{ccc}
    \tilde{x}_{N_1-L+2}^{(1)}&\ldots& \tilde{x}_{N_1}^{(1)}\\
    \tilde{x}_{N_2-L+2}^{(2)}&\ldots& \tilde{x}_{N_2}^{(2)}\\
    \vdots&\vdots&\vdots\\
    \tilde{x}_{N_s-L+2}^{(s)}&\ldots& \tilde{x}_{N_s}^{(s)}
    \end{array}
    \right),
\eea
$\nu^2=\suml_{j=1}^r \pi(U_j)^2$.
If $\nu^2<1$, then the recurrent column forecast is uniquely defined  and can be calculated by the formula
\be
\label{EQ:Forec:ContMSSAL}
    R_N = \bfZ
{\cR}_{L}, \qquad \mbox{where} \qquad
{\cR}_{L}=\frac{1}{1-\nu^2}\suml_{j=1}^r\pi(U_j) \last{U_j}\;\;\in \spaceR^{L-1}.
\ee
Note that (\ref{EQ:Forec:ContMSSAL}) implies that the forecasting of
all individual series is made using the same LRR 
(with coefficients ${\cR}_{L}$), which is generated by the whole system of series.

\paragraph{Row forecast}
\label{ssec:KCont}

Introduce the vectors of the last $K_p-1$ values of the reconstructed signals
\bea
   Z^{(p)}=\big(\tilde{x}_{N-K_p+2}^{(p)},\ldots, \tilde{x}_{N_p}^{(p)}\big)^\top, \quad p=1,\ldots,s,
\eea
    and define a vector $Z \in {\spaceR}^{K - s}$ and an $s \times r$ matrix $\bfS$ as
\bea
    Z=\left(
    \begin{array}{c}
    Z^{(1)}\\
    Z^{(2)}\\
    \vdots\\
    Z^{(s)}
    \end{array}
    \right)
    ,\qquad
    \bfS = [\bfmu(V_1):\ldots:\bfmu(V_r)].
\eea

If the inverse matrix $(\bfI_{s}-\bfS\bfS^\top)^{-1}$ exists and $r\leq K - s$,
then the recurrent row forecast exists and can be calculated by the
formula
\be
\label{EQ:Forec:ContMSSAK}
     R_N=\bfcR_K Z,
\ee
where $\bfcR_K = (\bfI_{s}-\bfS\bfS^\top)^{-1}\,\bfS\llast{\bfV}^\top$.  Note
that (\ref{EQ:Forec:ContMSSAK}) implies that the forecasting of the
individual signals is made using the linear relations which are different for
different series. The forecasting value generally depends on the last values of
all the time series in the system of series.

\subsubsection{Vector MSSA forecasting}

Denote $\last{\cL}^{\mathrm{col}}=\sspan(\last{U}_1, \ldots, \last{U}_r)$ and
$\llast{\cL}^{\mathrm{row}}=\sspan(\llast{V}_1, \ldots, \llast{V}_r)$.  Let
$\Pi^{\mathrm{col}}$ be the orthogonal projector of $\spaceR^{L-1}$ on
$\last\cL^{\mathrm{col}}$ and $\Pi^{\mathrm{row}}$ be the orthogonal projector
of $\spaceR^{K-s}$ on $\llast\cL^{\mathrm{row}}$.

The rows of $\widehat\bfX$ are the projections of rows of the trajectory
matrix $\bfX$ on $\cL^{\mathrm{row}}$, while the columns of $\widehat\bfX$ are the
projections of columns of the trajectory matrix $\bfX$ on
$\cL^{\mathrm{col}}$.

The explicit form of the matrices of the column and row projectors can be found
in \citet[formula (4)]{Golyandina.Osipov2007}.  However, the calculation by this
formula is time-consuming. The fast algorithms for vector forecasting are presented in
Section~\ref{sec:implement_forecast}.

\paragraph{Column forecast}

We mentioned that for a given subspace ($\cL^{\mathrm{col}}$ in our case) the
column forecast is performed independently for each time series. Define the
linear operator $\cP_{\mathrm{Vec}}^{\mathrm{col}}:\spaceR^L\mapsto
\cL^{\mathrm{col}}$ by the formula
\be
  \label{eq:PGcolumn}
  \cP_{\mathrm{Vec}}^{\mathrm{col}}Z=
  \left(\!\!\!\begin{array}{c}\Pi^{\mathrm{col}} \first{Z}  \\  \cR_L^\top \first{Z}\end{array}\!\!\!\right).
\ee

Let us formulate the \emph{vector forecasting algorithm} for $j$th series.
\begin{enumerate}
    \item
    In the notation above,  define  the vectors $Z_j \in \spaceR^{L}$ as follows:
    \be
    \label{eq:V_FORcolumn}
    Z_j=\left\{
        \begin{array}{ll}
            \what{X}_j^{(p)} &{\rm for\;} \; j=1,\ldots,K_p,\\
            \cP_{\mathrm{Vec}}^{\mathrm{col}} Z_{j-1}&{\rm for\;} \; j=K_p+1,\ldots,K_p+M+L-1.\\
        \end{array}
    \right.
    \ee
    \item By constructing the matrix $\bfZ=[Z_1:\ldots:Z_{K_p+M+L-1}]$ and making
    its diagonal averaging we obtain the series $z_1,\ldots,z_{N_p+M+L-1}$.
    \item The values $(\tilde{x}_{N_p+1}^{(p)},\ldots,\tilde{x}_{N_p+M}^{(p)}) = (z_{N_p+1},\ldots,z_{N_p+M})$ form the $M$ terms of the forecast.
\end{enumerate}

\paragraph{Row forecast}

Define the linear operator
$\cP_{\mathrm{Vec}}^{\mathrm{row}}:\spaceR^K\mapsto \cL^{\mathrm{row}}$ by the formula
\be
  \label{eq:PGrow}
  \cP_{\mathrm{Vec}}^{\mathrm{row}}Z = A,
\ee
such that $\llast{A}=\Pi^{\mathrm{row}} \ffirst{Z}$ and $\bfmu(A)= \bfcR_K \ffirst{Z}$.

Let us formulate the \emph{vector forecasting algorithm}.
\begin{enumerate}
    \item In the notation above, define the vectors $Z_i \in \spaceR^K$ as follows: \be
    \label{eq:V_FORrow}
    Z_i=\left\{
        \begin{array}{ll}
            \what{Y}_i &{\rm for\;} \; i=1,\ldots,L,\\
            \cP_{\mathrm{Vec}}^{\mathrm{row}} Z_{i-1}&{\rm for\;} \; i=L+1,\ldots,L+M+K^*-1,\\
        \end{array}
    \right.
    \ee
    where $K^*=\max(K_p, p=1,\ldots,s)$.
    \item By constructing the matrix $\bfZ=[Z_1:\ldots:Z_{L+M+K^*-1}]^{\top}$ and
    making MSSA Reconstruction step we obtain the series
    $z_1^{(p)},\ldots,z_{N_p+M+K^*-1}^{(p)}$, $p=1,\ldots,s$.
    \item The values $(\tilde{x}_{N_p+1}^{(p)},\ldots,\tilde{x}_{N_p+M}^{(p)}) = (z_{N_p+1}^{(p)},\ldots,z_{N_p+M}^{(p)})$ form the $M$ terms of the forecast.
\end{enumerate}

\begin{remark}
\label{rem:danger_vect}
For the $M$-step ahead vector forecast, $M+K^*-1$ new lagged vectors for row
forecasting and $M+L-1$ ones for column forecasting are constructed. The reason
for this is to make the $M$~step forecast inheriting the $(M-1)$~step forecast
with no redrawing.  This specific of the vector forecasting provides its
stability and accuracy if the accurately extracted component of finite rank is
forecasted, that is, if long-term forecast is appropriate. Otherwise, long-term
vector forecasting can be wrong and even the result of short-term vector
forecasting can be also wrong for large $K^*$ or $L$ correspondingly.
\end{remark}

%% file: multi_package_typicalcode.tex
\subsection{Package}
\label{sec:mssa_package}
\subsubsection{Typical code}
Here we demonstrate how the MSSA decomposition of a system of time series can be
performed by means of the Rssa package.  Since the analysis and forecasting for
one-dimensional time series by Rssa are thoroughly described in
\citet{Golyandina.Korobeynikov2013}, we put more attention on the difference.

In Section~\ref{sec:ssa_typical}, we decomposed the one-dimensional series FORT
(sales of fortified wines). Here we add one more series, sales of dry wines
(shortly DRY), for simultaneous analysis.

For loading the data we use the code from Fragment \ref{frag:wines_input}.

\begin{fragment}[FORT and DRY: Reconstruction]
\label{frag:wineFortDry_rec}
\input{fragments/wineFortDry_rec.tex}
\end{fragment}

\bfgh
        \begin{center}
        \includegraphics[width=16 cm]{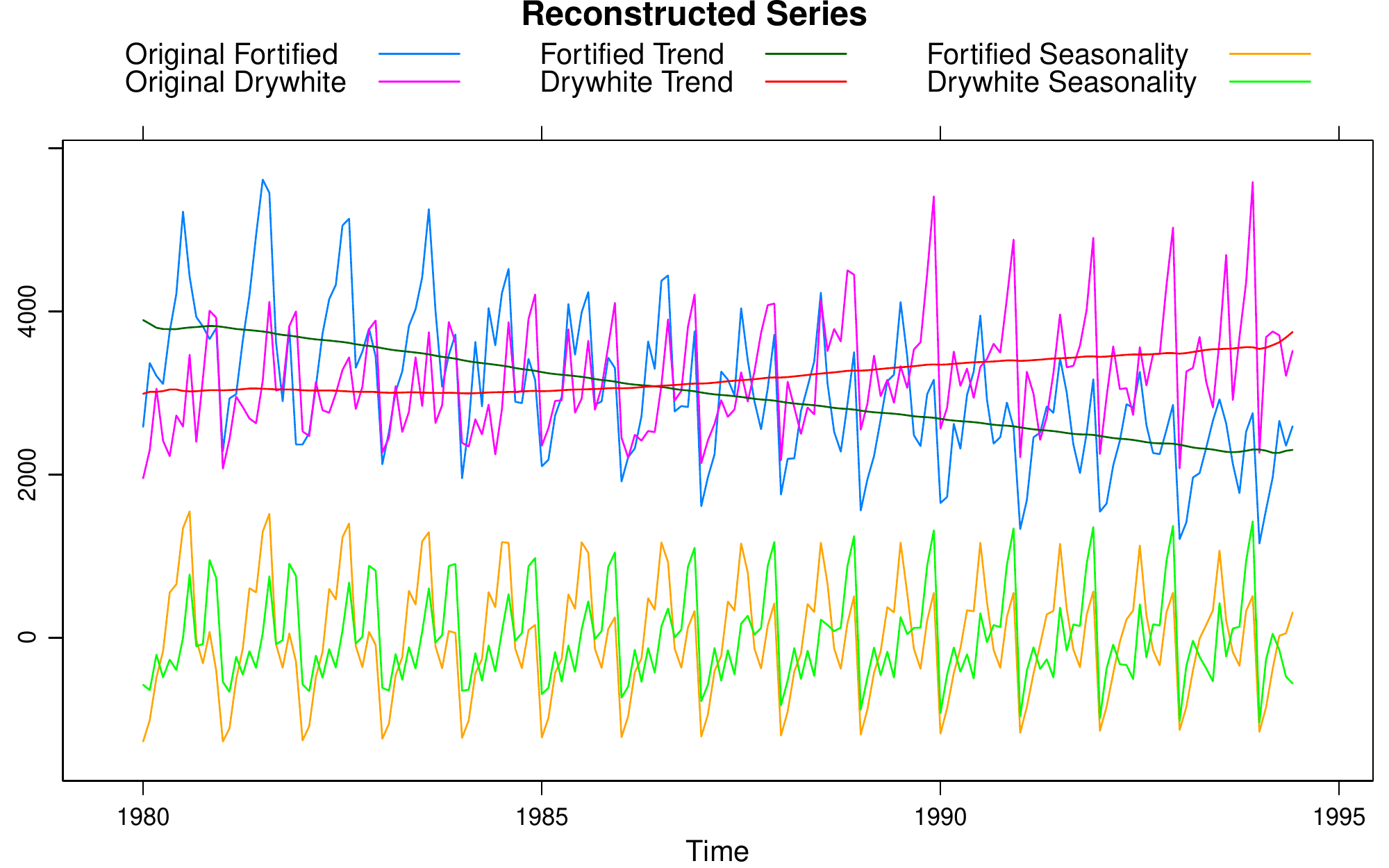}
        \end{center}
        \caption{FORT and DRY: Reconstructed trend and seasonality.}
        \label{fig:wineFortDry_rec}
\efg

Fragment~\ref{frag:wineFortDry_rec} contains a typical code for simultaneous
extraction of the trend and seasonality (compare with
Fragment~\ref{frag:fort_rec}).  An evident difference is in the indicated value
of parameter \code{kind} in the \code{ssa} function.  A more significant
difference is related to plotting of the results.  For multivariate series there
is in a sense a matrix of series, where one index is the number of the series in the
system, while the second index corresponds to the number of the component in the
decomposition.  The \code{plot} function for the reconstruction object allows to
indicate, which subset (which slice) of this matrix one wants to depict by means
of the parameter \code{slice}. The parameter \code{slice} consists of the list
of numbers of series and numbers of decomposition components.

The code for component identification in MSSA is very similar to that in SSA,
compare Fragment~\ref{frag:wineFortDry_identific} and
Fragment~\ref{frag:fort_identific}. The difference consists in the structure of
the factor vectors; however, the factor vectors are not necessary for
identification. Figure~\ref{fig:wineFortDry_1d} (compare Figure~\ref{fig:fort_1d}) shows
that the trend is described by ET1 and ET6, which is slightly mixed with
seasonality. Figure~\ref{fig:wineFortDry_2d} (compare Figure~\ref{fig:fort_2d})
demonstrates the pairs of ETs that are related to seasonality.

Since the implemented methods of parameter estimation are based on eigenvectors
only, they can be applied to eigenvectors in MSSA in exactly
the same way as in the one-dimensional case.

\begin{fragment}[FORT and DRY: Identification]
\label{frag:wineFortDry_identific}
\input{fragments/wineFortDry_identific.tex}
\end{fragment}

 \bfgh
        \begin{center}
        \includegraphics[width=120 mm]{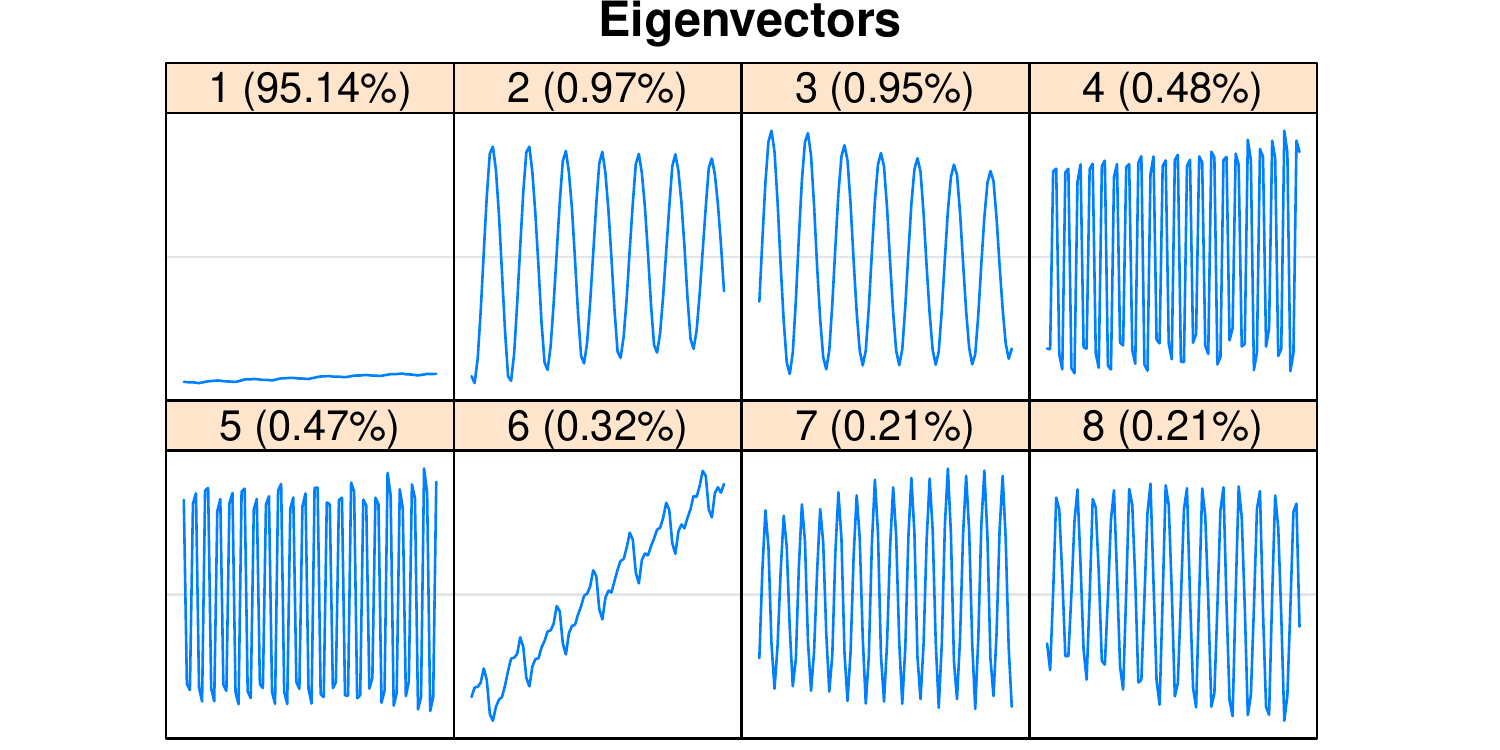}
        \end{center}
        \caption{FORT and DRY: 1D graphs of eigenvectors.}
        \label{fig:wineFortDry_1d}
 \efg
 \bfgh
        \begin{center}
        \includegraphics[width=120 mm]{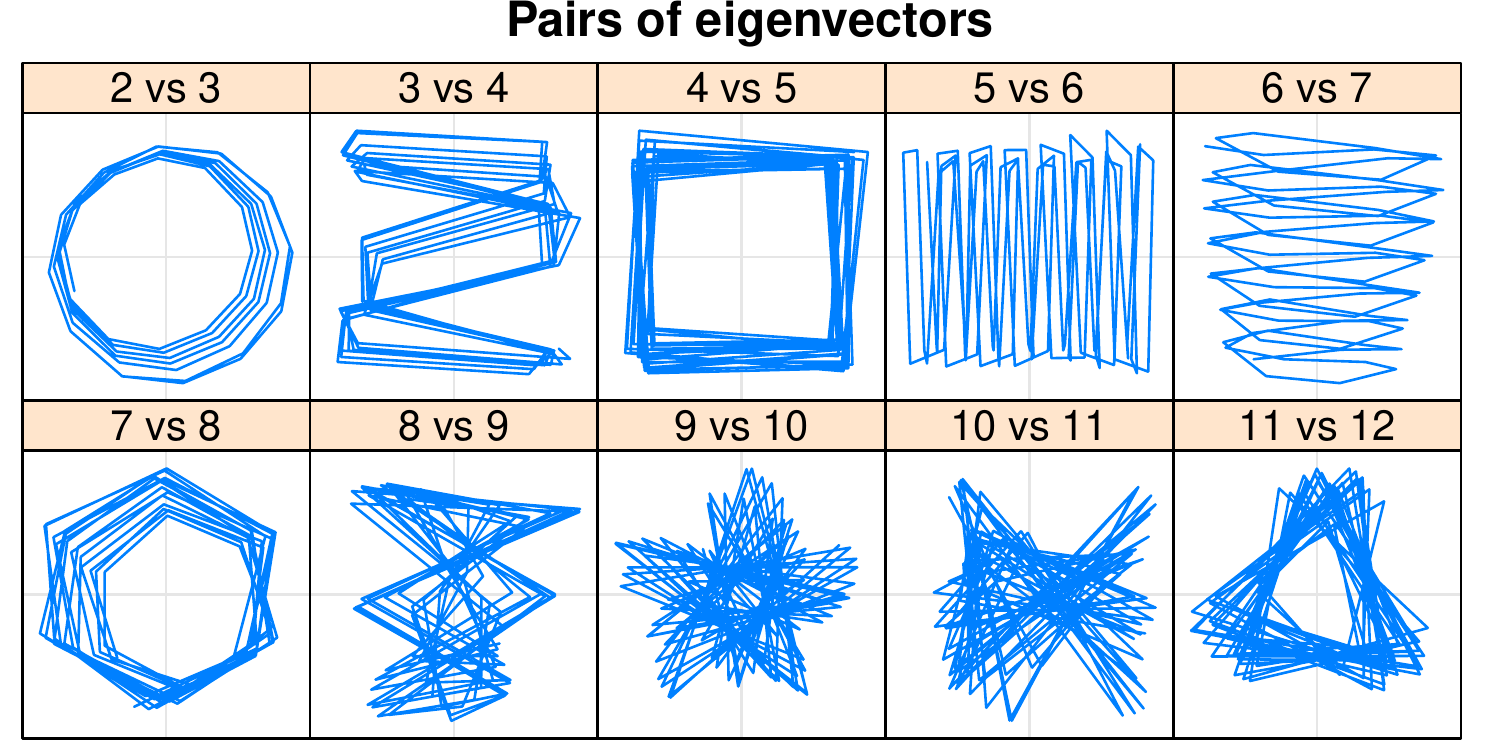}
        \end{center}
        \caption{FORT and DRY: 2D scatterplots of eigenvectors.}
        \label{fig:wineFortDry_2d}
 \efg
 \bfgh
        \begin{center}
        \includegraphics[width=80 mm]{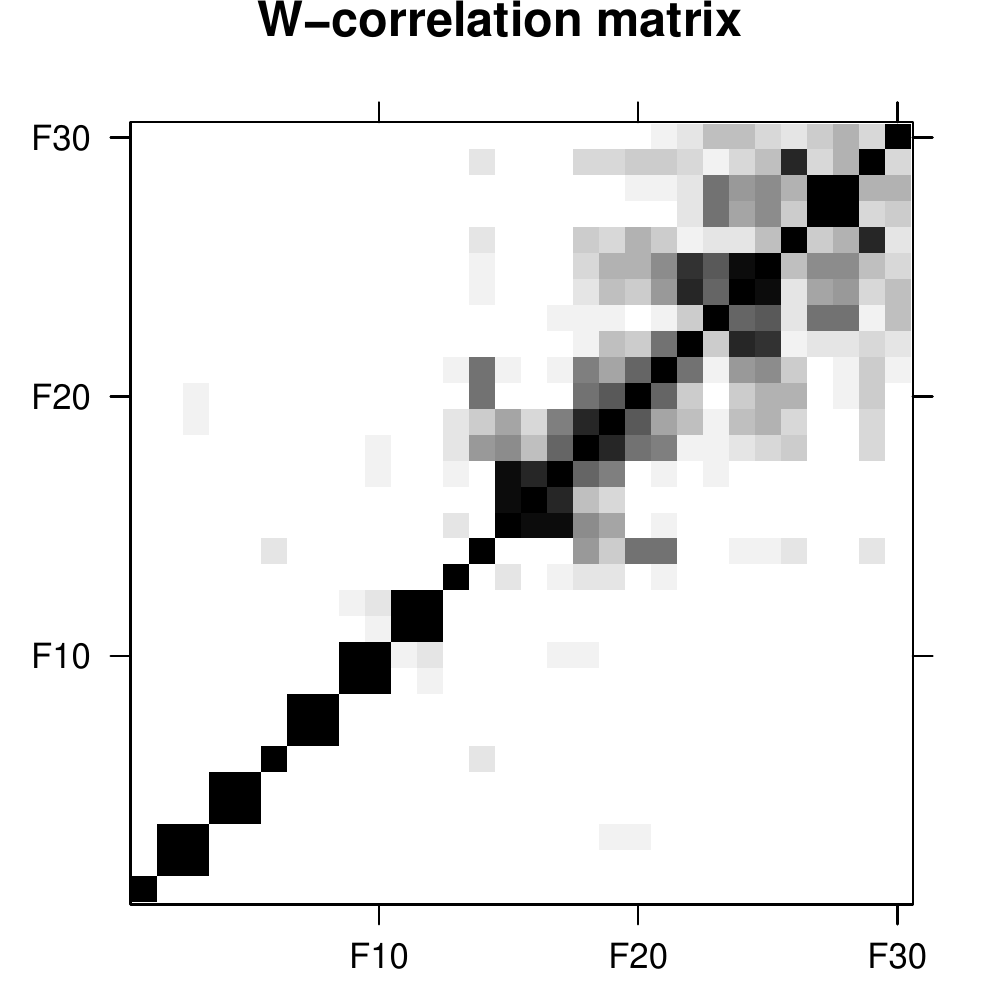}
        \end{center}
        \caption{FORT and DRY: Weighted correlations.}
        \label{fig:wineFortDry_wcor}
\efg

The code for forecasting is also very similar to that in SSA, compare
Fragment~\ref{frag:wineFortDry_forecast} and Fragment~\ref{frag:fort_forecast}.

\begin{fragment}[FORT and DRY: Forecast]
\label{frag:wineFortDry_forecast}
\input{fragments/wineFortDry_forecast.tex}
\end{fragment}

 \bfgh
        \begin{center}
        \includegraphics[width=120 mm]{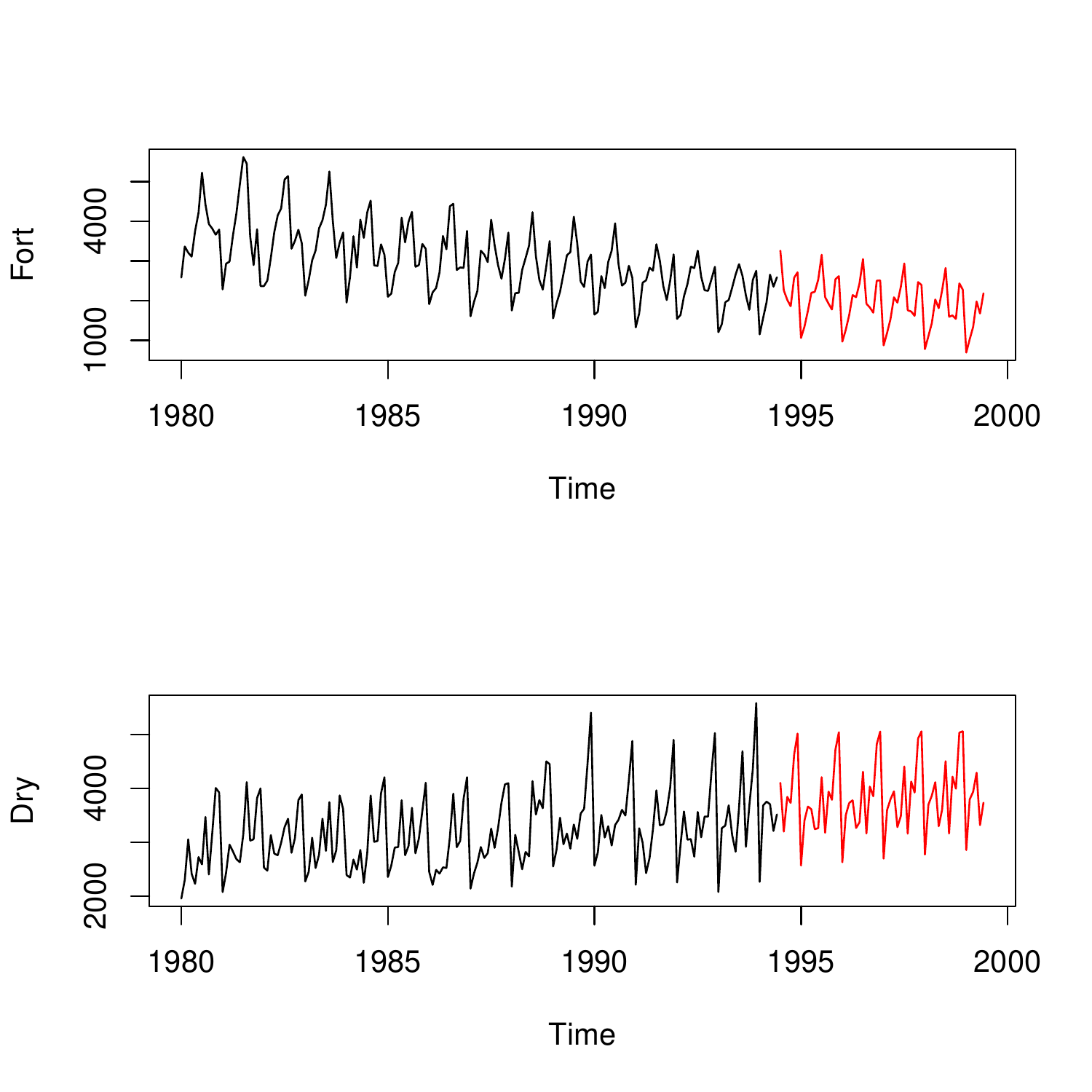}
        \end{center}
        \caption{FORT and DRY: Forecast of the signal.}
        \label{fig:wineFortDry_forecast}
\efg

The results of MSSA analysis are similar to the results of SSA analysis.
However, the separability is slightly worse 
(compare $\bfw$-correlations between signal components in Figures~\ref{fig:fort_wcor}~and~\ref{fig:wineFortDry_wcor}). 
Therefore, the methods for
improvement of separability can be very useful \citep{Golyandina.Shlemov2013}.
Note that the mixture of the signal components is not important for signal
forecasting.

%% file: fragments/wineFortDry_rec.tex
\begin{CodeChunk}
\begin{CodeInput}

> wineFortDry <- wine[, c("Fortified", "Drywhite")]
> L <- 84
> s.wineFortDry <- ssa(wineFortDry, L = L, kind = "mssa")
> r.wineFortDry <- reconstruct(s.wineFortDry,
+                              groups = list(Trend = c(1, 6),
+                                            Seasonality = c(2:5, 7:12)))
> plot(r.wineFortDry, add.residuals = FALSE,
+      plot.method = "xyplot",
+      superpose = TRUE, auto.key = list(columns = 3))
\end{CodeInput}

\end{CodeChunk}

%% file: fragments/wineFortDry_identific.tex
\begin{CodeChunk}
\begin{CodeInput}

> plot(s.wineFortDry, type = "vectors", idx = 1:8)
> plot(s.wineFortDry, type = "paired", idx = 2:11, plot.contrib = FALSE)
> parestimate(s.wineFortDry, groups = list(2:3, 4:5), method = "esprit-ls")
$F1
   period     rate   |    Mod     Arg  |     Re        Im
   12.128  -0.004789 |  0.99522   0.52 |  0.86463   0.49283
  -12.128  -0.004789 |  0.99522  -0.52 |  0.86463  -0.49283
$F2
   period     rate   |    Mod     Arg  |     Re        Im
    4.007  -0.001226 |  0.99877   1.57 |  0.00279   0.99877
   -4.007  -0.001226 |  0.99877  -1.57 |  0.00279  -0.99877
> plot(wcor(s.wineFortDry, groups = 1:30), scales = list(at = c(10, 20, 30)))
\end{CodeInput}

\end{CodeChunk}

%% file: fragments/wineFortDry_forecast.tex
\begin{CodeChunk}
\begin{CodeInput}

> f.wineFortDry <- rforecast(s.wineFortDry, groups = list(1, 1:12),
+                            len = 60, only.new = TRUE)
> par(mfrow = c(2, 1))
> plot(cbind(wineFortDry[, "Fortified"], f.wineFortDry$F2[, "Fortified"]),
+      plot.type = "single",
+      col = c("black", "red"), ylab = "Fort")
> plot(cbind(wineFortDry[, "Drywhite"], f.wineFortDry$F2[, "Drywhite"]),
+      plot.type = "single",
+      col = c("black", "red"), ylab = "Dry")
> par(mfrow = c(1, 1))
\end{CodeInput}

\end{CodeChunk}

%% file: multi_package_comments.tex
\subsubsection{Comments}
\paragraph{Formats of input and output data}
While the representation of a one dimensional time series in \proglang{R} is
pretty obvious, there are multiple possible ways of defining the multivariate
time series. Let us outline some common choices.

\begin{itemize}
    \item Matrix with separate series in the columns. Optionally, additional
    time structure like in \code{mts} objects, can be embedded.
    \item Matrix-like (e.g., a \code{data.frame}) object with series in the
    columns. In particular, \code{data.frame} would be a result of reading the
    series from the file via \code{read.table} function.
    \item List of separate time series objects (e.g., a \code{list} of \code{ts}
    or \code{zoo} objects).
\end{itemize}

Also, the time scales of the individual time series can be normalized via head
or tail padding with \code{NA} (for example, as a result of the \code{ts.union}
call), or specified via time series attributes. Or, everything can be mixed
all together.

The package is designed to allow any of the input cases outlined above and
produces the reconstructed series in the same format. All the attributes, names
of the series, \code{NA} padding, etc. is carefully preserved. For forecasted
series, the time scale attributes for several known time series objects
(e.g., \code{ts}) are inferred automatically where possible.

The examples in the Fragments~\ref{frag:wineFortDry_rec},
\ref{frag:mssa_comparison} provide an overview of the possible input series
formats.

\paragraph{Plotting specifics}
Keep in mind that the default (\code{'native'}) plotting method for
reconstruction objects may or may not be suitable for multivariate time series
plotting. For example, it provides many useful plotting possibilities for
\code{ts} and \code{mts} objects, but might totally be unusable in case of
\code{data.frame} objects, because it will just call the \code{pairs} function
on the resulting data frame in the end.

\paragraph{Summary for \code{ssa} object}
The \code{summary} for SSA object (see Fragment~\ref{frag:wineFortDry_summary})
in MSSA case coincides with the one for SSA up to slight differences.

\begin{fragment}[FORT and DRY: Summary]
\label{frag:wineFortDry_summary}
\input{fragments/wineFortDry_summary.tex}
\end{fragment}

Here one can see the individual series lengths (excluding the \code{NA}
padding), the window length, the selected default SVD decomposition method and
the number of computed eigentriples. No factor vectors are computed, they will be
recomputed on the fly when necessary.

\paragraph{Efficient implementation}
All ideas from the one-dimensional case can be extended to the multivariate
case.  In one-dimensional case, the complexity is determined by the series
length $N$ and the window length $L$, and the worst case corresponds to $L\sim
K\sim N/2$ with overall complexity of $O(L^{3} + L^{2}K) = O(N^{3})$.

In the multidimensional case (for the sake of simplicity, assume that all the
series have equal lengths $N$), the worst case corresponds to $L\sim K\sim
s(N+1)/(s+1)$, that is, the order of complexity is the same $O(N^{3})$ but the
constant can be considerably larger.  Therefore, the achieved speedup can be
much stronger than that in the one-dimensional case.

Note that the Multichannel SSA can be viewed as a special case of Shaped 2D-SSA
(see Section~\ref{sec:shaped_special}) and the current implementation in the
package uses this under the hood.


%% file: fragments/wineFortDry_summary.tex
\begin{CodeChunk}
\begin{CodeInput}

> summary(s.wineFortDry)
Call:
ssa(x = wineFortDry, L = L, kind = "mssa")
Series length: 174, 174,	Window length: 84,	SVD method: eigen
Computed:
Eigenvalues: 50,	Eigenvectors: 50,	Factor vectors: 0
Precached: 0 elementary series (0 MiB)
Overall memory consumption (estimate): 0.03734 MiB
\end{CodeInput}

\end{CodeChunk}

%% file: multi_examples_decomp.tex
\subsection{Examples}
\label{sec:examples_mssa}
\subsubsection{Factor vectors}
Factor vectors in MSSA have length $K$ and consist of stacked vectors of length
$K_p$, $p=1,\ldots,s$, related to each series. Therefore, it is natural to
depict them as a system of $s$ vectors.  Factor vectors are not necessarily
contained in the \code{ssa} object.  In particular, the \code{s.wineFortDry}
object created in Fragment~\ref{frag:wineFortDry_rec} has $0$ factor vectors,
what can be checked by the \code{summary(s.wineFortDry)} function.  Note that if
one uses \code{svd.method="svd"} (and \code{propack}), then factor vectors will be calculated
automatically.  In order to get factor vectors corresponding to the eigenvectors contained
in \code{ssa}, the user can call the \code{calc.v} function.

Fragment~\ref{frag:wineFortDry_factor} shows the difference between eigenvectors
and factor vectors for small window length. The result for $L=24$ is depicted in
Figure~\ref{fig:wineFortDry_factorvseigen}.  The eigenvector in
Figure~\ref{fig:wineFortDry_factorvseigen} captures the common behavior (which is
almost constant) in the timescale of two years, while the factor vector is divided
into parts reflecting individual features of the series, compare with trends in
Figure~\ref{fig:wineFortDry_rec} which are repeated in
Figure~\ref{fig:wineFortDry_factorvseigen}. Note that signs of the calculated
eigenvectors are random. For example, the first eigenvector is negative
here. Thereby, the factor vectors are similar to the reconstructed series with
opposite sign.

The choice of large window length $L=163$ also yields the trajectory matrix of
rank $24$, since then $K_p=176-163+1=12$ and therefore $K=2\cdot12=24$; however,
the common structure is captured by two eigentriples.  Note that in climatology
the SVD of the transposed trajectory matrix is traditionally considered
\citep{Hannachi.etal2007}.  Therefore, the eigenvectors $U_i$ correspond to
normalized extended principal components in \citet{Hannachi.etal2007},
while the factor vectors $V_i$ are called Extended Empirical Orthogonal Functions (EEOFs).

\begin{fragment}[FORT and DRY: Use of factor vectors in MSSA]
\label{frag:wineFortDry_factor}
\input{fragments/wineFortDry_factor.tex}
\end{fragment}

\bfgh
        \begin{center}
        \includegraphics[width=120 mm]{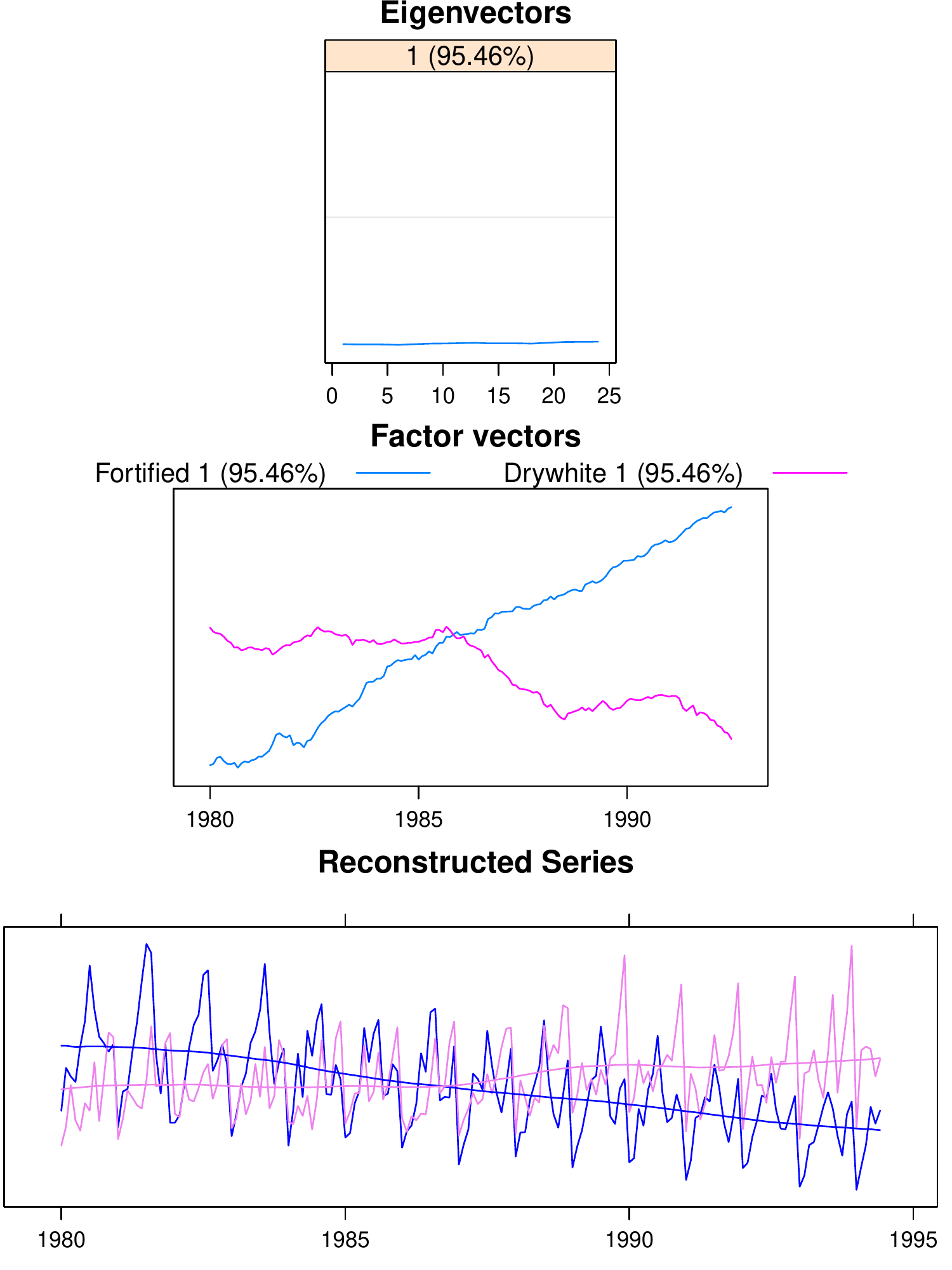}
        \end{center}
        \caption{FORT and DRY: $L=24$, relation between eigen- and factor vectors and reconstructed series.}
        \label{fig:wineFortDry_factorvseigen}
 \efg

\subsubsection{Preprocessing (normalization)}

Let the time series in the system be measured in different scales. In statistics,
this problem is typically decided by standardization of data. In SSA, centering
may not be an appropriate preprocessing. Therefore, two types
of preprocessing can be applied, conventional standardization and normalization,
that is, division on the square root of mean sum of squares.  The normalization
can be even more appropriate for positive series, since it changes only the
scale of data.

Let us consider Fortified and Ros\'{e} wine sales. Sales of Fortified wines are
of the order of thousands while sales of Ros\'{e} wines are of the order of tens and
hundreds.  Fragment~\ref{frag:wineFortRose_scales} shows how the scale influences
the reconstruction result.

\begin{fragment}[FORT and ROSE: Influence of series scales]
\label{frag:wineFortRose_scales}
\input{fragments/wineFortRose_scales.tex}
\end{fragment}

Figure~\ref{fig:wineFortRose_scales} demonstrates the result of a trend
reconstruction, where the trend was detected in the same way as before, that is, by means
of form of eigenvectors and weighted correlations.  The trend of the ROSE series
is more complicated. However, FORT overweighs the decomposition and the
eigentriples that refine the ROSE trend have very small weight and mix with the
common noise.  Therefore, the SSA processing with no normalization is worse for
analysis of the series ROSE with smaller scale.

\bfgh
        \begin{center}
        \includegraphics[width=120 mm]{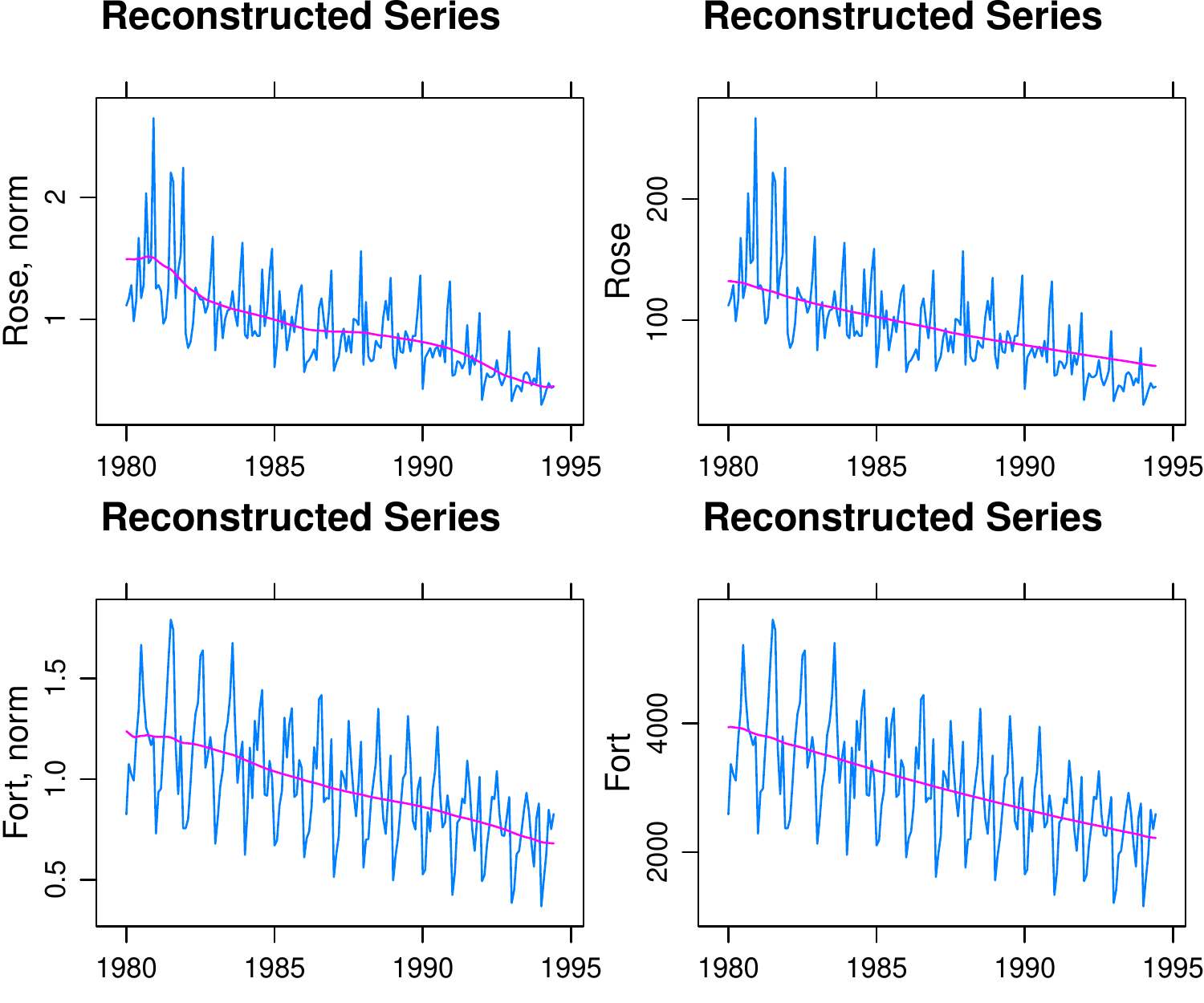}
        \end{center}
        \caption{FORT and ROSE: Trends with  normalization (ET1,12,14) and without (ET1).}
        \label{fig:wineFortRose_scales}
 \efg

\subsubsection{Forecasting of series with different lengths as filling in}
Typical code in Section~\ref{sec:mssa_package} was demonstrated for series of equal lengths.
However, the same code can be applied to series of different lengths.  The full
AustralianWine data are incomplete, there are no data for two months (point 175
and 176) for sales of Ros\'{e} wines and there are no data for the last 11
months of Total sales.

Let us perform the following actions: (A) fill in the missing data in ROSE, (B)
calculate the sum of sales of the present wines and (C) process this sum
together with the total series and fill in the missing data in Total series by
the simultaneous forecasting.

To use the analysis performed before, let us forecast the ROSE series together
with the FORT series for (A).  Fragment~\ref{frag:wineFortRose_forecast}
implements (A). Certainly, this can be done by a separate analysis of the ROSE
series and the SSA methods of filling in missing data, but
Figure~\ref{fig:wineFortRose_forecast} shows that the result of the used method is
quite accurate.

\begin{fragment}[FORT and ROSE: Forecast of the missing data in ROSE]
\label{frag:wineFortRose_forecast}
\input{fragments/wineFortRose_forecast.tex}
\end{fragment}

\bfgh
        \begin{center}
        \includegraphics[width=120 mm]{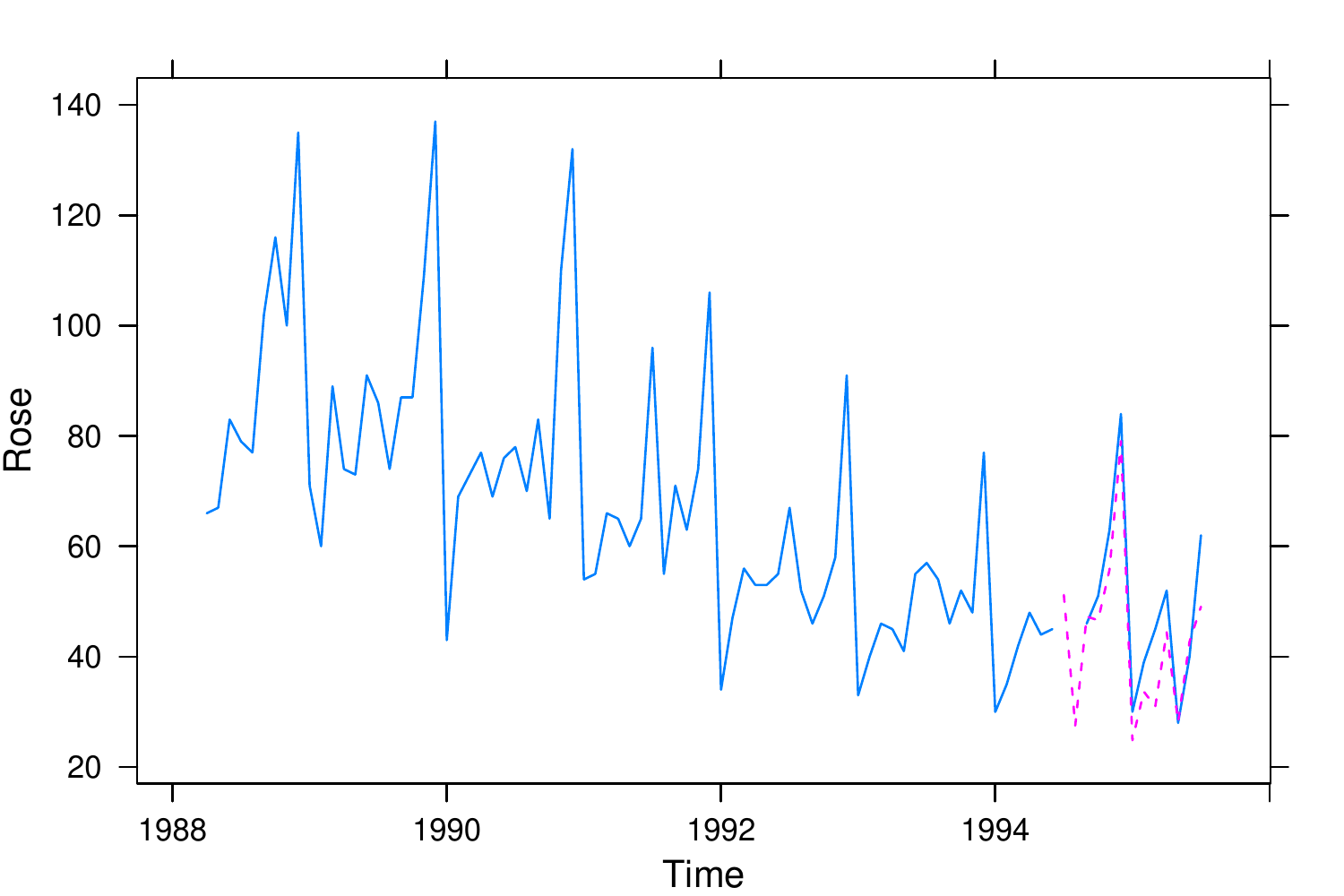}
        \end{center}
        \caption{FORT and ROSE: Forecast of ROSE in comparison with the existing data.}
        \label{fig:wineFortRose_forecast}
\efg

Fragment~\ref{frag:winetotal_forecast} implements (B) and (C).  Also, we compare
the forecast of total sales separately (TOTAL) and together with the available
sum of sales of main wines (MAINSALES).  We consider two simultaneous forecasts
together with changing the weight of the series MAINSALES from 1 to 100.
Figure~\ref{fig:winetotal_forecast} clearly demonstrates that for small
contribution of MAINSALES the simultaneous forecast is close to separate
forecast of Total, while for large weight of MAINSALES the simultaneous forecast
of TOTAL tend to have the form similar to the form of MAINSALES.  An additional
investigation is needed to choose the better version.

\begin{fragment}[TOTAL: Different ways of forecasting]
\label{frag:winetotal_forecast}
\input{fragments/winetotal_forecast.tex}
\end{fragment}

\bfgh
        \begin{center}
        \includegraphics[width=120 mm]{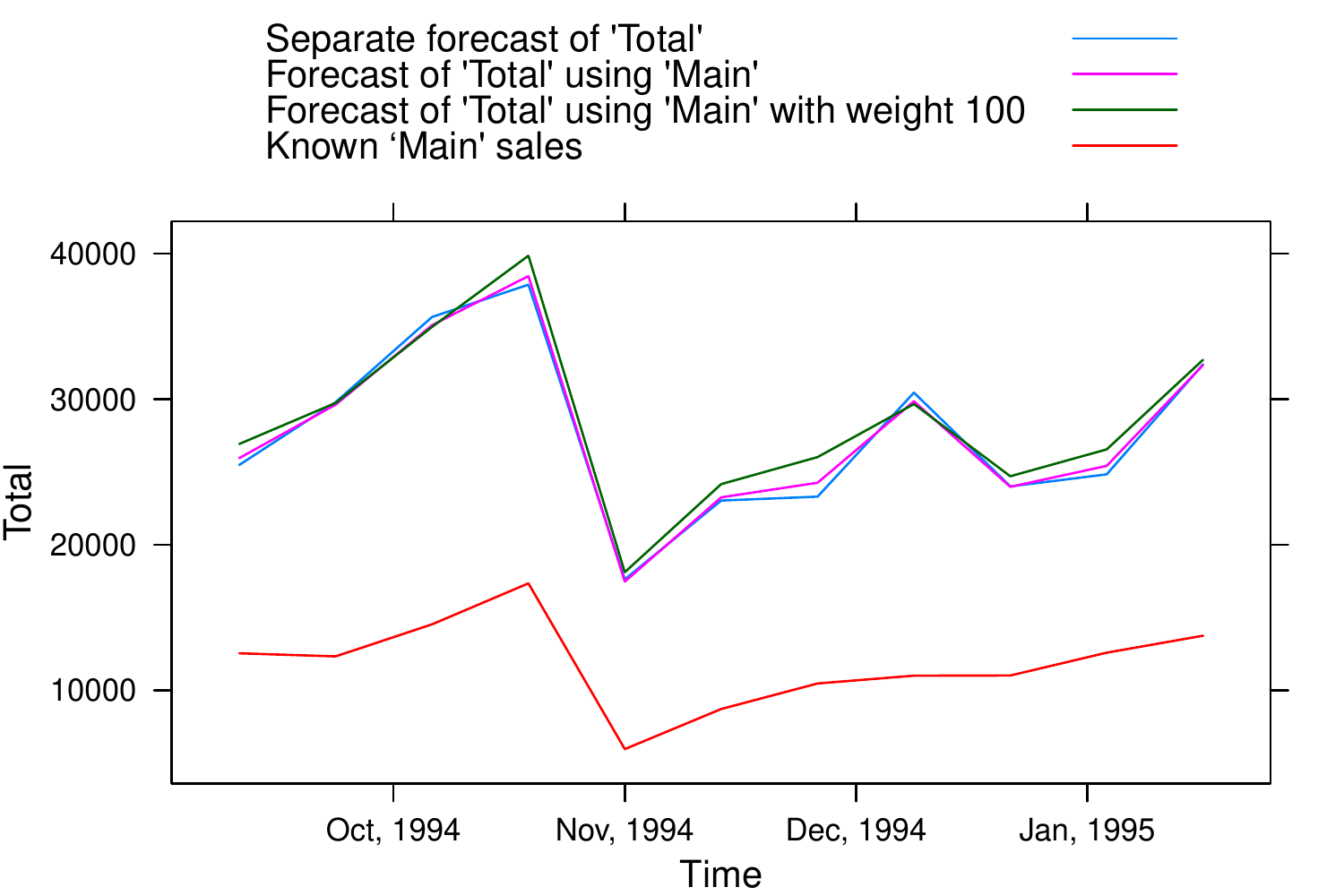}
        \end{center}
        \caption{TOTAL: Forecast of total sales, separate and with the available information.}
        \label{fig:winetotal_forecast}
 \efg

\subsubsection{Simultaneous decomposition of many series}
In this example, we consider a system of many time series and show that the
decomposition by MSSA helps to look at similar patterns of the series. We will
use the methodology described in the previous sections.

Let us consider the collection of $s=6$ series from the `AustralianWine'
dataset, which includes the series of wine sales considered in the typical code
from Section~\ref{sec:mssa_package}. A considerable part of this multivariate
series can be described as seasonality. Therefore, MSSA can have an advantage
over conventional SSA.

Since the time series have different scales, namely, the `Ros\'{e} wines' sales
have the order of 100 thousand of litres in month, while the sales of `Fortified
wines' are about 30 times larger, the time series should be transformed into the
same scale.  We choose the window length $L=163$, then $K_p=12$ and $K=84$ and therefore the number of
elementary components is equal to $84=\min(163,84)$. 
For the choice of $L$ that makes the number of elementary components larger,
there can be elementary components with approximately equal contribution
(there can be a lack of strong separability).
It appears that the decomposition with the choice $L=163$ is better than, say, 
a more detailed decomposition with $L=151$, $K_p=24$, since the choice $L=163$
helps to avoid mixture of components.

The identification of the trend (ET1,2,5) and the seasonality (ET3,4, 6--12) is
performed on the base of eigenvectors and uses the principles described in the
typical code from Section~\ref{sec:mssa_package}.  Fragment~\ref{frag:wine_full}
contains the code to get the reconstruction shown in
Figure~\ref{fig:wine_trends}~and~\ref{fig:wine_seasonality}.

\begin{fragment}[Wine sales: Simultaneous decomposition by MSSA]
\label{frag:wine_full}
\input{fragments/wine_full.tex}
\end{fragment}

\bfgh
\begin{center}
    \includegraphics[width=120 mm]{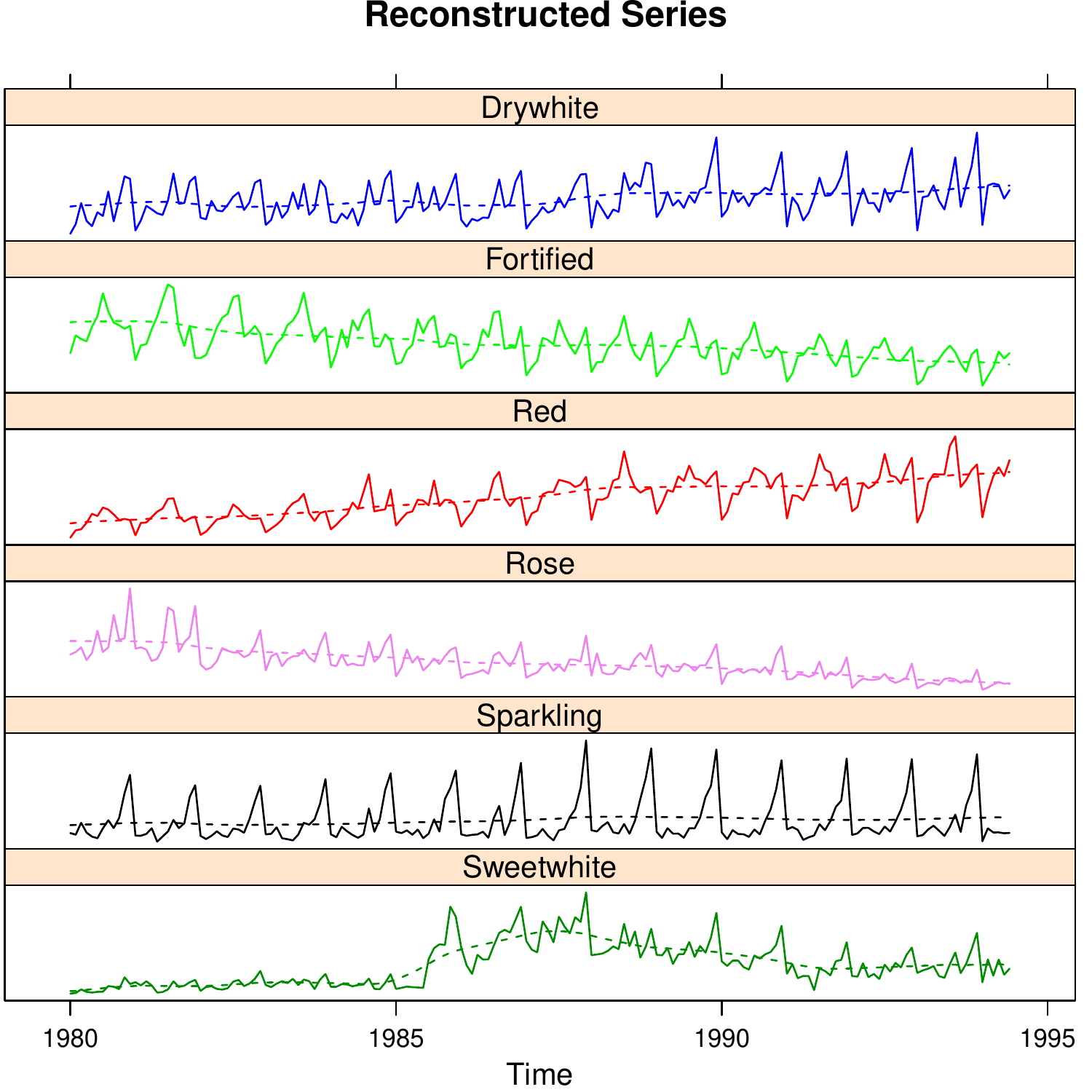}
\end{center}
\caption{Wine sales: Extraction of trends.}
\label{fig:wine_trends}
\efg
\bfgh
\begin{center}
    \includegraphics[width=120 mm]{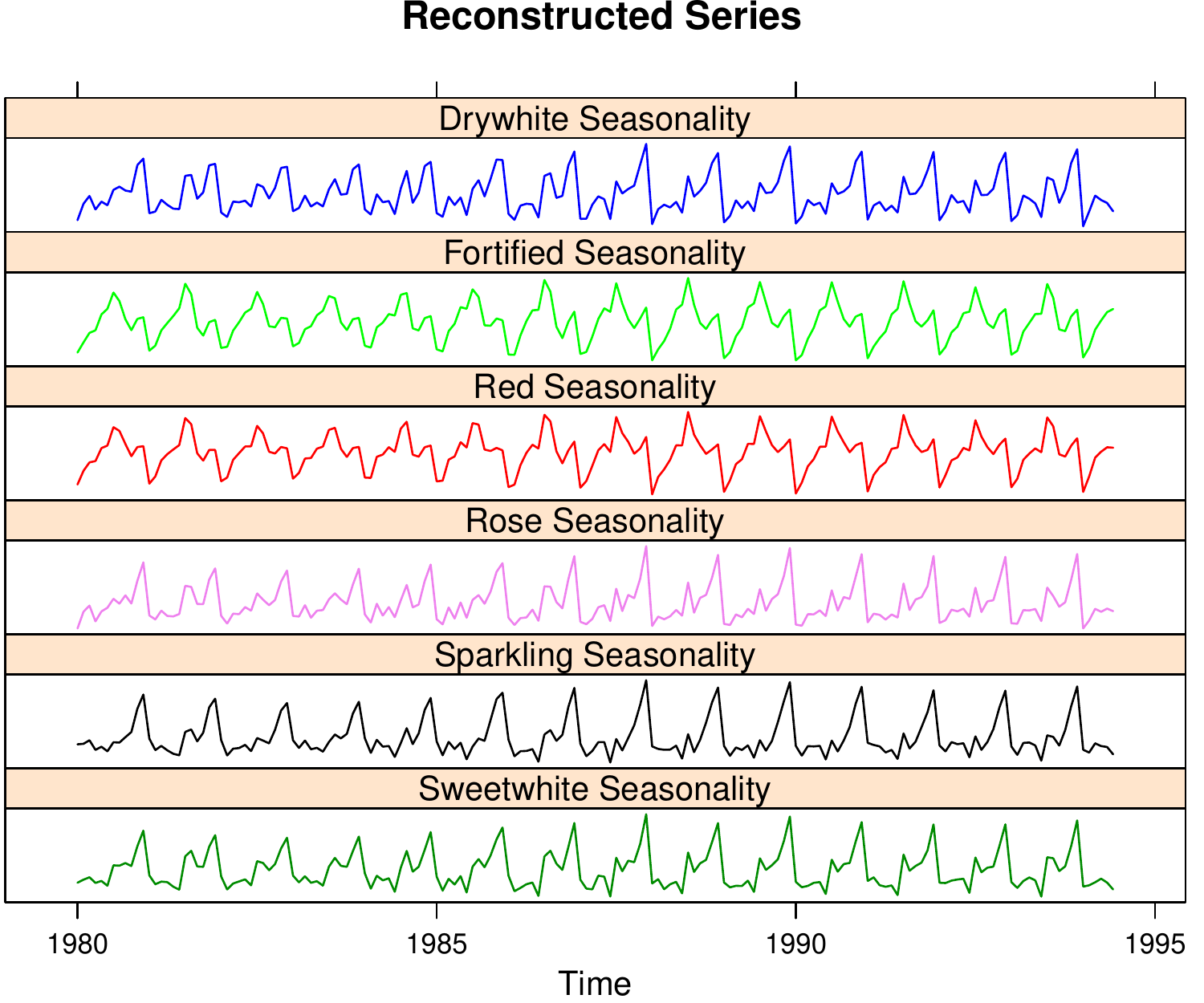}
\end{center}
\caption{Wine sales: Extraction of seasonality.}
\label{fig:wine_seasonality}
\efg

The reconstructed trends and seasonal components look adequate.  In addition,
the simultaneous processing of several time series is very convenient, since we
obtain similar time series components all at once. In particular, it is clearly
seen from Figure~\ref{fig:wine_seasonality} that the sale volumes of fortified
wines are maximal in June-July (that are winter months in Australia), while the
sale volumes of sparkling wines have a peak in December.

%% file: fragments/wineFortDry_factor.tex
\begin{CodeChunk}
\begin{CodeInput}

> library("lattice")
> L <- 24
> s.wineFortDrya <- ssa(wineFortDry, L = L, kind = "mssa")
> r.wineFortDrya <- reconstruct(s.wineFortDrya, groups = list(Trend = 1))
> tp1 <- plot(r.wineFortDrya,
+             add.residuals = FALSE, add.original = TRUE,
+             plot.method = "xyplot", aspect = 0.3, superpose = TRUE,
+             scales = list(y = list(draw = FALSE)),
+             auto.key = "", xlab = "",
+             col = c("blue", "violet", "blue", "violet"))
> tp2 <- plot(s.wineFortDrya, type = "vectors", vectors = "factor", idx = 1,
+             aspect = 0.5, superpose = TRUE,
+             scales = list(x = list(draw = TRUE), y = list(draw = FALSE)),
+             auto.key = list(columns = 2))
> tp3 <- plot(s.wineFortDrya, type = "vectors", vectors = "eigen", idx = 1,
+             aspect = 1,
+             scales = list(x = list(draw = TRUE), y = list (draw = FALSE)))
> plot(tp3, split = c(1, 1, 1, 3), more = TRUE)
> plot(tp2, split = c(1, 2, 1, 3), more = TRUE)
> plot(tp1, split = c(1, 3, 1, 3), more = FALSE)
\end{CodeInput}

\end{CodeChunk}

%% file: fragments/wineFortRose_scales.tex
\begin{CodeChunk}
\begin{CodeInput}

> wineFortRose <- wine[, c("Fortified", "Rose")]
> summary(wineFortRose)
   Fortified         Rose       
 Min.   :1154   Min.   : 30.00  
 1st Qu.:2372   1st Qu.: 66.00  
 Median :2898   Median : 87.00  
 Mean   :3010   Mean   : 93.01  
 3rd Qu.:3565   3rd Qu.:114.25  
 Max.   :5618   Max.   :267.00  
> norm.wineFortRosen <- sqrt(colMeans(wineFortRose^2))
> wineFortRosen <- sweep(wineFortRose, 2, norm.wineFortRosen, "/")
> L <- 84
> s.wineFortRosen <- ssa(wineFortRosen, L = L, kind = "mssa")
> r.wineFortRosen <- reconstruct(s.wineFortRosen,
+                                groups = list(Trend = c(1, 12, 14),
+                                              Seasonality = c(2:11, 13)))
> s.wineFortRose <- ssa(wineFortRose, L = L, kind = "mssa")
> r.wineFortRose <- reconstruct(s.wineFortRose,
+                               groups = list(Trend = 1,
+                                             Seasonality = 2:11))
> wrap.plot <- function(rec, component = 1, series, xlab = "", ylab, ...)
+   plot(rec, add.residuals = FALSE, add.original = TRUE,
+        plot.method = "xyplot", superpose = TRUE,
+        scales = list(y = list(tick.number = 3)),
+        slice = list(component = component, series = series),
+        xlab = xlab, ylab = ylab, auto.key = "", ...)
> trel1 <- wrap.plot(r.wineFortRosen, series = 2, ylab = "Rose, norm")
> trel2 <- wrap.plot(r.wineFortRosen, series = 1, ylab = "Fort, norm")
> trel3 <- wrap.plot(r.wineFortRose, series = 2, ylab = "Rose")
> trel4 <- wrap.plot(r.wineFortRose, series = 1, ylab = "Fort")
> plot(trel1, split = c(1, 1, 2, 2), more = TRUE)
> plot(trel2, split = c(1, 2, 2, 2), more = TRUE)
> plot(trel3, split = c(2, 1, 2, 2), more = TRUE)
> plot(trel4, split = c(2, 2, 2, 2))
\end{CodeInput}

\end{CodeChunk}

%% file: fragments/wineFortRose_forecast.tex
\begin{CodeChunk}
\begin{CodeInput}

> f.wineFortRosen <- rforecast(s.wineFortRosen, groups = list(1:14),
+                              len = 13, only.new = TRUE)[, "Rose"]
> f.wineFortRosen_long <- c(rep(NA, 174),
+                           norm.wineFortRosen["Rose"] * f.wineFortRosen)
> xyplot(AustralianWine[100:187, "Rose"] + f.wineFortRosen_long[100:187] ~
+          time(AustralianWine)[100:187],
+        type = "l", xlab = "Time", ylab = "Rose", lty = c(1, 2))
\end{CodeInput}

\end{CodeChunk}

%% file: fragments/winetotal_forecast.tex
\begin{CodeChunk}
\begin{CodeInput}

> FilledRoseAustralianWine <- AustralianWine
> FilledRoseAustralianWine[175:176, "Rose"] <- f.wineFortRosen_long[175:176]
> mainsales <- ts(rowSums(FilledRoseAustralianWine[, -1]))
> # Sum of sales of main wines
> total <- FilledRoseAustralianWine[, "Total"]
> L = 84
> s.totalmain1 <- ssa(list(mainsales[12:187], total[1:176]),
+                     L = L, kind = "mssa")
> f.totalmain1 <- rforecast(s.totalmain1, groups = list(1:14),
+                           len = 11, only.new = TRUE)
> s.totalmain2 <- ssa(list(100 * mainsales[12:187], total[1:176]),
+                     L = L, kind = "mssa")
> f.totalmain2 <- rforecast(s.totalmain2, groups = list(1:14),
+                           len = 11, only.new = TRUE)
> s.total <- ssa(total[1:176], L = L, kind = "1d-ssa")
> f.total <- rforecast(s.total, groups = list(1:14),
+                      len = 11, only.new = TRUE)
> xtime <- time(AustralianWine)[177:187]
> xtime.labels <- paste(month.abb[round(xtime * 12) %% 12 + 1],
+                       floor(xtime), sep = ", ")
> xyplot(f.total + f.totalmain1[[2]] + f.totalmain2[[2]] +
+          mainsales[177:187] ~ xtime,
+        type = "l", xlab = "Time", ylab = "Total",
+        scales = list(x = list(labels = xtime.labels)),
+        auto.key = list(text = c("Separate forecast of 'Total'",
+                   "Forecast of 'Total' using 'Main'",
+                   "Forecast of 'Total' using 'Main' with weight 100",
+                   "Known `Main' sales"),
+                        lines = TRUE, points = FALSE))
\end{CodeInput}

\end{CodeChunk}

%% file: fragments/wine_full.tex
\begin{CodeChunk}
\begin{CodeInput}

> L <- 163
> norm.wine <- sqrt(colMeans(wine[, -1]^2))
> winen <- sweep(wine[, -1], 2, norm.wine, "/")
> s.winen <- ssa(winen, L = L, kind = "mssa")
> r.winen <- reconstruct(s.winen,
+                        groups = list(Trend = c(1, 2, 5),
+                                      Seasonality = c(3:4, 6:12)))
> plot(r.winen, add.residuals = FALSE,
+      plot.method = "xyplot",
+      slice = list(component = 1), screens = list(colnames(winen)),
+      col = c("blue", "green", "red", "violet", "black", "green4"),
+      lty = rep(c(1, 2), each = 6),
+      scales = list(y = list(draw = FALSE)),
+      layout = c(1, 6))
> plot(r.winen, plot.method = "xyplot", add.original = FALSE,
+      add.residuals = FALSE, slice = list(component = 2),
+      col = c("blue", "green", "red", "violet", "black", "green4"),
+      scales = list(y = list(draw = FALSE)),
+      layout = c(1, 6))
\end{CodeInput}

\end{CodeChunk}

%% file: multi_examples_mssa_vs_ssa.tex
\subsubsection{Numerical comparison}

In this section, we demonstrate how the accuracy of MSSA is related to the
structure of the multivariate time series.  The aim is to compare accuracy for
separate analysis and forecasting of time series with simultaneous processing of
the series system.  We repeat the results from \citet{Golyandina.Stepanov2005}
and supplement them with new comparisons.  In particular, the comparison results
explain the choice of the default forecasting method.

In the study below, we consider the case $s=2$ and hence include CSSA into the
range of SSA methods we compare.

The investigated model examples include the least favorable and the most
favorable cases for MSSA as well as some cases well suited for the application
of CSSA.

Let us observe $(\tX^{(1)}, \tX^{(2)})=(\tH^{(1)}, \tH^{(2)})+(\tN^{(1)},
\tN^{(2)})$, where $(\tH^{(1)}, \tH^{(2)})$ is a two-dimensional signal
consisting of two harmonic time series, $\tN^{(1)}$ and $\tN^{(2)}$ are
realizations of independent Gaussian white noises.  Then we can use the standard
simulation procedure to obtain estimates of mean square errors (${\rm MSE}$) for
reconstruction and forecasting of $(\tH^{(1)}, \tH^{(2)})$ by the considered
above SSA methods.  Note that the resultant ${\rm MSE}$ is calculated as the
mean of ${\rm MSE}^{(1)}$ and ${\rm MSE}^{(2)}$ for $\tH^{(1)}$ and $\tH^{(2)}$
correspondingly.

We take the following parameters for the simulation of the time series: $N=71$,
the variance of all noise components is $\sigma^2=25$, the number of replications
is 10000.
We consider the following three versions of the signal $(\tH^{(1)}, \tH^{(2)})$.

{\bf Example A}
(the same periods, the difference between the phases is not equal to $\pi/2$):
\bea
    h_k^{(1)} = 30\cos(2\pi k/12), \quad h_k^{(2)} = 20\cos(2\pi k/12 + \pi/4), \quad k=1,\ldots,N.
\eea

{\bf Example B}
(the same periods and amplitudes; the difference between the phases is equal to $\pi/2$):
\bea
    h_k^{(1)} = 30\cos(2\pi k/12), \quad h_k^{(2)} = 30\cos(2\pi k/12 + \pi/2), \quad k=1,\ldots,N.
\eea

{\bf Example C}
(different periods):
\bea
    h_k^{(1)} = 30\cos(2\pi k/12), \quad h_k^{(2)} = 20\cos(2\pi k/8 + \pi/4), \quad k=1,\ldots,N.
\eea

The choice of these examples is determined by the fact that the dimensions of
the signal trajectory spaces (i.e., ranks) are different for different
extensions of SSA methods, see Table~\ref{tab:dim}.  For each example the rank
in bold font corresponds to the method with the best accuracy for this example.

\begin{table}[h]
\begin{center}
\begin{tabular}{|c|c|c|c|}
    \hline
    &Example A&Example B&Example C\\
    \hline
    SSA    & 2 & 2 & {\bf 2}\\
    MSSA    & {\bf 2} & 2 & 4\\
    CSSA    & 2 & {\bf 1} & 4\\
    \hline
\end{tabular}
\caption{Dimension of the signal trajectory space.}
\label{tab:dim}
\end{center}
\end{table}
The results of investigation for different window lengths $L$ are summarized in
Tables~\ref{tab:mssa:an}~and~\ref{tab:mssa:for}. The smallest value in each row
is marked by typing it in bold.  The 24 term-ahead forecast was performed.
\btbh
\begin{center}
\begin{tabular}{|l|c|c|c|c|c|c|c|}
    \hline
    Method (Ex.)  & $L=12$ & $L=24$ & $L=36$ & $L=48$ & $L=60$\\
    \hline
    SSA (Ex.A,B,\textbf{C})   & 3.22 &  \textbf{2.00} &  \textbf{2.00} &  \textbf{2.00} &  3.22\\
    MSSA (Ex.\textbf{A},B)   & 3.17 &  1.82 &  1.58 &  {\bf 1.46} &  1.97\\
    {MSSA} (Ex.C)   & 6.90 &  3.77 &  3.07 &  {\bf 2.87} &  3.84\\
    CSSA  (Ex.A)  & 3.23 &  \textbf{2.01} & {\bf 2.01} &  \textbf{2.01} &  3.23\\
    CSSA  (Ex.\textbf{B})  & 1.57 &  \textbf{1.00} & {\bf 1.00} &  \textbf{1.00} &  1.57\\
    CSSA  (Ex.C)  & 6.97 &  {4.05} & {\bf 3.81} &  {4.05} &  6.97\\
    \hline
  \end{tabular}
\caption{MSE of signal reconstruction.}
\label{tab:mssa:an}
\end{center}
\end{table}
\begin{table}[h]
\begin{center}
\begin{tabular}{|l|c|c|c|c|c|c|c|}
    \hline
    Example A    & $L=12$ & $L=24$ & $L=36$ & $L=48$ & $L=60$\\
    \hline
    \textbf{Recurrent} &&&&&\\
    MSSA-column   & 5.34  &  {\bf 3.60} &  {3.64} &  {3.66} &  4.38\\
    {MSSA-row}    & 6.01  &  4.20 &  3.75 & {\bf 3.27} &  3.93\\
    SSA       & 7.18  &  {\bf 5.55}&  6.23&  6.04&  8.00\\
    CSSA      & 7.40  &  {\bf 5.53}&  6.22&  6.41&  7.80\\
    \textbf{Vector} &&&&&\\
    MSSA-column    & 5.94  &  {3.72} &  3.56 &  \textbf{3.08} &  3.64\\
    {MSSA-row}    & 3.98  &   \textcolor{blue}{\textbf{2.98}} &  3.24 &  {3.12} &  4.20\\
    SSA       & 7.62  &   5.37&  5.82&  \textbf{5.12}&  6.53\\
    CSSA      & 7.83  &   5.42&  5.89&  \textbf{5.14}&  7.02\\
    \hline
    \hline
    Example C   & $L=12$ & $L=24$ & $L=36$ & $L=48$ & $L=60$\\
    \hline
    \textbf{Recurrent} &&&&&\\
    MSSA-column   & 25.42  &  {\bf 7.36} &  {7.47} &  {7.41} &  9.02\\
    {MSSA-row}    & 19.76  &  8.43 &  7.93 &  {\bf 6.54} &  8.28\\
    SSA       & 7.18  &  {\bf 5.55}&  6.23&  6.04&  8.00\\
    CSSA       & 38.29  &  {\bf 11.19}&  13.35&  13.08&  25.03\\
    \textbf{Vector} &&&&&\\
    MSSA-column    & 25.02  &  {7.50} &  7.50 &  \textbf{6.15} &  7.64\\
    {MSSA-row}    & 57.15  &  \textbf{6.01} &  6.96 &  {6.25} &  8.73\\
    SSA       & 7.62  &   5.37&  5.82&   \textcolor{blue}{\textbf{5.12}}&  6.53\\
    CSSA       & 38.43  &  {10.80}&  13.45&  {\textbf{10.19}}&  70.17\\
    \hline
    \hline
    Example B   & $L=12$ & $L=24$ & $L=36$ & $L=48$ & $L=60$\\
    \hline
    {CSSA} recurrent      & 3.45  &  {\bf 2.78}&  3.15&  3.13&  3.99\\
    {CSSA} vector      & 3.86  &  {2.69}&  2.93&   \textcolor{blue}{\textbf{2.55}}&  3.23\\
    \hline
\end{tabular}
\caption{MSE of signal forecast.}
\label{tab:mssa:for}
\end{center}
\end{table}

Comparison of Tables~\ref{tab:mssa:an} and \ref{tab:mssa:for} with
Table~\ref{tab:dim} clearly demonstrates the relation between the accuracy of
the signal reconstruction/forecast and the dimension of the signal trajectory
space.  For each example, the cells corresponding to the method with the best
reconstruction accuracy are shown in bold and the overall minimum is in
blue color.

Note that the reconstruction by SSA and CSSA is the same for window lengths $L$
and $N-L+1$ (12 and 60, 24 and 48 for the considered examples).  Reconstructions
by MSSA are different for different $L$. Also note that the SSA-trajectory
matrix has rank equal to $\min(L,N-L+1)$ and the rank is maximal for $L\approx
(N+1)/2$. The MSSA-trajectory matrix has rank equal to $\min(L,(N-L+1)s)$, where $s$
is the number of time series in the system. This rank is maximal for $L\approx
s(N+1)/(s+1)$. Although the maximality of the rank does not guarantee the minimality
of errors, this consideration means that the window length $L$ for better
separability might be larger than $(N+1)/2$. The simulations confirm this: the
minimum of the reconstruction error for MSSA is achieved at $L=48 = 72\times
2/3$.

The forecasting errors have much more complicated structure (see
\citet{Golyandina2010}). In particular, these errors for forecasting depend on
the reconstruction errors for the last time series points; therefore, the error
may have a dependence on $L$ which is different from that for the average
reconstruction errors.  The considered examples show that the vector forecast is
more accurate than the recurrent one and that the row MSSA forecast is slightly
more accurate than the column MSSA forecast.

The considered examples confirm the following assertions:
\begin{itemize}
    \item The accuracy of the SSA-based methods is closely related to the
    structure of the signal trajectory spaces generated by these methods.  MSSA
    has an advantage if time series from the system include matched components.
    \item Optimal window lengths for analysis and forecasting can differ.  The
    accuracy of forecast is related to the accuracy of reconstruction; however,
    this relation is not straightforward.
    \item The vector forecast with the best window length is more accurate than
    the recurrent forecast. However, it is not always fulfilled when comparing
    the accuracy of methods for the same window length.  Note that this is probably valid for
    forecasting of well-separated signal of finite rank only, see
    Remark~\ref{rem:danger_vect} for an explanation.
    \item The recommendations for the choice of window length (larger or smaller
    than the half of the time series length) for recurrent forecasting are in a
    sense opposite to that for the vector forecasting.
    \item For row and column forecasting (SSA and CSSA forecasting methods are
    particular cases of column forecasting) the recommendations are also
    opposite. It is not surprising, since $L$ and $K$ are swapped.
\end{itemize}

Fragment~\ref{frag:mssa_comparison}
demonstrates how the numbers from the tables can be obtain by means of
\pkg{Rssa} for MSSA and CSSA analysis and vector forecasting applied to Example
A.

\begin{fragment}[Simulation for reconstruction and forecasting accuracy estimation]
\label{frag:mssa_comparison}
\input{fragments/mssa_comparison.tex}
\end{fragment}

%% file: fragments/mssa_comparison.tex
\begin{CodeChunk}
\begin{CodeInput}

> N <- 71
> sigma <- 5
> Ls <- c(12, 24, 36, 48, 60)
> len <- 24
> signal1 <- 30 * cos(2*pi * (1:(N + len)) / 12)
> signal2 <- 30 * cos(2*pi * (1:(N + len)) / 12 + pi / 4)
> signal <- cbind(signal1, signal2)
> R <- 10 #10000 in simulations for the paper
> mssa.errors <- function(Ls) {
+   f1 <- signal1[1:N] + rnorm(N, sd = sigma)
+   f2 <- signal2[1:N] + rnorm(N, sd = sigma)
+   f <- cbind(f1, f2)
+   err.rec <- numeric(length(Ls)); names(err.rec) <- Ls
+   err.for <- numeric(length(Ls)); names(err.for) <- Ls
+   for (l in seq_along(Ls)) {
+     L <- Ls[l]
+     s <- ssa(f, L = L, kind = "mssa")
+     rec <- reconstruct(s, groups = list(1:2))[[1]]
+     err.rec[l] <- mean((rec - signal[1:N, ])^2)
+     pred <- vforecast(s, groups = list(1:2), direction = "row",
+                       len = len, drop = TRUE)
+     err.for[l] <- mean((pred - signal[-(1:N), ])^2)
+   }
+   list(Reconstruction = err.rec, Forecast = err.for)
+ }
> mres <- replicate(R, mssa.errors(Ls))
> err.rec <- rowMeans(simplify2array(mres["Reconstruction", ]))
> err.for <- rowMeans(simplify2array(mres["Forecast", ]))
> print(err.rec)
      12       24       36       48       60 
2.869683 1.587789 1.248881 1.153730 1.855115 
> print(err.for)
      12       24       36       48       60 
2.671251 2.578059 1.501565 2.595378 4.564218 
> #######################
> signal <- signal1 + 1i*signal2
> cssa.errors <- function(Ls) {
+   f1 <- signal1[1:N] + rnorm(N, sd = sigma)
+   f2 <- signal2[1:N] + rnorm(N, sd = sigma)
+   f <- f1 + 1i*f2
+   err.rec <- numeric(length(Ls)); names(err.rec) <- Ls
+   err.for <- numeric(length(Ls)); names(err.for) <- Ls
+ 
+   for (l in seq_along(Ls)) {
+     L <- Ls[l]
+     s <- ssa(f, L = L, kind = "cssa", svd.method = "svd")
+     rec <- reconstruct(s, groups = list(1:2))[[1]]
+     err.rec[l] <- mean(abs(rec - signal[1:N])^2)
+     pred <- vforecast(s, groups = list(1:2), len = len,
+                       drop = TRUE)
+     err.for[l] <- mean(abs(pred - signal[-(1:N)])^2)
+   }
+   list(Reconstruction = err.rec, Forecast = err.for)
+ }
> cres <- replicate(R, cssa.errors(Ls))
> err.rec <- rowMeans(simplify2array(cres["Reconstruction", ]))
> err.for <- rowMeans(simplify2array(cres["Forecast", ]))
> print(err.rec)
      12       24       36       48       60 
7.349316 4.298144 4.101666 4.298144 7.349316 
> print(err.for)
      12       24       36       48       60 
24.67425 13.60116 14.54819 11.72135 15.86380 
\end{CodeInput}

\end{CodeChunk}

%% file: 2dssa_algorithm_2dssa.tex
In this section, we consider the extension of the SSA algorithm for
decomposition of two-dimensional data.  This extension has the name \emph{2D
singular spectrum analysis} (or \emph{2D-SSA} for short).
For 2D-SSA, the data object $\tX$ is a \emph{two-dimensional data array} of size
$\Nx \times \Ny$ (or simply an $\Nx \times \Ny$ real-valued matrix), represented
as $\tX = \tX_{\Nx,\Ny}=(x_{ij})_{i,j=1}^{\Nx,\Ny}$.  A typical example of a
2D-array is a digital 2D monochrome image.

2D-SSA was proposed as an extension of SSA in \citet{Danilov.Zhigljavsky1997}, and
was further developed in \citet{Golyandina.Usevich2010,Rodriguez-Aragon.Zhigljavsky10SII-Singular}.
However, its first usage can traced back to the work of \citet{Ade83SP-Characterization}
on texture analysis (this work was also continued recently, see \citet{Monadjemi04-Towards}).
Related decompositions can be found in methods for processing of seismological data
\citep{Trickett08conf-F}. Finally, as with SSA-like methods for time series,
the 2D-SSA decomposition is the base of subspace-based parameter estimation methods
for sums of two-dimensional complex exponentials (see, e.g., \citet{Rouquette.Najim01Estimation}).

A major drawback of the methods based on 2D-SSA decomposition was its computational
complexity. The \pkg{Rssa} package contains an efficient implementation of the 2D-SSA
decomposition and reconstruction, which overcomes this deficiency.

\subsection{2D-SSA algorithm}
For a matrix $\rmA \in \spaceR^{M\times N}$ (or $\spaceC^{M\times N}$), we denote by
$\mvec(\rmA) \in \spaceR^{MN}$ (or $\spaceC^{MN}$) its column-major vectorization.
For a vector $A \in \spaceR^{MN}$ (or $\spaceC^{MN}$), we define its $M$~\emph{devectorization}
as the matrix $\mvec_M^{-1}(A) = \rmB \in \spaceR^{M\times N}$ (or $\spaceC^{M\times N}$)
that satisfies $\mvec(\rmB) = A$. We mostly use the notation from the
paper \citet{Golyandina.Usevich2010}.

\subsubsection{The embedding operator}
Since the general scheme of the 2D-SSA algorithm is described in
Section~\ref{sec:common}, we need to define only the embedding operator
$\trajmat{\TWODSSA}(\tX)=\bfX$.

The parameters of the method are the two-dimensional {\em window sizes}
$(\Lx,\Ly)$, which are bounded as $1\leq \Lx\leq \Nx$, $1\leq \Ly \leq \Ny$ and
$1<\Lx \Ly<\Nx \Ny$. For convenience, we also denote $\Kx = \Nx -\Lx + 1$, $\Ky
= \Ny -\Ly + 1$.  As in the general scheme of the algorithms, we define $L =
\Lx\Ly$ (the number of rows of $\bfX$) and $K = \Kx\Ky$ (the
number of columns of $\bfX$).

Consider all possible $L_x\times L_y$ submatrices of $\tX$ (2D sliding
windows). For $k = 1,\ldots, \Kx$ and $l = 1,\ldots, \Ky$, we define by
$\tX^{(L_x,L_y)}_{k,l} = (x_{i,j})_{i=k,j=l}^{\Lx+k-1,\Ly+l-1}$ the $\Lx\times\Ly$
submatrix shown in Figure~\ref{fig:2dssa_algorithm_window}. Note that
the $x$ axis is oriented to the bottom, and the $y$ axis is oriented to the right;
the origin is the upper left corner. We use this orientation because it is consistent
with the standard mathematical indexing of matrices \citep{Golyandina.Usevich2010}.

\bfgh
\centering
\input{2dssa_algorithm_window}\caption{Moving 2D windows.}
\label{fig:2dssa_algorithm_window}
\efg

Then the trajectory matrix is defined as
\be
\label{eq:block_hankel_cols}
\trajmat{\TWODSSA} (\tX) = \bfX = [ X_{1} : \ldots : X_{\Kx\Ky} ],
\ee
where the columns are vectorizations of $\Lx \times \Ly$  submatrices:
\[
X_{k+(l-1)\Kx} = \mvec (\tX^{(\Lx,\Ly)}_{k,l})
\]

\subsubsection{Hankel-block-Hankel structure}
The trajectory matrix (\ref{eq:block_hankel_cols}) has the following structure \citep{Golyandina.Usevich2010}:
\be
\label{eq:block_hankel_matrix}
\bfX = \trajmat{\TWODSSA} (\tX) =
\left(\begin{array}{lllll}
\bfH_1        & \bfH_2 \quad      & \bfH_3\quad & \dots  & \bfH_{\Ky}   \\
\bfH_2        & \bfH_3            & \bfH_4      & \dots  & \bfH_{\Ky+1}     \\
\bfH_3        & \bfH_4            & \!\!\adots& \;\;\adots         & \;\vdots       \\
\;\vdots      & \;\vdots &\!\!\adots &    \;\;\adots      & \;\vdots       \\
\bfH_{\Ly}  & \bfH_{\Ly+1}\quad   & \dots & \dots         & \bfH_{\Ny}
\end{array}
\right),
\ee
where each $\bfH_j$ is an $\Lx \times \Kx$ Hankel matrix constructed from
 $\tX_{:,j}$ (the $j$th column of the 2D array $\tX$).  More precisely,
$\bfH_j = \trajmat{SSA}(\tX_{:,j})$, where $\trajmat{SSA}$ is defined in
(\ref{eq:ssa_embedding}).  The matrix (\ref{eq:block_hankel_cols}) is called
\emph{Hankel-block-Hankel} (shortened to \emph{HbH}), since it is block-Hankel with Hankel blocks.

\subsubsection{Trajectory space}
From (\ref{eq:block_hankel_cols}) we have that the trajectory space is the
linear space spanned by the $\Lx \times \Ly$ submatrices of $\tX$. Therefore,
the eigenvectors $U_i$ also can be viewed as vectorized $\Lx \times \Ly$
arrays. Their devectorizations are denoted by $\Psi_i = \mvec^{-1}_{\Lx}
(U_i)$.
Similarly, the rows of $\bfX$ are vectorizations of the $(\Kx,\Ky)$ submatrices
\be
\label{eq:block_hankel_rows}
\bfX = [ X^{1} : \ldots : X^{\Lx\Ly}]^{\top},\quad X^{k+(l-1)\Lx} = \mvec (\tX^{(\Kx,\Ky)}_{k,l}),
\ee
where $X^{j}$ is the $j$th row of the matrix $\tX$. Therefore, the factor
vectors $V_i$ also can be viewed as $\Kx \times \Ky$ arrays. Their
devectorizations are denoted by $\Phi_i = \mvec^{-1}_{\Kx} (U_i)$.

\subsubsection{Comments}
\begin{enumerate}
    \item The algorithm of 2D-SSA coincides with the algorithm of MSSA for time series of the same length when 
    $\Lx=1$ or $\Ly=1$ \citep{Golyandina.Usevich2010}.
    This idea will be extended later on in Section~\ref{sec:Shaped2DSSA}.

    \item The arrays of finite rank in 2D-SSA (i.e., the arrays such that
    $\trajmat{\TWODSSA}$ has finite rank) are sums of products of polynomials,
    exponentials and cosines, similarly to the one-dimensional case. More
    details can be found in Section~\ref{sec:app_2dssa_theory}.
\end{enumerate}

%% file: 2dssa_algorithm_window.tex
\begin{tikzpicture}
    \draw (0,0) rectangle (6,4);
    \draw (2,1.25) rectangle (4.5,2.75);
    \draw (0,4) node[anchor=north east]  { $1$};
    \draw (0,4) node[anchor=south west]  {$1$};
    \draw (0,0) node[anchor=south east]  {$\Nx$};
    \draw (6,4) node[anchor=south east]  {$\Ny$};

    \draw[dotted, very thick]  (2.125,4) node[above]  {$l$} -- (2.125,2.625);
    \draw[dotted, very thick] (0,2.625) node[left]  {$k$} -- (2.125,2.625);
    \draw [<->] (2,1.15) -- node[below] {$\Ly$} (4.5,1.15) ;
    \draw [<->] (4.65,1.25) -- node[right] {$\Lx$} (4.65,2.75);
    
    \draw (3.25,2) node {$\tX^{(\Lx,\Ly)}_{k,l}$};
  \end{tikzpicture}

%% file: 2dssa_package.tex
\subsection{Package}
\label{sec:2dssa_package}
\subsubsection{Typical code}
Here we demonstrate a typical decomposition of 2D images with the package
\pkg{Rssa}. We follow the example in Section~\ref{sec:ssa_typical}, but stress
on the differences that appear in 2D case.

As an example, we use the image of Mars by Pierre Thierry,  from the
tutorial\footnote{{\tt  http://www.astrosurf.com/buil/iris/tutorial8/doc23\_us.htm}}
of the free IRIS software (see the data description in the
package). The image is of size $258 \times 275$, 8-bit grayscale, values from
$0$ to $255$. The input code for this image can be found in
Fragment~\ref{frag:Mars_input} (the image is included in the package
\pkg{Rssa}).
\begin{fragment}[Mars: Input]
\label{frag:Mars_input}
\input{fragments/Mars_input.tex}
\end{fragment}
We would like to decompose this image with a $25\times 25$ window (see the example
in the paper \citet{Golyandina.Usevich2010}).  Easy calculations show that even
these small window sizes would give rise to a $625 \times 58734$ trajectory
matrix.  In Fragment~\ref{frag:Mars_25_dec_svd} with \code{svd.method = "svd"}, we
intentionally comment the call to the SSA function, because we do not recommend to
use it, unless the trajectory matrix is very small.
\begin{fragment}[Mars: Decomposition with \code{svd.method = "svd"}]
\label{frag:Mars_25_dec_svd}
\input{fragments/Mars_25_dec_svd.tex}
\end{fragment}
A remedy for this could be calculation of just the matrix $\bfX\bfX^{\top}$,
and computing its eigendecomposition (this approach was
taken in \citet{Golyandina.Usevich2010} and other papers). In the package
\pkg{Rssa}, this is implemented in \code{svd.method = "eigen"}, see
Fragment~\ref{frag:Mars_25_dec_eigen}.
\begin{fragment}[Mars: Decomposition with \code{svd.method = "eigen"}]
\label{frag:Mars_25_dec_eigen}
\input{fragments/Mars_25_dec_eigen.tex}
\end{fragment}
However, for larger window sizes this approach quickly becomes impractical,
because the complexity of the full eigendecomposition grows at least as $O(L^3)$.
Therefore, in the \pkg{Rssa} package the method \code{nutrlan} is used by default.
This gives a considerable speedup even for moderate window
sizes ($25\times 25$), as demonstrated by Fragment~\ref{frag:Mars_25_dec}.  Note
that for the 2D-SSA decomposition, \code{kind = "2d-ssa"} should be used.
\begin{fragment}[Mars: Decomposition]
\label{frag:Mars_25_dec}
\input{fragments/Mars_25_dec.tex}
\end{fragment}
Fragment~\ref{frag:Mars_25_rec} shows a typical reconstruction code for 2D-SSA.
\begin{fragment}[Mars: Reconstruction]
\label{frag:Mars_25_rec}
\input{fragments/Mars_25_rec.tex}
\end{fragment}
The reconstruction results are shown in Figure~\ref{fig:Mars_25_rec}.
\bfgh
        \begin{center}
        \includegraphics[width=12 cm]{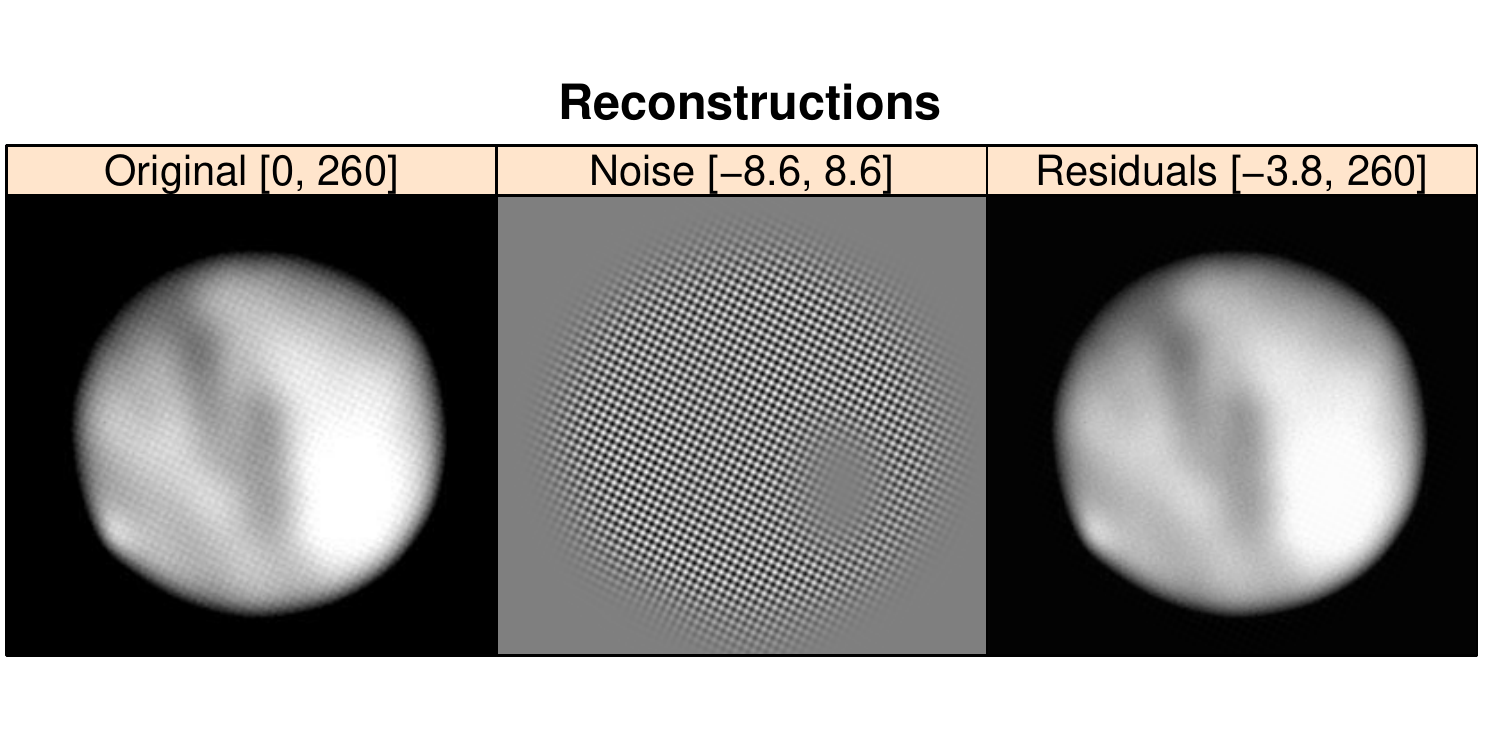}
        \end{center}
        \caption{Mars: Separated periodic noise, $(\Lx,\Ly)=(25,25)$.}
        \label{fig:Mars_25_rec}
\efg

The grouping for this decomposition was made, as in \citet{Golyandina.Usevich2010}, based on the following information:
\begin{itemize}
\item eigenarrays $\Psi_i$ (see Figure~\ref{fig:Mars_25_ident_psi}), and
\item the matrix  of $\bfw$~correlations (see Figure~\ref{fig:Mars_25_ident_wcor}).
\end{itemize}
Fragment~\ref{frag:Mars_25_ident} shows the corresponding code.
\begin{fragment}[Mars: Identification]
\label{frag:Mars_25_ident}
\input{fragments/Mars_25_ident.tex}
\end{fragment}
\bfgh
        \begin{center}
        \includegraphics[width=10 cm]{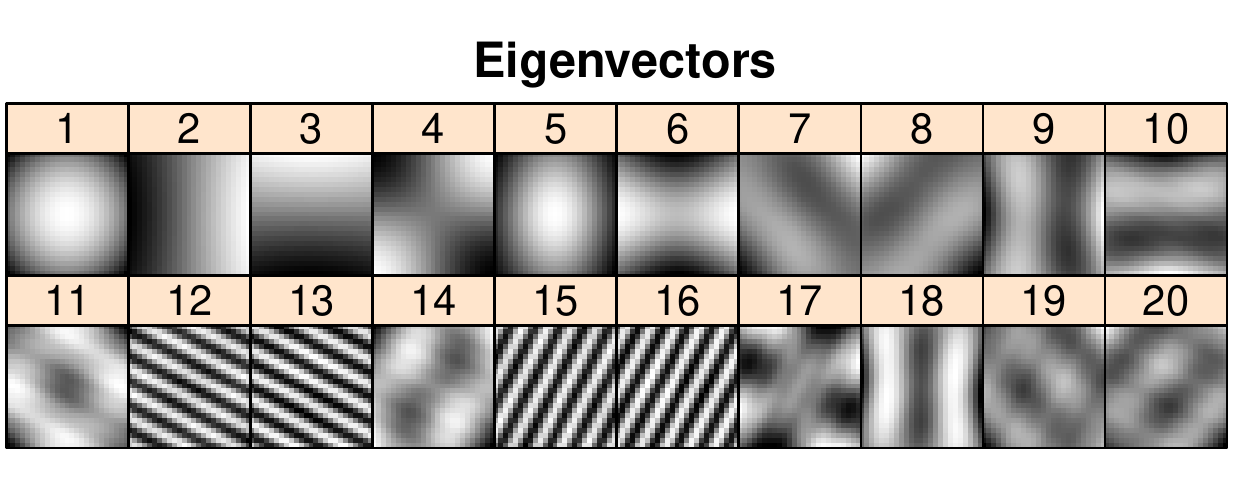}
        \end{center}
        \caption{Mars: Eigenarrays, $(\Lx,\Ly)=(25,25)$.}
        \label{fig:Mars_25_ident_psi}
\efg
\bfgh
        \begin{center}
        \includegraphics[width=8 cm]{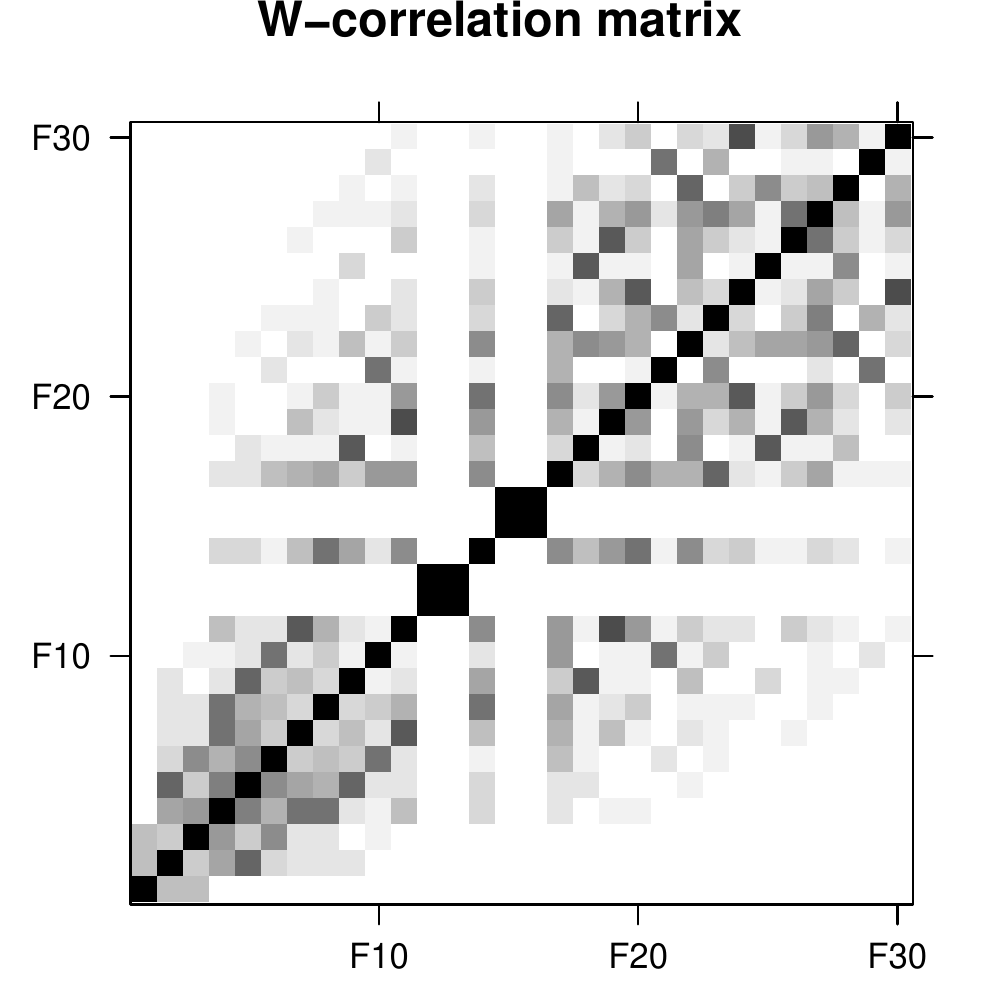}
        \end{center}
        \caption{Mars: $\bfw$~Correlations, $(\Lx,\Ly)=(25,25)$.}
        \label{fig:Mars_25_ident_wcor}
\efg

Next, we try more challenging window sizes $(\Lx,\Ly) = (160,80)$, where the
trajectory matrix is of size $12800 \times 19404$. In this case,
\code{svd.method = "eigen"} would take very long time. However, with the default
method \code{svd.method = "nutrlan"} the computation of 50 first eigenvectors
can be done in about a second, see Fragment~\ref{frag:Mars_160_80_rec}. The
results of the reconstruction are shown in Figure~\ref{fig:Mars_160_80_rec}.
\begin{fragment}[Mars: Reconstruction]
\label{frag:Mars_160_80_rec}
\input{fragments/Mars_160_80_rec.tex}
\end{fragment}
\bfgh
\begin{center}
    \includegraphics[width=12 cm]{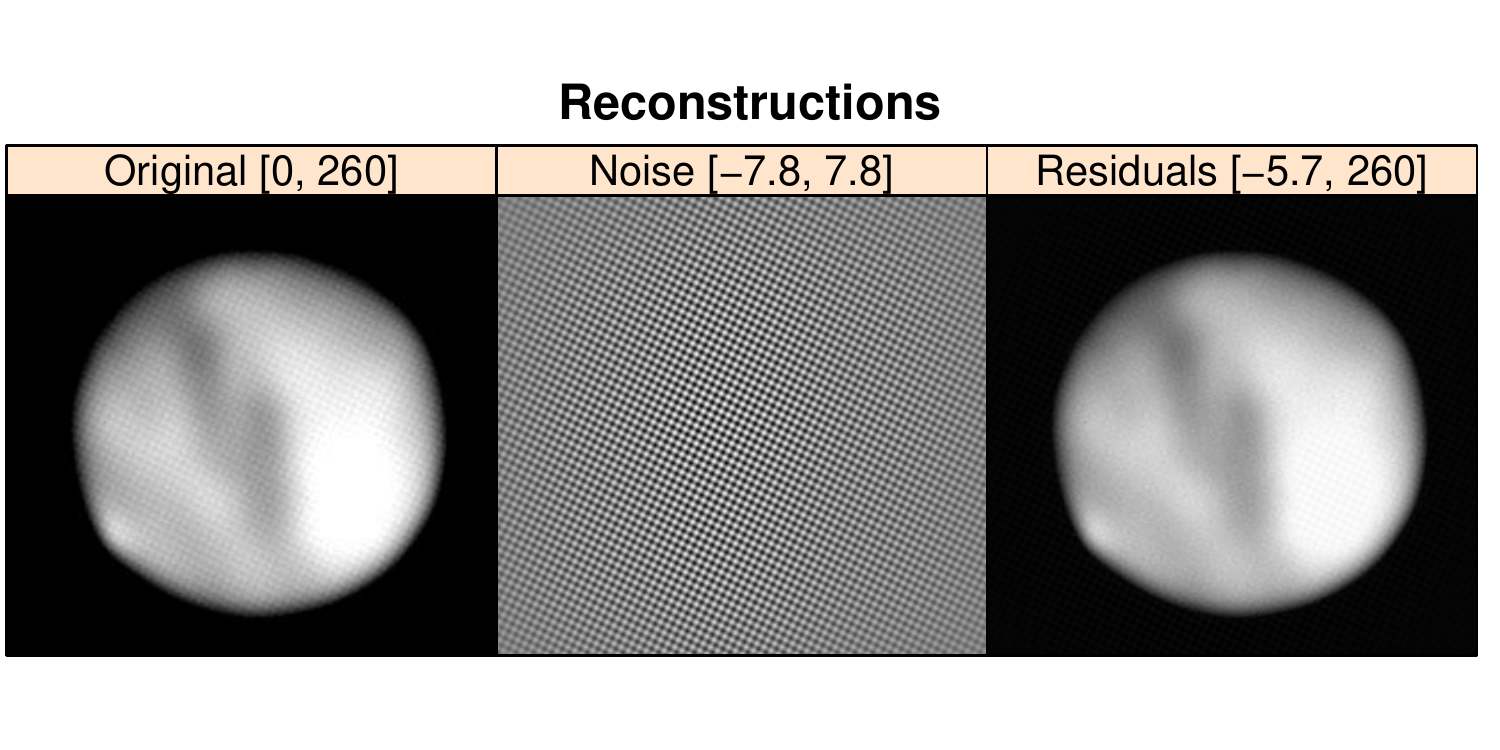}
\end{center}
\caption{Mars: Reconstruction, $(\Lx,\Ly)=(160,80)$.}
\label{fig:Mars_160_80_rec}
\efg

From Figure~\ref{fig:Mars_160_80_rec} we see that in the case of large window
sizes, the extracted periodic noise in not modulated (compare to
Figure~\ref{fig:Mars_25_rec}).  This can be interpreted as follows. We chose the
window sizes as $(160, 80)$ (approximately $(0.6 \Nx, 0.3 \Ny)$) for which the
separability of signal and noise in the parametric model should be the better
than for small window sizes (see
\citet{Golyandina2010}). On the other hand, if we choose smaller window sizes
(for example, $25 \times 25$), then the 2D-SSA decomposition acts more like
smoothing.

\subsubsection{Comments}
\paragraph{Formats of input and output data}
The input for 2D SSA is assumed to be a matrix (or an object which can be
coerced to a matrix).

\paragraph{Plotting specifics}
All the plotting routines by default use the raster representation (via
\code{useRaster = TRUE} argument provided to the \pkg{lattice} plotting
functions). It most cases it does not make sense to turn the raster mode off,
since the input is a raster image in any case. However, not all the graphical
devices support this mode.

\paragraph{Efficient implementation}
Most of the ideas from the one-dimensional case can be either transferred
directly or generalized to the 2D case. The overall computational complexity of
the direct implementation of 2D-SSA is $O(L^{3} + K L^{2})$  and thus 2D-SSA
can be prohibitively time consuming even for moderate image and window sizes.
(Recall that $L = \Lx\Ly$ and $K = \Kx\Ky$.) The ideas presented in
Section~\ref{ssec:qhmatmul} coupled with Lanczos-based truncated SVD
implementations \citep{Larsen98, Yamazaki08, Korobeynikov2010} allow to
dramatically reduce the computational complexity down to $O(kN\log{N} + k^{2}N)$,
where $N = \Nx\Ny$ and $k$ denotes the number
of desired eigentriples. Therefore, the achieved speedup can be much higher
than that for the SSA and MSSA cases.

Note that the Lanczos-based methods have significant overhead for
small trajectory matrices, so that in this case other SVD methods should be used.
For \code{svd.method="eigen"}, the matrix $\bfX\bfX^{\top}$ is
computed in $O(LN\log{N})$ flops using the fast matrix-vector multiplication from Section~\ref{ssec:qhmatmul}.
Therefore, the total complexity of the decomposition method is $O(LN\log{N}+L^3)$,
which makes the method applicable for small $L$ and moderate $N$.


%% file: fragments/Mars_input.tex
\begin{CodeChunk}
\begin{CodeInput}

> library("Rssa")
> data("Mars")
\end{CodeInput}

\end{CodeChunk}

%% file: fragments/Mars_25_dec_svd.tex
\begin{CodeChunk}
\begin{CodeInput}

> # ssa(Mars, kind = "2d-ssa", L = c(25, 25), svd.method = "svd")
\end{CodeInput}

\end{CodeChunk}

%% file: fragments/Mars_25_dec_eigen.tex
\begin{CodeChunk}
\begin{CodeInput}

> print(system.time(ssa(Mars, kind = "2d-ssa", L = c(25, 25),
+                       svd.method = "eigen")))
   user  system elapsed 
  3.886   0.061   3.843 
\end{CodeInput}

\end{CodeChunk}

%% file: fragments/Mars_25_dec.tex
\begin{CodeChunk}
\begin{CodeInput}

> print(system.time(s.Mars.25 <- ssa(Mars, kind = "2d-ssa", L = c(25, 25))))
   user  system elapsed 
  0.495   0.011   0.519 
\end{CodeInput}

\end{CodeChunk}

%% file: fragments/Mars_25_rec.tex
\begin{CodeChunk}
\begin{CodeInput}

> r.Mars.25 <- reconstruct(s.Mars.25,
+                          groups = list(Noise = c(12, 13, 15, 16)))
> plot(r.Mars.25, cuts = 255, layout = c(3, 1))
\end{CodeInput}

\end{CodeChunk}

%% file: fragments/Mars_25_ident.tex
\begin{CodeChunk}
\begin{CodeInput}

> plot(s.Mars.25, type = "vectors", idx = 1:20,
+      cuts = 255, layout = c(10, 2),
+      plot.contrib = FALSE)
> plot(wcor(s.Mars.25, groups = 1:30), scales = list(at = c(10, 20, 30)))
\end{CodeInput}

\end{CodeChunk}

%% file: fragments/Mars_160_80_rec.tex
\begin{CodeChunk}
\begin{CodeInput}

> print(system.time(s.Mars.160.80 <-
+   ssa(Mars, kind = "2d-ssa", L = c(160, 80))))
   user  system elapsed 
  0.642   0.026   0.587 
> r.Mars.160.80.groups <- list(Noise = c(36, 37, 42, 43))
> r.Mars.160.80 <- reconstruct(s.Mars.160.80, groups = r.Mars.160.80.groups)
> plot(r.Mars.160.80, cuts = 255, layout = c(3, 1))
\end{CodeInput}

\end{CodeChunk}

%% file: 2dssa_examples_smoothing.tex
\subsection{Examples}
\label{sec:2dssa_examples}
\subsubsection{Adaptive smoothing}
2D-SSA can be also used for adaptive smoothing of two-dimensional data. It was
used for this purpose in \citet{Golyandina.etal07IJED-Filtering} for smoothing
digital terrain models (DTM) and in
\citet{Holloway.etal11PCB-Gene,Golyandina.etal12PCS-Measuring} for smoothing
spatial gene expression data.

Similarly to the example in \citet{Golyandina.etal07IJED-Filtering}, we consider
an image extracted from the SRTM database. The test DTM of a region in South Wales, UK,
is extracted by the function \code{getData} of the package \pkg{raster}
\citep{Hijmans2013}. The DTM is $80 \times 100$, and it includes the Brecon
Beacons national park. The point $(\Nx,1)$ lies in a heighbourhood of
Port Talbot, $(\Nx,\Ny)$ --- in a neighbourhood of Newport, and $(1,\Ny)$ is
near Whitney-on-Wye.  In Fragment~\ref{frag:2dssa_brecon_dec}, we decompose the
image with a small window, and plot the eigenvectors and the matrix of
$\bfw$~correlations in Figure~\ref{fig:2dssa_brecon_ident}.
\begin{fragment}[Brecon Beacons: Decomposition]
\label{frag:2dssa_brecon_dec}
\input{fragments/2dssa_brecon_dec.tex}
\end{fragment}
\bfgh
        \begin{center}
\parbox{9cm}{\includegraphics[width=9 cm]{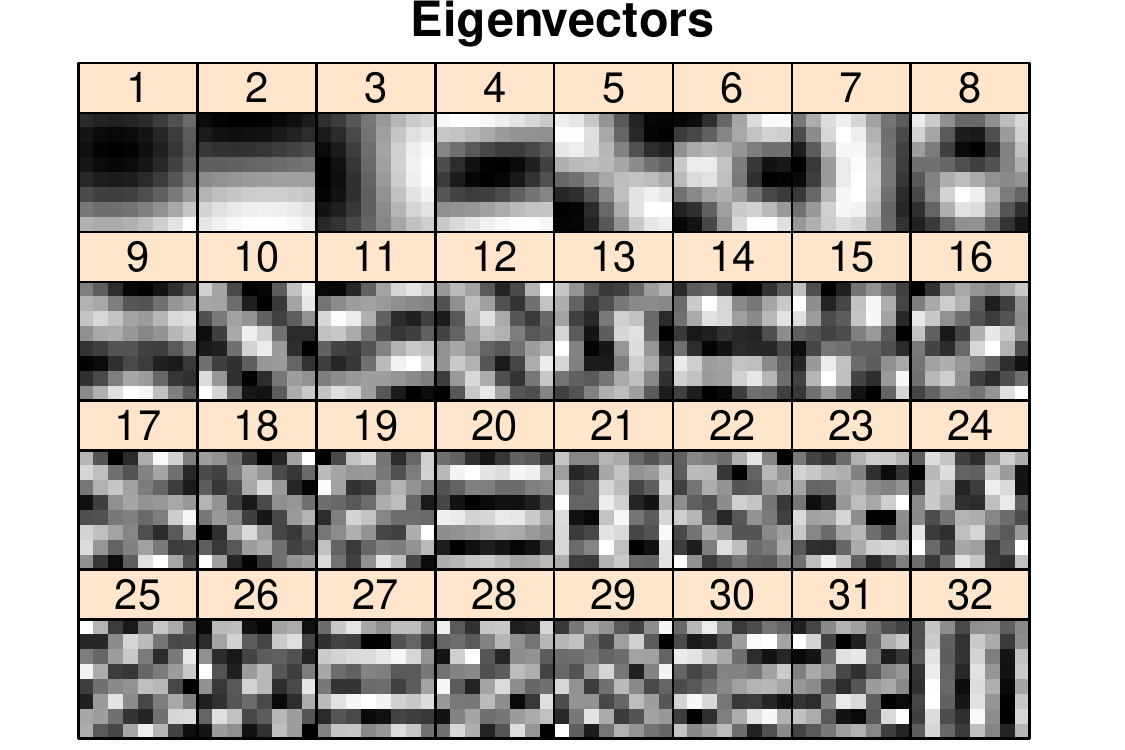}}%
\quad
\parbox{6cm}{\includegraphics[width=6 cm]{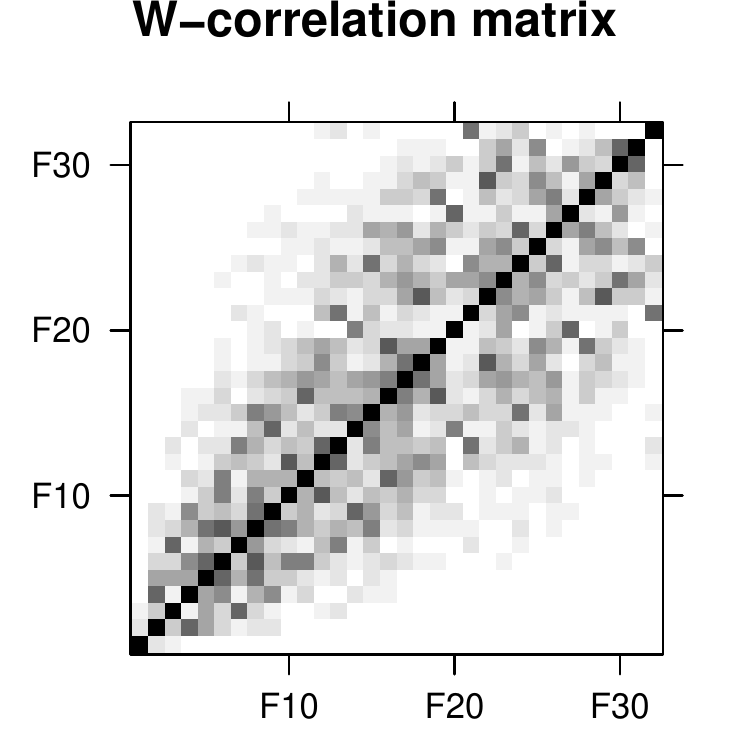}}%
        \end{center}
        \caption{Brecon Beacons: $8 \times 8$ windows, eigenarrays and $\bfw$~correlations.}
        \label{fig:2dssa_brecon_ident}
\efg
Next, we reconstruct the image with components with $I_1 = \{1,\ldots,3\}$, $I_2
= \{4,\ldots,8\}$, and $I_3 = \{9,\ldots,17\}$. The grouping was chosen based on
eigenarrays to gather frequencies of similar scale in the same components.

We take cumulative sums of reconstructed components $\wtilde{\tY}_1 =
\wtilde{\tX}_1$, $\wtilde{\tY}_1 = \wtilde{\tX}_1 + \wtilde{\tX}_2$, and
$\wtilde{\tY}_1 = \wtilde{\tX}_1 + \wtilde{\tX}_2 + \wtilde{\tX}_3$ (this is a
convenient way to compute reconstructed components for sets $J_1 = I_1$, $J_2 =
I_1 \cup I_2$, and $J_3 = I_1 \cup I_2\cup I_3$).  We plot the reconstructed
components in Figure~\ref{fig:2dssa_brecon_rec}. The cumulative components
$\wtilde{\tY}_k$ are shown in Figure~\ref{fig:2dssa_brecon_rec_cumsum} using the
parameter \code{cumsum} of the plotting function of \pkg{Rssa}.
\begin{fragment}[Brecon Beacons: Reconstruction]
\label{frag:2dssa_brecon_rec}
\input{fragments/2dssa_brecon_rec.tex}
\end{fragment}
\bfgh%
\begin{center}
\includegraphics[width=15cm]{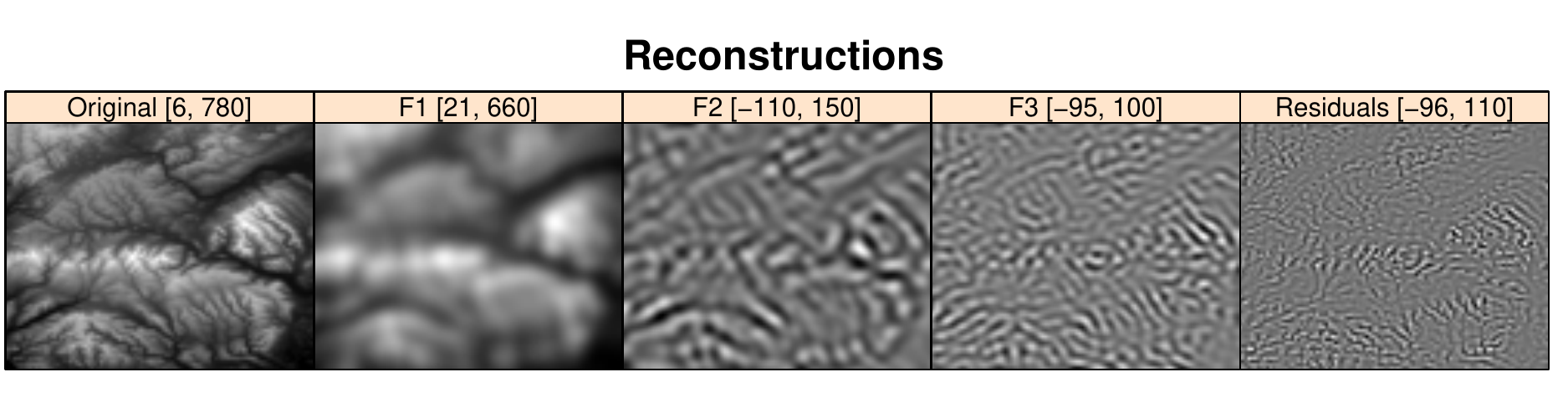}
\end{center}
        \caption{Brecon Beacons: $8 \times 8$ window, reconstructions ($\wtilde{\tX}_k$).}
        \label{fig:2dssa_brecon_rec}
\efg
\bfgh
        \begin{center}
\includegraphics[width=15cm]{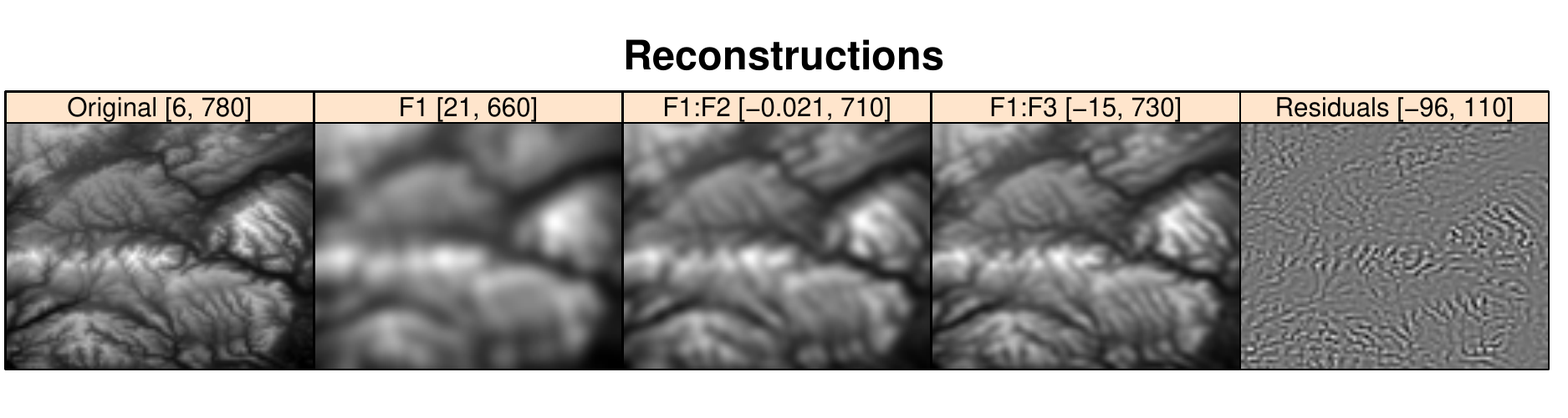}
        \end{center}
        \caption{Brecon Beacons: $8 \times 8$ window, cumulative reconstructions ($\wtilde{\tY}_k$).}
        \label{fig:2dssa_brecon_rec_cumsum}
\efg
From Figure~\ref{fig:2dssa_brecon_rec} and Figure~\ref{fig:2dssa_brecon_rec_cumsum}
it can be seen that the reconstructed components capture morphological features
of different scale \citep{Golyandina.etal07IJED-Filtering}. Cumulative
reconstructions represent smoothing of the original DTM of different resolution.

To illustrate the behaviour of smoothing, we plot the absolute values of the
centered discrete Fourier transforms (DFT) of $\tX -\wtilde{\tY}_k$ (residuals for cumulative reconstructions,
see Figure~\ref{fig:brecon_cumsum_dft}). The corresponding code can be found in
Fragment~\ref{frag:2dssa_brecon_rec_dft}. We also introduce some code for
plotting the arrays and computing their centered DFTs in Fragment~\ref{frag:2dssa_plots}.

\begin{fragment}[2D-SSA: plotting functions]
\label{frag:2dssa_plots}
\input{fragments/2dssa_plots.tex}
\end{fragment}

\begin{fragment}[Brecon Beacons: DFT of cumulative reconstructions]
\label{frag:2dssa_brecon_rec_dft}
\input{fragments/2dssa_brecon_rec_dft.tex}
\end{fragment}

In Figure~\ref{fig:brecon_cumsum_dft},
it is clearly seen that 2D-SSA reconstruction by leading components acts as
filter that preserves dominating frequencies (in this case, a low-pass filter).

\bfgh
\begin{center}
\parbox{4cm}{\includegraphics[width=4 cm]{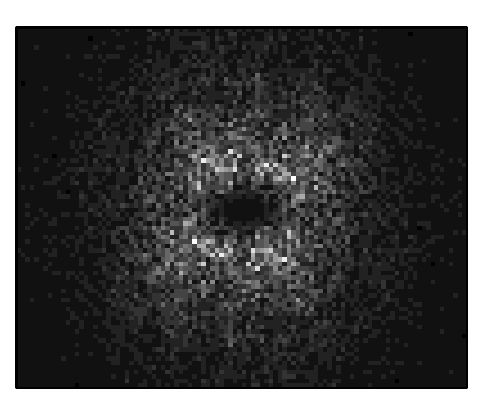}}%
\quad
\parbox{4cm}{\includegraphics[width=4 cm]{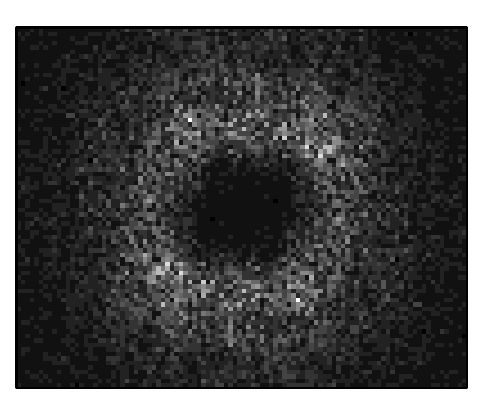}}%
\quad
\parbox{4cm}{\includegraphics[width=4 cm]{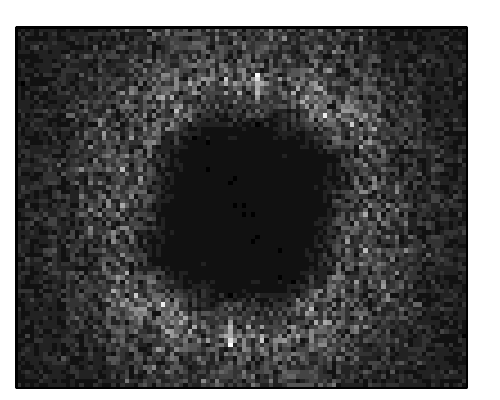}}%
        \end{center}
        \caption{Brecon Beacons: $8 \times 8$ window, absolute values of the DFT
                 of $\tX - \wtilde{\tY}_k$, $k=1,\ldots,3$.}
        \label{fig:brecon_cumsum_dft}
\efg

%% file: fragments/2dssa_brecon_dec.tex
\begin{CodeChunk}
\begin{CodeInput}

> library("raster")
> library("Rssa")
> UK <- getData("alt", country = "GB", mask = TRUE)
> brecon <- crop(UK, extent(UK, 1040, 1119, 590, 689))
> m.brecon <- as.matrix(brecon)
> # ssa does as.matrix inside, but cannot see as.matrix.raster
> s.brecon <- ssa(m.brecon, kind = "2d-ssa", L = c(8, 8),
+                 svd.method = "eigen")
> plot(s.brecon, type = "vectors", idx = 1:32,
+      cuts = 255, layout = c(8, 4), plot.contrib = FALSE)
> plot(wcor(s.brecon, groups = 1:32), scales = list(at = c(10, 20, 30)))
\end{CodeInput}

\end{CodeChunk}

%% file: fragments/2dssa_brecon_rec.tex
\begin{CodeChunk}
\begin{CodeInput}

> r.brecon <- reconstruct(s.brecon, groups = list(1:3, 4:8, 9:17))
> plot(r.brecon, cuts = 255, layout = c(5, 1), 
+      par.strip.text = list(cex = 0.75))
> plot(r.brecon, cuts = 255, layout = c(5, 1), 
+      par.strip.text = list(cex = 0.75), type = "cumsum", at = "free")
> # Convert reconstructed objects back to raster, if necessary
> brecon.F1 <- raster(r.brecon$F1, template = brecon)
> brecon.F2 <- raster(r.brecon$F2, template = brecon)
\end{CodeInput}

\end{CodeChunk}

%% file: fragments/2dssa_plots.tex
\begin{CodeChunk}
\begin{CodeInput}

> plot2d <- function(x) {
+   regions <- list(col = colorRampPalette(grey(c(0, 1))));
+   levelplot(t(x[seq(nrow(x), 1, -1), ]), aspect = "iso",
+             par.settings = list(regions = regions), colorkey = FALSE, 
+             scales = list(draw = FALSE, relation = "same"), 
+             xlab = "", ylab = "")
+ }
> centered.mod.fft <- function(x) {
+   N <- dim(x)
+   shift.exp <- exp(2i*pi * floor(N/2) / N)
+   shift1 <- shift.exp[1]^(0:(N[1] - 1))
+   shift2 <- shift.exp[2]^(0:(N[2] - 1))
+   Mod(t(mvfft(t(mvfft(outer(shift1, shift2) * x)))))
+ }
\end{CodeInput}

\end{CodeChunk}

%% file: fragments/2dssa_brecon_rec_dft.tex
\begin{CodeChunk}
\begin{CodeInput}

> library("lattice")
> plot2d(centered.mod.fft(m.brecon-r.brecon$F1))
> plot2d(centered.mod.fft(m.brecon-r.brecon$F1-r.brecon$F2))
> plot2d(centered.mod.fft(m.brecon-r.brecon$F1-r.brecon$F2-r.brecon$F3))
\end{CodeInput}

\end{CodeChunk}

%% file: 2dssa_examples_esprit.tex
\subsubsection{Parameter estimation}
\label{sec:app_2dssa_theory_esprit}
2D-SSA decomposition is also used in subspace-based methods of parameter
estimation.  Let $\tX = \tS + \tR$, where $\tS$ is an array of finite rank and
$\tR$ is the residual.  If the signal and noise are (approximately) separable,
then the matrix $\widehat{U} \in \bbR^{\Lx\Ly \times r}$ of the basis
eigenvectors approximates the original signal subspace of $\tS$.

Some of the methods of 2D-ESPRIT type are implemented in \pkg{Rssa}. The methods
are based on computing a pair of shift matrices for $x$ and $y$ directions and
their joint diagonalization \citep{Rouquette.Najim01Estimation}. Currently, two
methods of joint diagonalization are implemented in \pkg{Rssa}
2D-ESPRIT (from \citet{Rouquette.Najim01Estimation}) and 2D-MEMP with
improved pairing step (see
\citet{Rouquette.Najim01Estimation,Wang.etal05IToSP-Comments}).

We continue the example of Mars from
Fragment~\ref{frag:Mars_160_80_rec}.  This example demonstrates advantages of
\pkg{Rssa} because of a possibility to choose large window sizes, which was
always considered to be problematic ESPRIT-type methods.
Fragment~\ref{frag:Mars_160_80_esprit} shows the corresponding code that outputs
the estimated exponentials.  In Figure~\ref{fig:Mars_160_80_esprit}, the estimated
pairs of complex exponentials $(\lambda_k, \mu_k)$, are shown on separate complex
plane plots and the points for $\lambda_k$ and $\mu_k$ have the
same color (for the same $k$). 
 
\begin{fragment}[Mars: Parameter estimation with 2D-ESPRIT]
\label{frag:Mars_160_80_esprit}
\input{fragments/Mars_160_80_esprit.tex}
\end{fragment}
\bfgh
        \begin{center}
        \includegraphics[width=8 cm]{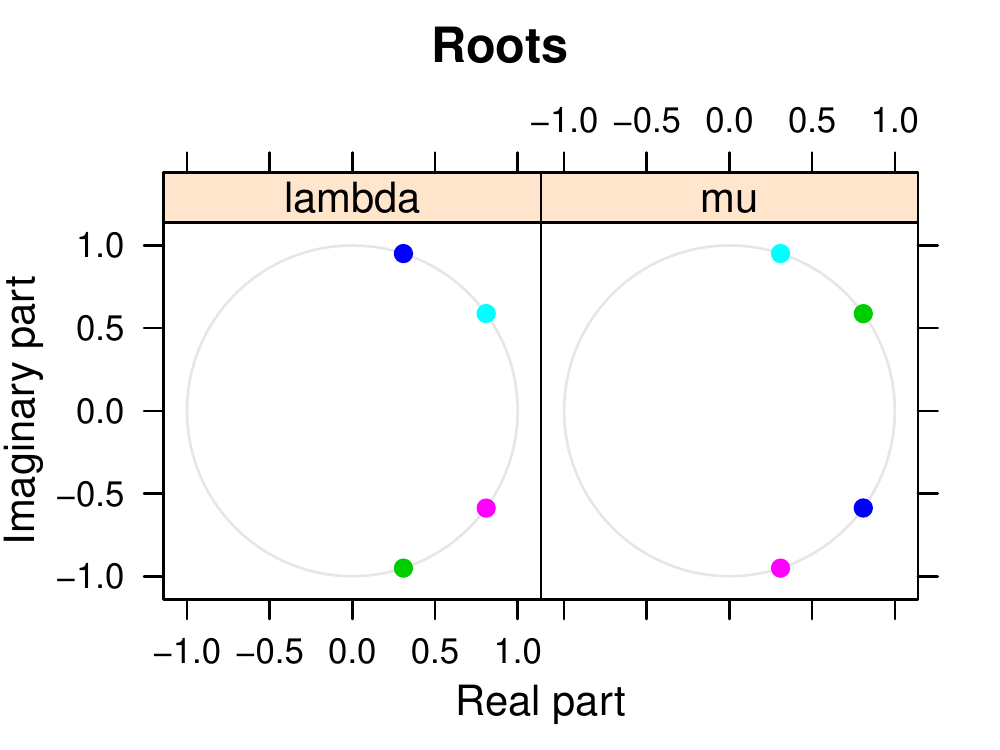}
        \end{center}
        \caption{Mars: Parameter estimation with 2D-ESPRIT, $(\Lx,\Ly)=(160,80)$.}
        \label{fig:Mars_160_80_esprit}
\efg

In Fragment~\ref{frag:Mars_160_80_esprit} and Figure~\ref{fig:Mars_160_80_esprit},
it can be seen that each pair $(\lambda_k, \mu_k)$ has its conjugate
counterpart $(\lambda_{k'}, \mu_{k'}) = \conjug{(\lambda_k, \mu_k)}$
(where $\conjug{\cdot}$ denotes the complex conjugation). Indeed,
it is the case for $k=1, k'=2$, and for $k=3, k'=4$. Therefore, the periodic
noise is a sum of two planar sines, as explained in
Section~\ref{sec:app_2dssa_theory}. This is also confirmed by the plots of
eigenarrays in Figure~\ref{fig:Mars_160_80_psi}.
\bfgh
        \begin{center}
        \includegraphics[width=6 cm]{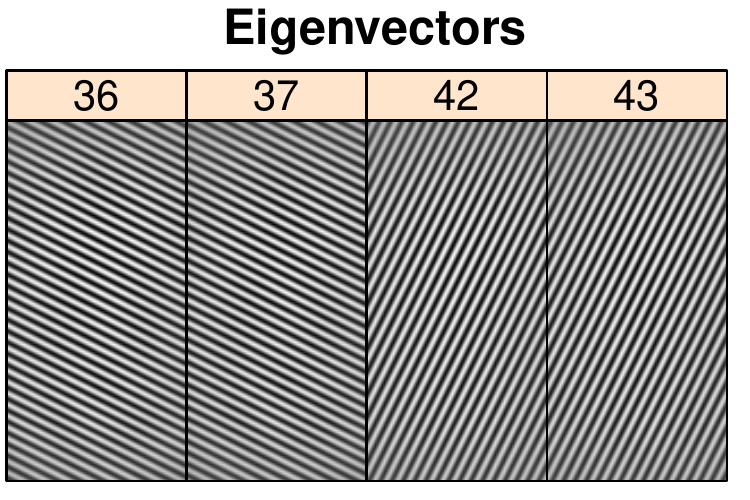}
        \end{center}
        \caption{Mars: Eigenarrays corresponding to periodic noise, $(\Lx,\Ly)=(160,80)$.}
        \label{fig:Mars_160_80_psi}
\efg

%% file: fragments/Mars_160_80_esprit.tex
\begin{CodeChunk}
\begin{CodeInput}

> pe.Mars.160.80 <- parestimate(s.Mars.160.80, groups = r.Mars.160.80.groups)
> # Overview
> print(pe.Mars.160.80)
x: period     rate   | y: period     rate
   -5.000  -0.000169 |    10.003  -0.000111 
    5.000  -0.000169 |   -10.003  -0.000111 
    9.995   0.000175 |     4.999  -0.000093 
   -9.995   0.000175 |    -4.999  -0.000093 
> # Detailed output
> print(pe.Mars.160.80[[1]])
   period     rate   |    Mod     Arg  |     Re        Im
   -5.000  -0.000169 |  0.99983  -1.26 |  0.30906  -0.95087
    5.000  -0.000169 |  0.99983   1.26 |  0.30906   0.95087
    9.995   0.000175 |  1.00017   0.63 |  0.80897   0.58814
   -9.995   0.000175 |  1.00017  -0.63 |  0.80897  -0.58814
> print(pe.Mars.160.80[[2]])
   period     rate   |    Mod     Arg  |     Re        Im
   10.003  -0.000111 |  0.99989   0.63 |  0.80905   0.58755
  -10.003  -0.000111 |  0.99989  -0.63 |  0.80905  -0.58755
    4.999  -0.000093 |  0.99991   1.26 |  0.30879   0.95103
   -4.999  -0.000093 |  0.99991  -1.26 |  0.30879  -0.95103
> plot(pe.Mars.160.80, col = c(11, 12, 13, 14))
> plot(s.Mars.160.80, type = "vectors", idx = r.Mars.160.80.groups$Noise,
+      cuts = 255, layout = c(4, 1), plot.contrib = FALSE)
\end{CodeInput}

\end{CodeChunk}

%% file: shaped_algorithm.tex
\emph{Shaped 2D-SSA} (or \emph{ShSSA} for short) is a generalization of 2D-SSA,
which allows an arbitrary shape of the input array and window.  
This feature considerably extends the range of real-life applications, since 
it makes it possible to decompose parts of the image with different 
structures separately, exclude the areas with corrupted data, analyze images
with gaps, decompose non-rectangular images, etc. 
In ShSSA, not all the values of the rectangular image need be specified, and the sliding
window is not necessarily rectangular (see Figure~\ref{fig:shaped_algorithm_window}).

Formally speaking, a \emph{shape} $\mathfrak{B}$ is a bounded subset of
$\spaceN^2$ (a set of two-dimensional natural indices). A
\emph{$\mathfrak{B}$-shaped array} is a partially indexed array $\tX =
\tX_{\mathfrak{B}} = (x_{(i,j)})_{(i,j) \in \mathfrak{B}}$.

For two-dimensional indices $\LsE = (\LsE_x,\LsE_y)$ and $\KsE = (\KsE_x,\KsE_y)$
we define a shifted sum
\[
\LsE \shs \KsE = (\LsE_x+\KsE_x-1, \LsE_y+\KsE_y-1).
\]
We also define a shifted Minkowski sum of two shapes $\mathfrak{A}$ and $\mathfrak{B}$
as
\[
\mathfrak{A} \shs \mathfrak{B} = \{\alpha\shs\beta \,|\, \alpha \in \mathfrak{A}, \beta \in \mathfrak{B}\}.
\]

\subsection{Construction of the trajectory matrix}
The input data for the algorithm is an $\Ns$-shaped array $\tX = \tX_{\Ns} =
(x_\alpha)_{\alpha \in \Ns}$, where $\Ns \subseteq \{1,\ldots,\Nx\} \times
\{1,\ldots,\Ny\}$.  The parameter of the algorithm is a \emph{window shape} $\Ls
\subseteq \{1,\ldots,\Lx\} \times \{1,\ldots,\Ly\}$, given as $\Ls = \{\LsE_1,
\ldots, \LsE_L\}$, where $\LsE_i \in \spaceN^2$ are ordered in lexicographical
order (i.e., the order in which the elements $x_\alpha$ would appear in the
vectorized rectangular array $\mvec (\tX_{\{1,\ldots,\Nx\} \times
    \{1,\ldots,\Ny\}})$).

\bfgh
\centering
\input{shaped_algorithm_window.tex}
\caption{Moving shaped windows.}
\label{fig:shaped_algorithm_window}
\efg

\subsubsection{The embedding operator}
For $\KsE \in \spaceN^2$ we define a shifted $\Ls$-shaped subarray as
$\tX_{\Ls\shs\{\KsE\}} = (x_{\alpha})_{\alpha\in\Ls\shs\{\KsE\}}$ (see
Figure~\ref{fig:shaped_algorithm_window}).  The index $\KsE$ is a position of the
origin for the window. Consider the set of all $K$ possible origin positions for
$\Ls$-shaped windows: \be\label{eq:sh_Ks} \Ks = \{\KsE \in \spaceN^2 \,|\, \Ls
\shs \{\KsE\} \subseteq \Ns \}.  \ee We assume that $\Ks = \{\KsE_1, \ldots,
\KsE_K\} \subset \spaceN^2$, where $\KsE_j$ are ordered in lexicographical
order. Then the trajectory matrix is constructed as follows: \be
\label{eq:quasi_hankel_cols}
\trajmat{ShSSA} (\tX) := \bfX = [ X_{1} : \ldots : X_{K} ],
\ee
where the columns 
\[
X_{j} = (x_{\LsE_i\shs \KsE_j})^l_{i=1}
\]
are vectorizations of the shaped submatrices $\tX_{\Ls\shs\{\KsE_j\}}$.

\subsubsection{Quasi-Hankel matrix}
The trajectory matrix is exactly the \emph{quasi-Hankel} matrix 
\citep{Mourrain.Pan00Joc-Multivariate} constructed from the sets
$\Ls,\Ks\subset\spaceN^2$: \be\label{eq:qh_mat} \bfX = \trajmat{ShSSA} (\tX) =
\begin{pmatrix}
x_{\LsE_1\shs\KsE_1}  & x_{\LsE_1\shs\KsE_2}  & \dots  & x_{\LsE_1\shs\KsE_K} \\
x_{\LsE_2\shs\KsE_1}  & x_{\LsE_2\shs\KsE_2}  & \dots  & x_{\LsE_2\shs\KsE_K}  \\
\vdots               & \vdots               &        & \vdots               \\
x_{\LsE_L\shs\KsE_1}  & x_{\LsE_L\shs\KsE_2}  & \dots  & x_{\LsE_L\shs\KsE_K}  \\
\end{pmatrix}.  \ee Note that \citet{Mourrain.Pan00Joc-Multivariate}
use the conventional sum of indices instead of the shifted sum $\shs$, because their
definition of natural numbers $\spaceN$ includes $0$.

\subsubsection{Comments}
If $\Ns = \{1,\ldots,\Nx\} \times \{1,\ldots,\Ny\}$ and $\Ls = \{1,\ldots,\Lx\}
\times \{1,\ldots,\Ly\}$ (therefore, $\Ks = \{1,\ldots,\Kx\} \times
\{1,\ldots,\Ky\}$), then the matrix $\trajmat{ShSSA} (\tX)$ coincides with
$\trajmat{\TWODSSA} (\tX)$.  Therefore, the ShSSA decomposition with the rectangular
window and input array coincides with the 2D-SSA decomposition.

\bfgh
\centering
\input{shaped_algorithm_areas.tex}
\caption{Construction of $\Ks$.}
\label{fig:shaped_algorithm_areas}
\efg

In Figure~\ref{fig:shaped_algorithm_areas}, we demonstrate the case of general
shapes. The set $\Ks$ corresponds to all possible positions of the left upper
corner of the bounding box around the shaped window (depicted by green dots in
Figure~\ref{fig:shaped_algorithm_areas}). It also corresponds to the maximal set
$\Ks$ such that $\Ls \shs \Ks \subseteq \Ns$.  We describe an algorithm for
computation of the set $\Ks$ in Section~\ref{sec:implementation}.

It may happen that some of the elements of the original shaped array $\tX_{\Ns}$
do not enter in the trajectory matrix (for example, the elements $(3,8)$ and $(4,8)$, shown
in gray in Figure~\ref{fig:shaped_algorithm_areas}).  In this case, the ShSSA
analysis applies only to the $\Ns'$-shaped subarray $\tX_{\Ns'}$, where $\Ns'$ is
the set of all indices that enter in the trajectory matrix. (In fact, $\Ns' =
\Ks \shs \Ls$ .)

\subsubsection{Column and row spaces}
From the construction of the trajectory matrix, the trajectory space is the
linear space spanned by all $\Ls$-shaped subarrays and the eigenvectors $U_i$
also can be viewed as $\Ls$-shaped subarrays. These arrays, as in 2D-SSA, are
called eigenarrays.

Analogously, each row $X^{i}$ of $\bfX$ is a vectorized (in lexicographical order)
$\Ks$-shaped array $\tX_{\LsE_i + \KsE-1}$. This is a $\Ks$-shaped subarray of
$\tX$ starting from the element $\LsE_i$. Therefore, the factor vectors $V_i$
also can be viewed as $\Ks$-shaped subarrays. These arrays are called
factor arrays.

%% file: shaped_algorithm_window.tex
 \begin{tikzpicture}
    \draw[color = gray] (0,0) rectangle (6,4);
    \draw[color= gray] (2,1.25) rectangle (4.5,2.75);
    \draw (0,4) node[anchor=north east]  { $1$};
    \draw (0,4) node[anchor=south west]  {$1$};
    \draw (0,0) node[anchor=south east]  {$\Nx$};
    \draw (6,4) node[anchor=south east]  {$\Ny$};

    \draw (0,1) -- (0,3) -- (0.25,3) -- (0.25,3.5) -- (0.75, 3.5) -- (0.75,3.75) --
          (1,3.75) -- (1,4) -- (5.5,4) -- (5.5, 3.75) -- (5.75, 3.75) -- (5.75,3.25) --
          (6, 3.25) -- (6,1.25) -- (5.75,1.25) --(5.75,1) -- (5.75,0.75) -- (5.5,0.75) --
          (5.5,0.5) -- (5.25,0.5) -- (5.25,0) -- (1,0) -- (1,0.25) -- (0.75,0.25) --
          (0.75,0.5) -- (0.5,0.5) -- (0.5,0.75) -- (0.25,0.75) -- (0.25,1) -- (0,1);

    \draw (2,1.5) -- (2,2.25) -- (2.25,2.25) -- (2.25, 2.5) -- (2.5,2.5) -- (2.5,2.75) --
          (4,2.75) -- (4,2.5) -- (3.75,2.5) --(3.75,2.25) -- (4.5, 2.25) -- (4.5,1.75) --
          (4.25,1.75) -- (4.25,1.5) -- (4,1.5) -- (4,1.25) -- (2.5,1.25) -- (2.5,1.5) --
          (2,1.5);

    \draw[dotted, very thick]  (2.125,4) node[above]  {$l$} -- (2.125,2.625);
    \draw[dotted, very thick] (0,2.625) node[left]  {$k$} -- (2.125,2.625);
    \draw [<->] (2,1.15) -- node[below] {$\Ly$} (4.5,1.15) ;
    \draw [<->] (4.65,1.25) -- node[right] {$\Lx$} (4.65,2.75);
    
    \draw (3.25,2) node {$\tX_{\Ls + \{(k,l)\}}$};
  \end{tikzpicture}

%% file: shaped_algorithm_areas.tex
 \begin{tikzpicture}
    \draw[color = gray, fill=lightgray] (0,0) rectangle (4,3);
    \draw[color= gray, fill=lightgray] (0.5,1.5) rectangle (1.5,3);
    \draw (0,3) node[anchor=north east]  {\small$1$};
    \draw (0,3) node[anchor=south west]  {\small$1$};
    \draw (0,0) node[anchor=south east]    {\small$6$};
    \draw (4,3) node[anchor=south east]  {\small$8$};

    \draw[color=black, help lines, line width=.1pt] (0,0)
      grid[xstep=0.5cm, ystep=0.5cm] (4,3);
    \draw[fill=white] (0,2) -- (0.5,2) -- (0.5,2.5) -- (1,2.5) -- (1,3) -- (2.5,3) --
          (2.5,2.5) -- (3.5,2.5) -- (3.5,2) -- (4,2) -- (4,1) -- 
          (3.5,1) -- (3.5, 0.5) -- (2.5,0.5) --
          (2.5,0) -- (1,0) -- (1,0.5) -- (0.5,0.5) -- (0.5,1) -- (0,1) -- (0,2);

     \draw[fill=gray] (0.75,2.75) circle(.05);

    \draw[fill=green] (0.75,2.75) circle(.05);
    \draw[color=black,fill=lightgray] (-7.5,1.5)
     rectangle (-6.5,3);

    \draw[color = gray, fill=lightgray] (-5,1) rectangle (-1.5,3);
    \draw[color=black, help lines, line width=.1pt] (-5,1)
      grid[xstep=0.5cm, ystep=0.5cm] (-1.5,3);
        \draw[color = green, fill=white] (-5,2.5) -- (-4.5,2.5) -- (-4.5,3) -- (-3,3) --
          (-3,2.5) -- (-2,2.5) -- (-2,1.5)  -- (-3,1.5) --
          (-3,1) -- (-4,1) -- (-4,1.5) -- (-4.5,1.5) -- (-4.5,2) -- (-5,2) -- (-5,2.5);
    \draw[fill=green] (-4.25,2.75) circle(.05);
    \draw[fill=green] (-3.75,2.75) circle(.05);
    \draw[fill=green] (-3.25,2.75) circle(.05);
    \draw[fill=green] (-4.75,2.25) circle(.05);
    \draw[fill=green] (-4.25,2.25) circle(.05);
    \draw[fill=green] (-3.75,2.25) circle(.05);
    \draw[fill=green] (-3.25,2.25) circle(.05);
    \draw[fill=green] (-2.75,2.25) circle(.05);
    \draw[fill=green] (-2.25,2.25) circle(.05);
    \draw[fill=green] (-4.25,1.75) circle(.05);
    \draw[fill=green] (-3.75,1.75) circle(.05);
    \draw[fill=green] (-3.25,1.75) circle(.05);
    \draw[fill=green] (-2.75,1.75) circle(.05);
    \draw[fill=green] (-2.25,1.75) circle(.05);
    \draw[fill=green] (-3.75,1.25) circle(.05);
    \draw[fill=green] (-3.25,1.25) circle(.05);
    
    \draw[fill = lightgray!50] (3.5,1) rectangle (4,1.5);
    \draw[fill = lightgray!50] (3.5,1.5) rectangle (4,2);
    \draw[color = red, thick, fill=white] (-7.5,1.5) -- (-7.5,2.5) -- (-7,2.5) -- (-7,3) --
          (-6.5,3) -- (-6.5,1.5) -- (-7.5,1.5);

    \draw[color = red, thick] (0.5,1.5) -- (0.5,2.5) -- (1,2.5) -- (1,3) --
          (1.5,3) -- (1.5,1.5) -- (0.5,1.5);

    \draw (-7.5,3) node[anchor=north east]  {\small$1$};
    \draw (-7.5,3) node[anchor=south west]  {\small$1$};
    \draw (-7.5,1.5) node[anchor=south east]    {\small$3$};
    \draw (-6.5,3) node[anchor=south east]  {\small$2$};

    \draw (-5,3) node[anchor=north east]  {\small$1$};
    \draw (-5,3) node[anchor=south west]  {\small$1$};

    \draw (-5,1) node[anchor=south east]    {\small$4$};
    \draw (-1.5,3) node[anchor=south east]  {\small$7$};

    \draw (-5.75,2) node {$\shs$};
    \draw (-0.75,2) node {$=$};

    \draw (-7,2) node {$\Ls$};
    \draw (-3.5,2) node {$\Ks$};
    \draw (2,1.5) node {$\Ns'$};
  \end{tikzpicture}

%% file: shaped_package.tex
\subsection{Package}
\subsubsection{Typical code}
We repeat the experiment from Section~\ref{sec:2dssa_package}
(noise removal from the image of Mars), but using the Shaped
2D-SSA. Therefore, the code for loading the image is the same as in
Fragment~\ref{frag:Mars_input}.

The array shape can be specified in two different ways:
\begin{itemize}
\item by passing the \code{NA} values in the input array (these elements are excluded), or
\item by specifying the parameter \code{mask} --- a logical $\Nx \times \Ny$ array (the indicator of $\Ns$).
\end{itemize}
If both shape specifications are present, their intersection is considered.  The
shape of the window is typically passed as an $\Lx \times \Ly$ logical array
(\code{wmask}).  The shapes can be also specified by a command \code{circle}, as
shown in Fragment~\ref{frag:Mars_shaped_dec}.
\begin{fragment}[Mars: Mask specification and decomposition]
\label{frag:Mars_shaped_dec}
\input{fragments/Mars_shaped_dec.tex}
\end{fragment}
In Figure~\ref{fig:Mars_shaped_mask} one can see both types of masks and the
combined mask. 
\bfgh
        \begin{center}
\parbox{3cm}{\includegraphics[width=3 cm]{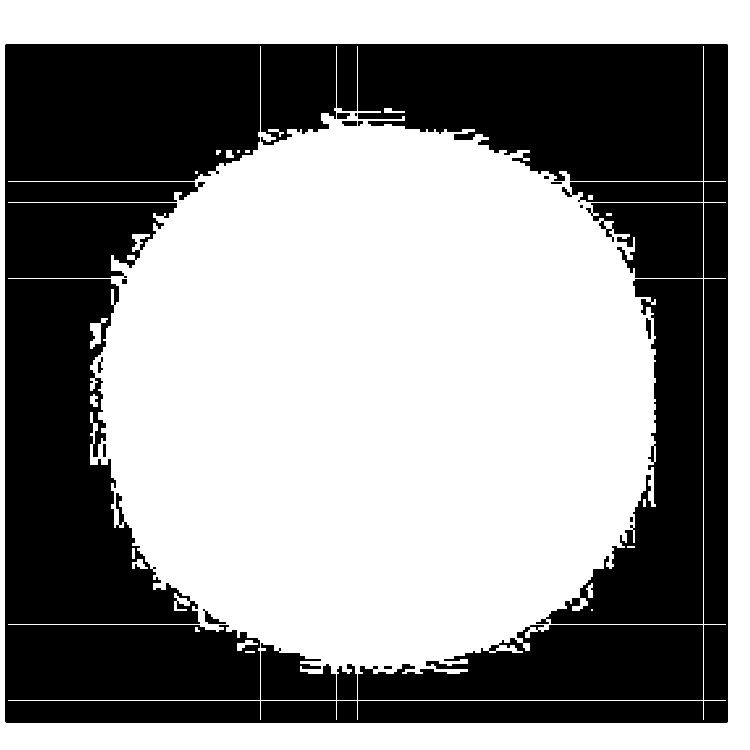}}%
\qquad
\parbox{3cm}{\includegraphics[width=3 cm]{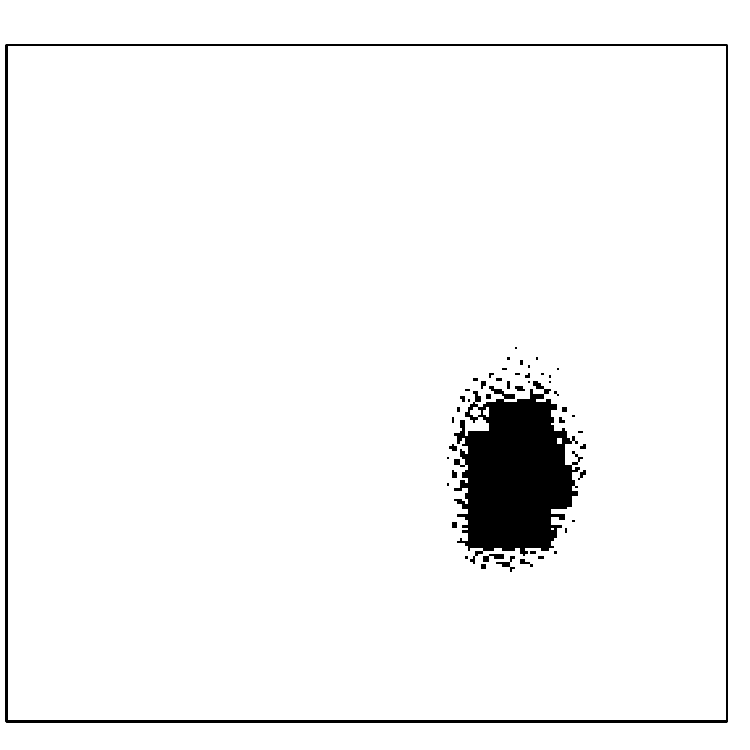}}%
\qquad
\parbox{3cm}{\includegraphics[width=3 cm]{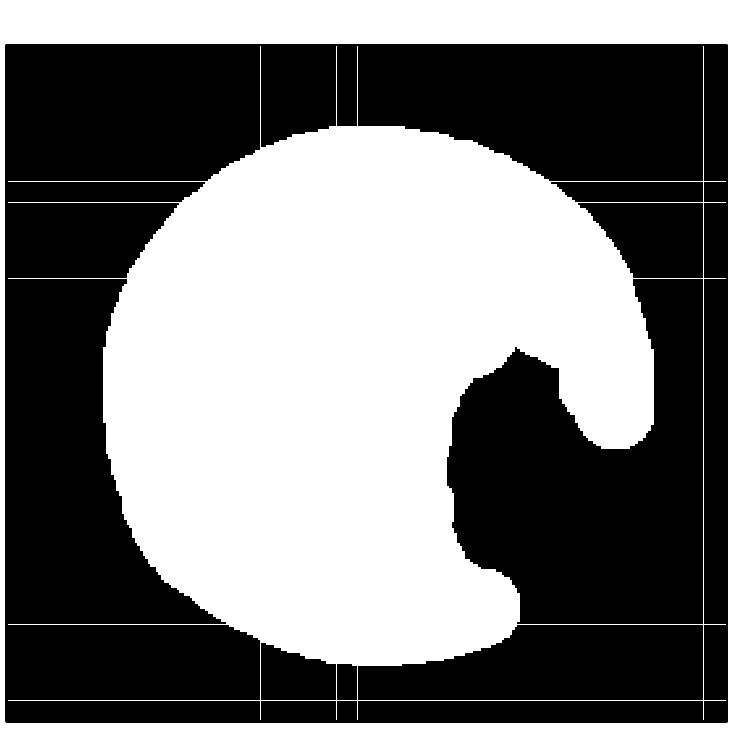}}%
        \end{center}
        \caption{Mars masks specification. Left: specified by \code{NA}, center:
            the parameter \code{mask}, right: resulting mask.
            White squares --- \code{TRUE}, black squares --- \code{FALSE}.}
        \label{fig:Mars_shaped_mask}%
\efg

Fragment~\ref{frag:Mars_shaped_rec} shows a typical reconstruction code for ShSSA.
\begin{fragment}[Mars: Reconstruction]
\label{frag:Mars_shaped_rec}
\input{fragments/Mars_shaped_rec.tex}
\end{fragment}
The reconstruction results are shown in Figure~\ref{fig:Mars_shaped_rec}.  In
Figure~\ref{fig:Mars_shaped_rec} we can see that the elements are reconstructed
only inside the resulting mask, however the original array is drawn for all
available elements (except the \code{NA} values).
The grouping for this decomposition was made based on the following information:
\begin{itemize}
\item eigenarrays (see Figure~\ref{fig:Mars_shaped_ident_psi}), and
\item the matrix of matrix of $\bfw$-correlations (see Figure~\ref{fig:Mars_shaped_ident_wcor}).
\end{itemize}
Fragment~\ref{frag:Mars_shaped_ident} shows the code that reproduces Figures~\ref{fig:Mars_shaped_ident_psi} and
\ref{fig:Mars_shaped_ident_wcor}.
\begin{fragment}[Mars: Identification]
\label{frag:Mars_shaped_ident}
\input{fragments/Mars_shaped_ident.tex}
\end{fragment}
\bfgh
        \begin{center}
        \includegraphics[width=12 cm]{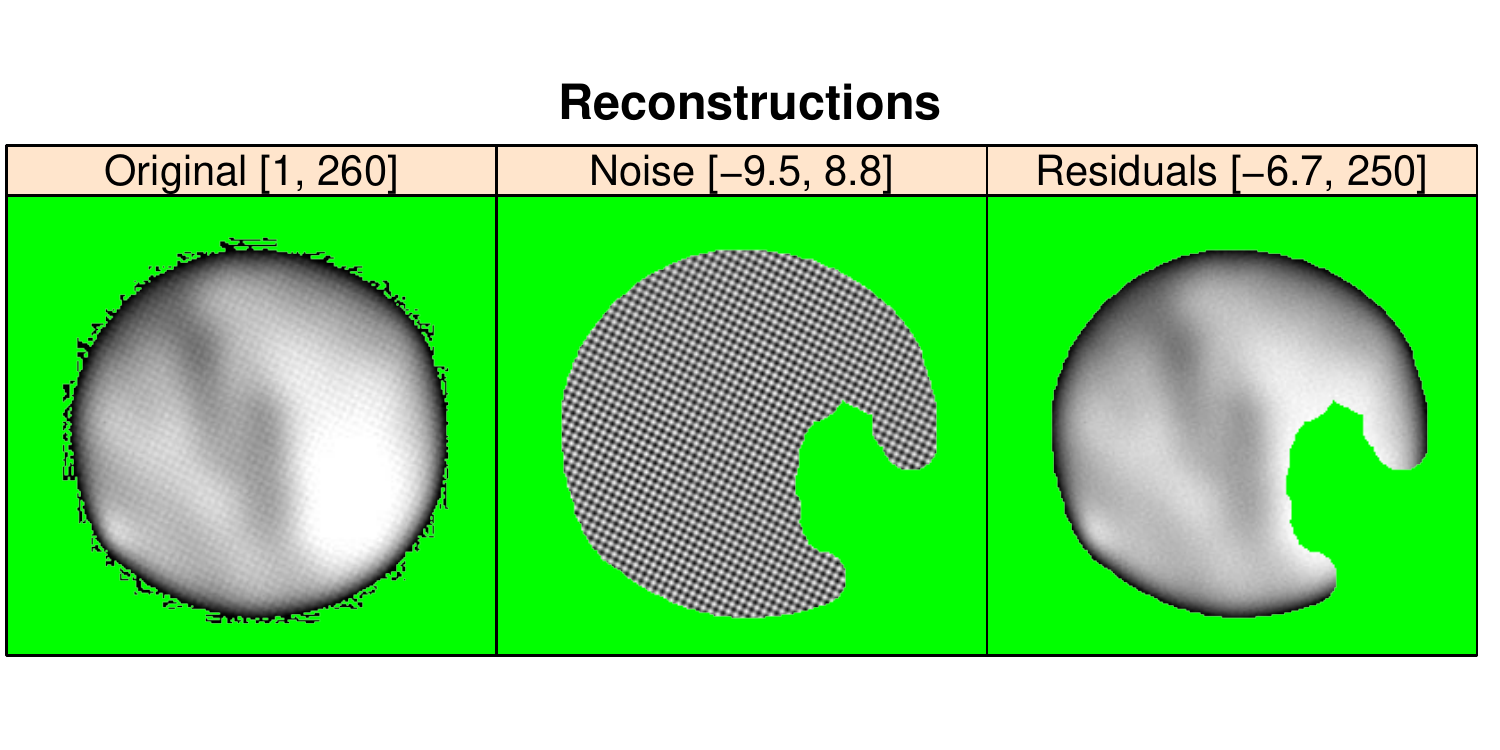}
        \end{center}
        \caption{Mars: Reconstruction, ShSSA.}
        \label{fig:Mars_shaped_rec}
\efg
\bfgh
        \begin{center}
        \includegraphics[width=10 cm]{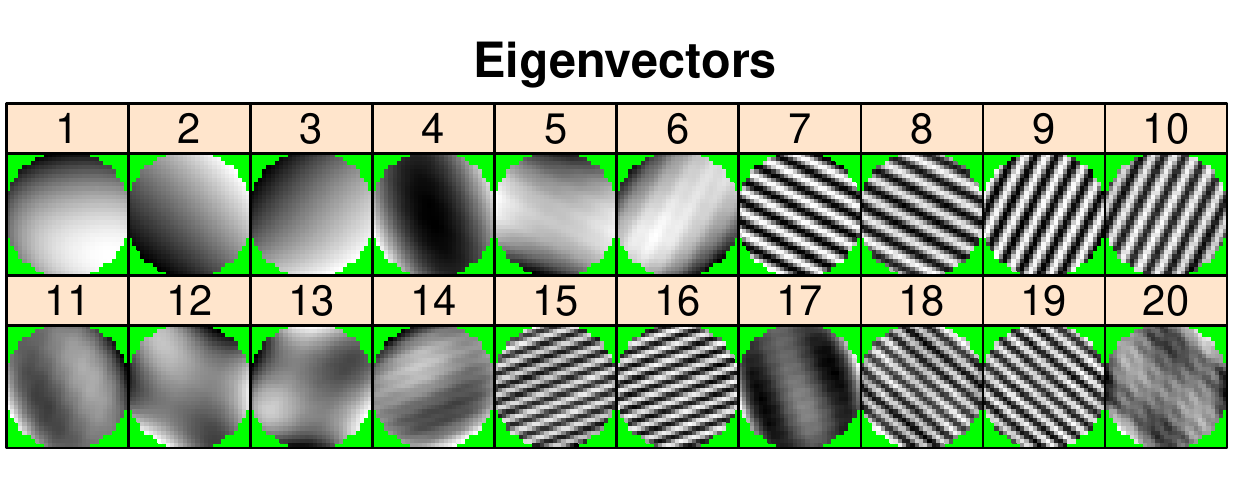}
        \end{center}
        \caption{Mars: Eigenarrays, ShSSA.}
        \label{fig:Mars_shaped_ident_psi}
\efg
\bfgh
        \begin{center}
        \includegraphics[width=8 cm]{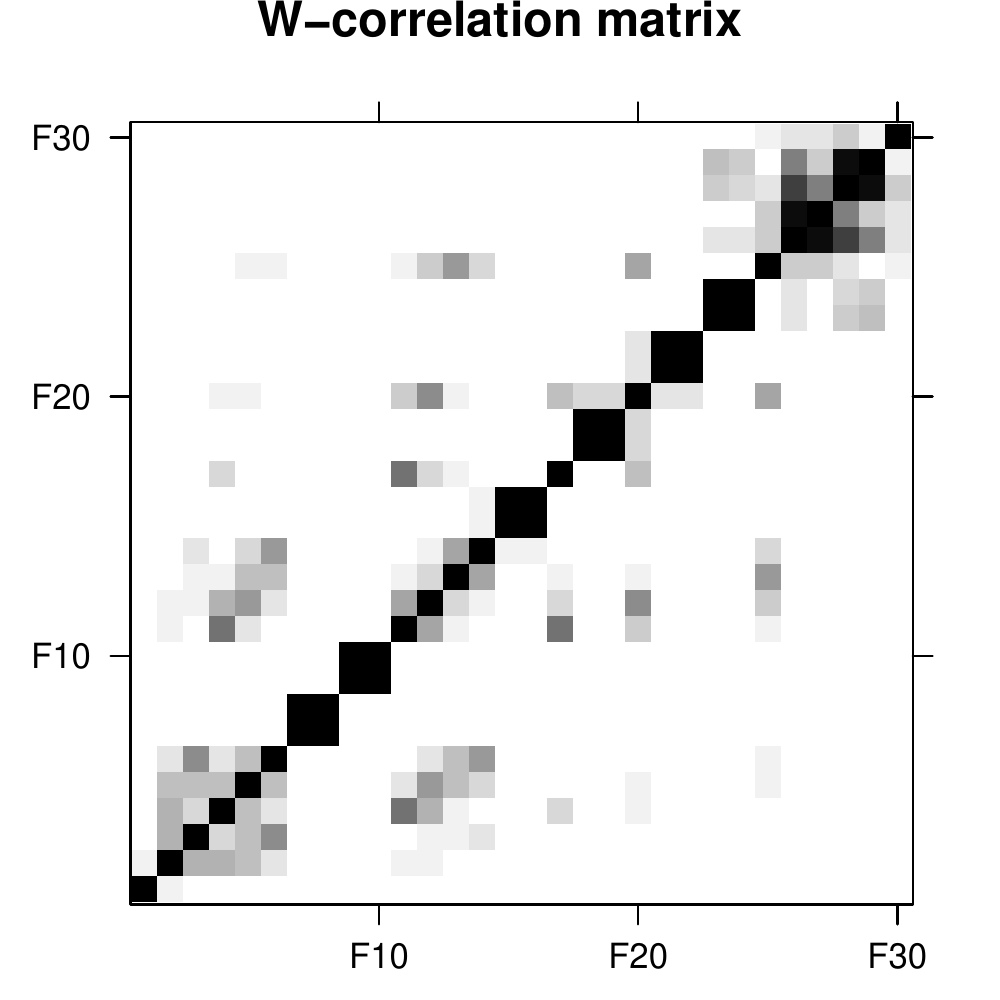}
        \end{center}
        \caption{Mars: $\bfw$~Correlations, ShSSA.}
        \label{fig:Mars_shaped_ident_wcor}
\efg

The quality of the texture extraction and therefore of the image recovery by Shaped SSA 
(Figure~\ref{fig:Mars_shaped_rec}) is considerably better than that performed by 2D-SSA 
(Figure~\ref{fig:Mars_25_rec}). The improvement of reconstruction accuracy is explained by an
edge effect that is caused by a sharp drop of intensity near the boundary of Mars. In Figure~\ref{fig:Mars_rect_vs_shaped}, we compare magnified reconstructed images for 2D-SSA and ShSSA. 
In the left subfigure,  a green shadow is shown for the background area in order to indicate the Mars boundary. In the right subfigure, light green color corresponds to \code{NA}.
The code that reproduces Figure~\ref{fig:Mars_rect_vs_shaped} is shown in Fragment~\ref{frag:Mars_rect_vs_shaped}.

\bfgh
        \begin{center}
        \includegraphics[width=12 cm]{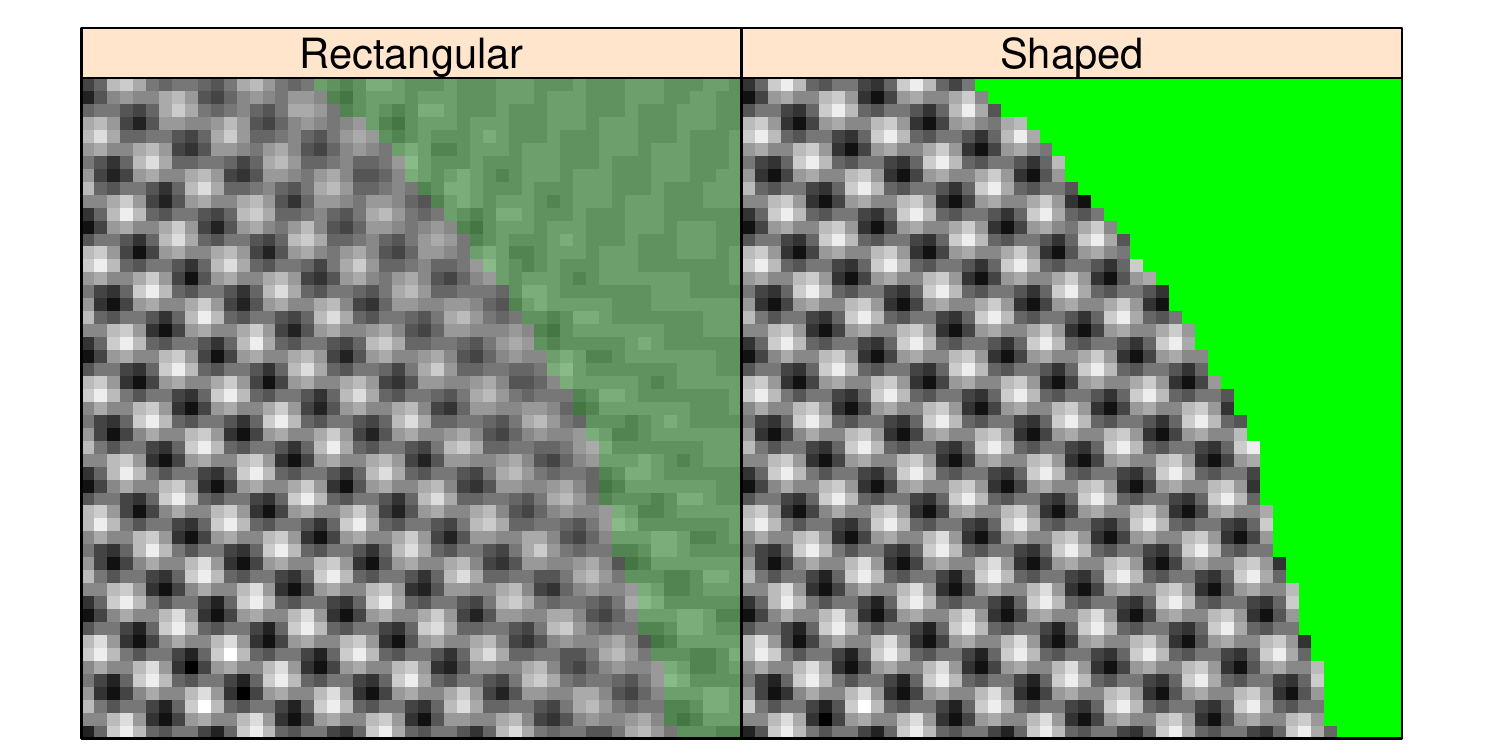}
        \end{center}
        \caption{Mars: comparison of texture reconstructions by 2D-SSA and ShSSA.}
        \label{fig:Mars_rect_vs_shaped}
\efg

\begin{fragment}[Mars: magnified reconstructions by 2D-SSA and ShSSA]
\label{frag:Mars_rect_vs_shaped}
\input{fragments/Mars_rect_vs_shaped.tex}
\end{fragment}

\subsubsection{Comments}
\paragraph{Efficient implementation}
In \pkg{Rssa}, the Shaped 2D-SSA decomposition shares \code{kind = "2d-ssa"} with 
the 2D-SSA decomposition. However, the use of input arrays with non-trivial shapes
would incur additional projection operation during the computations (see
Section~\ref{ssec:qhmatmul}). 

Ordinary 2D-SSA can be considered as a special case of
Shaped 2D-SSA with rectangular window; in this case, the mask covers the whole
image. The package optimizes for this common case and no projections are
performed when it is known that they are trivial and thus we have ordinary
2D-SSA. Therefore, the computational complexity of the method is the same regardless
how the full mask was specified: providing \code{NULL} to \code{wmask} and
\code{mask} arguments or making the masks trivial. This is a convenient behavior
which simplifies, e.g., batch processing of the images with shapes automatically
induced by the input images.

\paragraph{Special window shapes}
The package provides convenient interface for setting several special forms of
the window. This is implemented via special expressions which can be passed to
\code{wmask} argument:
\begin{itemize}
    \item \code{wmask = circle(R)} specifies circular mask of radius \code{R}.
    \item \code{wmask = triangle(side)} specifies the mask in the form of
        isosceles right-angled triangle with cathetus \code{side}. The right angle
        lays on the top left corner of the bounding box of the mask.
\end{itemize}


%% file: fragments/Mars_shaped_dec.tex
\begin{CodeChunk}
\begin{CodeInput}

> mask.Mars.0 <- (Mars != 0)
> mask.Mars.1 <- (Mars != 255)
> Mars[!mask.Mars.0] <- NA
> print(system.time(s.Mars.shaped <-
+   ssa(Mars, kind = "2d-ssa",
+       mask = mask.Mars.1, wmask = circle(15))))
   user  system elapsed 
  0.673   0.015   0.684 
> mask.Mars.res <- (s.Mars.shaped$weights > 0)
> plot2d(mask.Mars.0)
> plot2d(mask.Mars.1)
> plot2d(mask.Mars.res)
\end{CodeInput}

\end{CodeChunk}

%% file: fragments/Mars_shaped_rec.tex
\begin{CodeChunk}
\begin{CodeInput}

> r.Mars.shaped <- reconstruct(s.Mars.shaped,
+                              groups = list(Noise = c(7, 8, 9, 10)))
> plot(r.Mars.shaped, cuts = 255, layout = c(3, 1), fill.color = "green")
\end{CodeInput}

\end{CodeChunk}

%% file: fragments/Mars_shaped_ident.tex
\begin{CodeChunk}
\begin{CodeInput}

> plot(s.Mars.shaped, type = "vectors", idx = 1:20, fill.color = "green",
+      cuts = 255, layout = c(10, 2), plot.contrib = FALSE)
> plot(wcor(s.Mars.shaped, groups = 1:30),
+      scales = list(at = c(10, 20, 30)))
\end{CodeInput}

\end{CodeChunk}

%% file: fragments/Mars_rect_vs_shaped.tex
\begin{CodeChunk}
\begin{CodeInput}

> Mars.sh <- r.Mars.shaped$Noise
> Mars.rect.sh <- Mars.rect <- r.Mars.25$Noise
> Mars.rect.sh[is.na(Mars.sh)] <- NA
> library("latticeExtra")
> p.part.rect <- plot2d(Mars.rect[60:110, 200:250]) +
+                  layer(panel.fill(col = "green", alpha = 0.2), under = FALSE) +
+                  plot2d(Mars.rect.sh[60:110, 200:250])
> p.part.shaped <- plot2d(r.Mars.shaped[[1]][60:110, 200:250]) +
+                    layer(panel.fill(col = "green"), under = TRUE)
> plot(c(Rectangular = p.part.rect, Shaped = p.part.shaped))
\end{CodeInput}

\end{CodeChunk}

%% file: shaped_variants.tex
\subsection{Special cases of Shaped 2D-SSA}
\label{sec:shaped_special}
In this section, we show that all the considered variants of SSA are in
fact special cases of ShSSA, for carefully chosen array and mask.

First of all, SSA can be easily embedded in ShSSA.  Consider an $N \times 1$
array and $L \times 1$ window (or, alternatively $1 \times N$ array and $1
\times L$ window). Next, as it was discussed in the previous section,
2D-SSA is a special case of ShSSA. Finally, in \citet{Golyandina.Usevich2010} it was mentioned that
MSSA with equal time series lengths is a special case of 2D-SSA. In what follows,
we extend this construction to time series with different lengths.

\subsubsection{MSSA: 2D packing}
Consider a multivariate time series, that is, a collection
$\{\tX^{(p)}=(x_j^{(p)})_{j=1}^{N_p},\;\; p=1, \ldots, s\}$ of $s$ time series
of length $N_p$, $p=1,\ldots,s$.  We construct the shaped array $\tX_\Ns$ such
that
\[
\Ns = \{1,\ldots,N_1\} \times \{1\} \cup
\{1,\ldots,N_s\} \times \{s\} \subset \{1,\ldots,\max_p N_p \} \times \{1,\ldots, s\}.
\]
The window is taken to be $\Ls = \{1,\ldots,L\} \times \{1\}$. The construction
of this array is shown on Figure~\ref{fig:shaped_variants_mssa_2d}, also with the
window $\Ls$ and the shape $\Ks$. It is easy to verify that
$\bfX = \trajmat{ShSSA}(\tX)$ coincides with $\trajmat{MSSA}(\tX)$ defined in \eqref{eq:mssa_embedding}.
Therefore, the rows of $\bfX$ are vectorizations of the $\Ks$-shaped subarrays (see
Figure~\ref{fig:shaped_variants_mssa_2d}).

\bfgh
\centering
\begin{center}
\parbox{3cm}{\input{shaped_variants_mssa_2d.tex}}%
\qquad
\parbox{3cm}{\input{shaped_variants_mssa_2d_ks.tex}}%
        \end{center}
\caption{MSSA: 2D packing.}
\label{fig:shaped_variants_mssa_2d}
\efg

\subsubsection{MSSA: 1D packing}
Now we consider an alternative packing of MSSA.  From the same set of series we
construct a $N'\times 1$ (or $1\times N'$) array $\tX$, where $N' = N + (s-1)$
(recall that $N =\suml_{p=1}^s N_p$). The array consists of the time series plus
``separators'' between them that are not included in the array shape. The window is
taken to be $L \times 1$ (or $1 \times L$), depending on the arrangement
chosen. In Figure~\ref{fig:shaped_variants_mssa_1d} we show the horizontal variant
of packing.
\bfgh \centering
\input{shaped_variants_mssa_1d.tex}
\caption{MSSA: 1D packing.}
\label{fig:shaped_variants_mssa_1d}
\efg

\subsubsection{Mosaic Hankel matrices}
Mosaic Hankel matrix \citep{Heinig95LAA-Generalized} is a block matrix with
Hankel blocks. It can be considered as the most general one-dimensional (i.e., with
one-dimensional displacement) generalization of Hankel matrices.

Let $L_1,\ldots, L_s$ and $K_1,\ldots, K_t$ be integer vectors, and $\tX^{(i,j)}
\in \spaceR^{L_i + K_j -1}$ be time series. Then the mosaic Hankel matrix is
constructed as follows.
\[
\begin{pmatrix}
H_{L_1,K_1} (\tX^{(1,1)}) & \cdots & H_{L_1,K_t} (\tX^{(1,t)}) \\
\vdots      & \cdots & \vdots      \\
H_{L_s,K_1} (\tX^{(s,1)})& \cdots & H_{L_s,K_t} (\tX^{(p,t)})
\end{pmatrix}.
\]
Note that the sizes of the blocks may be different. The only requirement is that
they should match as a ``mosaic''.  The case of mosaic Hankel matrices
corresponds to several collections of multidimensional time series
\citep{Markovsky.Usevich13JCAM-Software}.

It is easy to construct mosaic Hankel matrices, based on 2D embedding of MSSA.
A $j$th block column is a transposed matrix $\trajmat{MSSA}$ for the collection
of time series $(\tX^{(1,j)}, \ldots,\tX^{(s,j)})$ and window length $L= K_j$.
Therefore, the mosaic Hankel matrix can be constructed by stacking shapes (with
separators) from Figure~\ref{fig:shaped_variants_mssa_2d} and replacing $\Ls$ with
$\Ks$ due to transposition.  The resulting construction of the shaped array is shown in
Figure~\ref{fig:shaped_variants_mosaic}.
\bfgh
\centering
\input{shaped_variants_mosaic.tex}
\caption{Shaped construction for mosaic Hankel matrices.}
\label{fig:shaped_variants_mosaic}
\efg

\subsubsection{M-2D-SSA}
Suppose that we have $s$ arrays $\tX^{(1)}, \ldots, \tX^{(s)} \in
\spaceR^{\Nx\times\Ny}$ and we would like to consider a variant of 2D-SSA where
the trajectory matrix is stacked from $s$ trajectory matrices (as in MSSA):
\be
\label{eq:embedding_m2dssa}
\trajmat{M\mbox{-}2D\mbox{-}SSA} (\tX^{(1)}, \ldots, \tX^{(s)}) =
\left[\trajmat{2D-SSA} (\tX^{(1)}) : \ldots : \trajmat{2D-SSA} (\tX^{(s)})\right].
\ee
In this case, the 2D-SSA-like decomposition will have the common basis of
eigenvectors, as in MSSA.  The trajectory matrices of the form \eqref{eq:embedding_m2dssa}
are used in 2D-SSA-based for comparison of images
\citep{Rodriguez-Aragon.Zhigljavsky10SII-Singular}.  These matrices are also 
used in recent methods of parallel magnetic resonance imaging
\citep{Uecker.etal13MRiM-ESPIRiT}.

The packing for the M-2D-SSA can be constructed in a similar way to the case of
mosaic Hankel matrices. An array $\tX'$ of size $\Nx \times (s\Ny + s -1)$ is
constructed from the arrays with one-element separators. The resulting array is
shown in Figure~\ref{fig:shaped_variants_m2dssa}.
\bfgh
\centering
\input{shaped_variants_m2dssa.tex}
\caption{Shaped construction for M-2D-SSA.}
\label{fig:shaped_variants_m2dssa}
\efg
In general, a similar construction can handle arrays of different sizes, shapes,
and shaped windows. Also note that in the extended array the original arrays
(both for Figure~\ref{fig:shaped_variants_m2dssa} and
Figure~\ref{fig:shaped_variants_mosaic}) may be arranged arbitrarily (for example
in a table-like planar arrangement). The only requirement is that they should be
separated.

%% file: shaped_variants_mssa_2d.tex
 \begin{tikzpicture}
    \draw[color = gray, fill=lightgray] (0,0) rectangle (2.5,4);
    \draw[fill=white] (2.5,0) -- (2,0) -- (2,1) -- (1.5,1) --
          (1.5,0.5) -- (1,0.5) -- (1,0) -- (0.5,0) -- (0.5,1.5) -- (0,1.5) --
          (0,4) -- (2.5,4) -- (2.5, 0);
\draw[color= red, thick] (0.5,1) rectangle (1,3);
    
    \draw (0.75,2) node {$\Ls$};

    \draw [<->] (1.15,1) -- node[right] {$L$} (1.15,3);
    \draw (0,4) node[anchor=south west] {\small$\tX^{(1)}$};
    \draw (2.5,4) node[anchor=south ]  {\small$\tX^{(s)}$};
  \end{tikzpicture}

%% file: shaped_variants_mssa_2d_ks.tex
 \begin{tikzpicture}
    \draw[color=gray, fill=lightgray] (0,0) rectangle (2.5,4);
    \draw[fill=white] (2.5,0) -- (2,0) -- (2,1) -- (1.5,1) --
          (1.5,0.5) -- (1,0.5) -- (1,0) -- (0.5,0) -- (0.5,1.5) -- (0,1.5) --
          (0,4) -- (2.5,4) -- (2.5, 0);

    \draw[color=green, thick] (2.5,1.5) -- (2,1.5) -- (2,2.5) -- (1.5,2.5) --
          (1.5,2) -- (1,2) -- (1,1.5) -- (0.5,1.5) -- (0.5,3) -- (0,3) --
          (0,4) -- (2.5,4) -- (2.5, 1.5);

    \draw (1.25,2.75) node {$\Ks$};

    \draw (0,4) node[anchor=south west] {\small$\tX^{(1)}$};
    \draw (2.5,4) node[anchor=south ]  {\small$\tX^{(s)}$};
  \end{tikzpicture}

%% file: shaped_variants_mssa_1d.tex
 \begin{tikzpicture}
    \draw[color = gray, fill=lightgray] (0,0) rectangle (12,0.5);

    \draw[color=black, fill=white] (0,0) rectangle (4,0.5);
    \draw[color=black, fill=white] (4.5,0) rectangle (7,0.5);
    \draw[color=black, fill=white] (9.5,0) rectangle (12,0.5);

    \draw[color= red, thick] (1,0) rectangle (3,0.5);
    \draw (2,0.25) node {$\Ls$};
    
    \draw [<->] (1,-0.15) -- node[anchor=north] {$L$} (3, -0.15);
    \draw (0,0.5) node[anchor=south west] {\small$\tX^{(1)}$};

    \draw (4.5,0.5) node[anchor=south west] {\small$\tX^{(2)}$};
    \draw (9.5,0.5) node[anchor=south  west]  {\small$\tX^{(s)}$};

    \draw (7.5,0.5) node[anchor=south ]  {\small$\cdots$};
  \end{tikzpicture}

%% file: shaped_variants_mosaic.tex
 \begin{tikzpicture}
 \begin{scope}
    \draw[color = black, fill=lightgray] (0,-1) rectangle (10.5,4);
    \draw (7,2) node {$\cdots$};
    \draw[fill=white] (2.5,0) -- (2,0) -- (2,1) -- (1.5,1) --
          (1.5,0.5) -- (1,0.5) -- (1,0) -- (0.5,0) -- (0.5,1.5) -- (0,1.5) --
          (0,4) -- (2.5,4) -- (2.5, 0);

    \draw[color=red, thick] (2.5,1.5) -- (2,1.5) -- (2,2.5) -- (1.5,2.5) --
          (1.5,2) -- (1,2) -- (1,1.5) -- (0.5,1.5) -- (0.5,3) -- (0,3) --
          (0,4) -- (2.5,4) -- (2.5, 1.5);

    \draw (1.25,2.75) node {$\Ls$};

    \draw (0,4) node[anchor=south west] {\small$\tX^{(1,1)}$};
    \draw (2.5,4) node[anchor=south ]  {\small$\tX^{(s,1)}$};    
  \end{scope}  
 \begin{scope}[xshift=3cm]

    \draw[fill=white] (2.5,-0.5) -- (2,-0.5) -- (2,0.5) -- (1.5,0.5) --
          (1.5,0) -- (1,0) -- (1,-0.5) -- (0.5,-0.5) -- (0.5,1) -- (0,1) --
          (0,4) -- (2.5,4) -- (2.5, -0.5);
    \draw (0,4) node[anchor=south west] {\small$\tX^{(1,2)}$};
    \draw (2.5,4) node[anchor=south ]  {\small$\tX^{(s,2)}$};
  \end{scope}  
  
  \begin{scope}[xshift=8cm]

    \draw[fill=white] (2.5,-1) -- (2,-1) -- (2,0) -- (1.5,0) --
          (1.5,-0.5) -- (1,-0.5) -- (1,-1) -- (0.5,-1) -- (0.5,0.5) -- (0,0.5) --
          (0,4) -- (2.5,4) -- (2.5, -1);
    \draw (0,4) node[anchor=south west] {\small$\tX^{(1,t)}$};
    \draw (2.5,4) node[anchor=south ]  {\small$\tX^{(s,t)}$};
  \end{scope}

  \end{tikzpicture}

%% file: shaped_variants_m2dssa.tex
 \begin{tikzpicture}
    \draw[color = gray, fill=lightgray] (0,0) rectangle (12,3);

    \draw[color=black, fill=white] (0,0) rectangle (2.5,3);
    \draw[color=black, fill=white] (3,0) rectangle (5.5,3);
    \draw[color=black, fill=white] (9.5,0) rectangle (12,3);

    \draw[color= red, thick] (1,1) rectangle (2,2.5);
    \draw (1.5,1.75) node {$\Ls$};
    \draw [<->] (1,0.85) -- node[anchor=north] {\small$\Ly$} (2, 0.85);
    \draw [<->] (0.85,1) -- node[anchor=east] {\small$\Lx$} (0.85,2.5);
    
    \draw (1.25,3) node[anchor=south] {\small$\tX^{(1)}$};

    \draw (4.25,3) node[anchor=south] {\small$\tX^{(2)}$};
    \draw (10.75,3) node[anchor=south]  {\small$\tX^{(s)}$};

    \draw (7.5,1.5) node {\small$\cdots$};
  \end{tikzpicture}

%% file: implement_hankel.tex
This section contains details of core algorithms implemented in the \pkg{Rssa} package. 
The algorithms discussed in this section are either absent or scarcely described in the 
literature. Although the implementation does not influence the interface of the package, 
this section is important for understanding why the package works correctly and effectively.

\subsection{Fast computations with Hankel matrices}
\label{ssec:hmatmul}
We start with a summary of the algorithms for multiplication of
Hankel matrices by vectors and rank-one hankelization. Although algorithms for
these operations were already proposed in \citet{Korobeynikov2010}, we provide here an alternative
description that uses another type of circulants.
The alternative description simplifies the algorithms compared to \citet{Korobeynikov2010} 
and also lays a foundation for Section~\ref{ssec:qhmatmul} on computations with quasi-Hankel matrices.
\subsubsection{Preliminaries}
In what follows, $A \odot B$ denotes the elementwise product of two vectors $A,B
\in \spaceC^{N}$, $\revv(A)$ denotes the reversion of the vector $A$, and
$\conjug{A}$ denotes the elementwise conjugation. We also denote by
$\unitvec{N}{j}$ the standard $j$th unit vector in $\spaceR^{N}$.

For a complex vector $X \in \spaceC^{N}$ we define its discrete Fourier
transform as \be\label{eq:fft_def} \fftw{N}(X) = \fftwone{N} X, \ee where
$\fftwone{N} \in \spaceC^{N\times N}$ is the Fourier matrix with elements
$(\fftwone{N} )_{k,l} = e^{-2 \pi \unit (k-1)(l-1) /N}$ (see
\citet{Korobeynikov2010}). The inverse Fourier transform is given by
\[
\fftw{N}^{-1}(X) = \frac{1}{N} \fftwone{N}^{*} X,
\]
where $\fftwone{N}^{*}$ is the Hermitian transpose of $\fftwone{N}$.

The \textit{Toeplitz circulant} constructed from the vector $X= (x_k)_{k=1}^{N}$
is, by definition,
\[
\tcirc{X} =
\begin{pmatrix}
x_1     & x_N     & x_{N-1}   & \cdots & x_3 & x_2 \\
x_2     & x_1     & x_{N}     & \cdots & x_4 & x_3 \\
\vdots  & \vdots  &\vdots     &   & \vdots &\vdots \\
x_{N-1} & x_{N-2} & x_{N-3}   & \cdots & x_1 & x_N \\
x_N     & x_{N-1} & x_{N-2}   & \cdots & x_2 & x_1 \\
\end{pmatrix}.
\]
The \textit{Hankel circulant} constructed from the vector $X$ is, by definition,
\[
\hcirc{X} =
\begin{pmatrix}
x_1     & x_2     & \cdots & x_{N-2}   & x_{N-1} & x_N \\
x_2     & x_3     & \cdots & x_{N-1}   & x_N     & x_1 \\
\vdots  & \vdots  &        &\vdots     & \vdots  &\vdots \\
x_{N-1} & x_{N}   & \cdots & x_{N-4}   & x_{N-3} & x_{N-2} \\
x_N     & x_{1}   & \cdots & x_{N-3}   & x_{N-2} & x_{N-1} \\
\end{pmatrix}.
\]

We will need the following relation between Hankel and Toeplitz circulants.
\bl\label{lem:h_t_elmat}
For any $j =1,\ldots,N$ and $\rmA \in \spaceC^{N}$ we have that
\[
{A}^{\top} \hcirc{\unitvec{N}{j}} = (\unitvec{N}{j})^{\top} \tcirc{A}.
\]
\el
\bproof
The vector ${A}^{\top} \hcirc{\unitvec{N}{j}}= (\hcirc{\unitvec{N}{j}} A)^\top$ is
the $j$th forward circular shift of the vector $\revv(A)^{\top}$.
Therefore, it coincides with the $j$th row of $\tcirc{A}$,
which is equal to $(\unitvec{N}{j})^{\top} \tcirc{A}$.
\eproof

\subsubsection{Matrix-vector multiplication}
It is well known \citep{Korobeynikov2010} that multiplication of the
Toeplitz circulant $\tcirc{X}$ by a vector $A \in\spaceC^N$ can be computed
using discrete Fourier transform as \be\label{eq:tcirc_mul_dft} \tcirc{X} A =
\fftw{N}^{-1}\big({\fftw{N}(X) \odot \fftw{N}(A)}\big).  \ee A similar property
holds for the $\hcirc{X}$. Since it is less common, we provide a short proof.
\bl For $X \in \spaceC^N$ and $A \in \spaceR^N$ we have that
\be\label{eq:hcirc_dft} \hcirc{X} A = \fftw{N}^{-1}\left(\fftw{N}(X) \odot
    \conjug{\fftw{N}(A)}\right).  \ee \el \bproof The circulant $C_2 =
\tcirc{(x_N, x_1, \ldots, x_{N-1})}$ can be obtained from $\hcirc{X}$ by
reversion of all rows. Therefore, the product of $\hcirc{X}$ by $A$ is equal to
\[
\begin{split}
\hcirc{X} A & = C_2 \revv{(A)} =
\fftw{N}\left({(x_N, x_1, \ldots, x_{N-1})^{\top}}\right) \odot \fftw{N}(\revv(A))\\
&= (\fftwone{N}X) \odot(e^{2\pi \unit/N} \fftwone{N}\revv(A))
= (\fftwone{N}X) \odot \conjug{\fftwone{N}A}.
\end{split}
\]
\eproof

For $\tX \in \spaceR^N$ and $K,L: N=K+L-1$, the Hankel matrix
$\trajmat{SSA}(\tX)$ is an $L \times K$ submatrix of $\hcirc{\tX}$.  Therefore,
multiplication by $\trajmat{SSA}(\tX)$ can be performed using the following
algorithm.
\begin{algorithm}
    \label{alg:hmul_u}
    Input: $V \in \spaceR^K$, $\tX \in \spaceR^N$. Output: $U =
    \trajmat{SSA}(\tX) V \in \spaceR^L$.
    \begin{enumerate}
        \item $V' \leftarrow \begin{pmatrix} V \\ 0_{L-1} \end{pmatrix}$
        \item $\widehat{V'} \leftarrow \fftw{N} (V')$
        \item $\widehat{X} \leftarrow \fftw{N} (\tX)$
        \item $U' \leftarrow \fftw{N}^{-1} (\widehat{X} \odot \conjug{\widehat{V'}})$
        \item $U \leftarrow (u'_1,\ldots, u'_L)^{\top}$
    \end{enumerate}
\end{algorithm}
In \citet{Korobeynikov2010}, two different algorithms were used for multiplication
by $\trajmat{SSA}(\tX)$ and its transpose. But Algorithm~\ref{alg:hmul_u} also
suits for multiplication by $\trajmat{SSA}^{\top}(\tX)$, because sizes of the
matrix are used only in the first step (padding of $V$ by zeros) and the last
step (truncating the result). Also, the DFT $\fftwone{N} \tX$ (step 3) can be
precomputed. But, in contrast to the approach of \citet{Korobeynikov2010}, the precomputed object
does not depend on $L$.

\subsubsection{Rank-one hankelization}
Next, we describe the algorithm for hankelization, which coincides with that
from \citet{Korobeynikov2010}.  But we describe the algorithm in a different way,
that lays a foundation for quasi-Hankelization in Section~\ref{ssec:qhmatmul}.

The hankelization operation computes for a given $\bfX \in \spaceR^{L \times K}$
the vector $\wtilde{\tX} \in \spaceR^N$  that minimizes the distance
$\|\trajmat{SSA}(\wtilde{\tX}) - \bfX\|_\frob^2$. It is well-known that the
hankelization of $\bfX$ is the vector $\wtilde{\tX} =
(\widetilde{x}_{j})_{j=1}^N$ of diagonal averages
\be\label{eq:hankeliz_def}
\widetilde{x}_{j} = \frac{\sum\limits_{k+l=j} (\bfX)_{k,l}}{w_j} =
\frac{\langle \bfX,
    \trajmat{SSA}(\unitvec{N}{j})\rangle_\frob}{\|\trajmat{SSA}(\unitvec{N}{j})\|^2_\frob}.
\ee
The denominators $w_j = \|\trajmat{SSA}(\unitvec{N}{j})\|^2_\frob$ are
exactly the weights in $\bfw$~correlations.  For rank-one matrices $\bfX =
UV^{\top} \in \spaceR^{L\times K}$, the numerator in (\ref{eq:hankeliz_def}) can
be expressed more compactly:
\[
\langle UV^{\top}, \trajmat{SSA}(\unitvec{N}{j})\rangle_\frob =
U^{\top} \trajmat{SSA}(\unitvec{N}{j}) V =
 (U')^{\top} \hcirc{\unitvec{N}{j}}V ' =
 (\unitvec{N}{j})^{\top} \tcirc{U'}V',
\]
where ${U'} = \begin{pmatrix} U \\ 0_{K-1} \end{pmatrix}$, ${V'}
= \begin{pmatrix} V \\ 0_{L-1} \end{pmatrix}$, and the last equality follows
from Lemma~\ref{lem:h_t_elmat}.

Then, the hankelization can be computed using the following algorithm.
\begin{algorithm}
    \label{alg:hankelization}
    \quad
    Input: $U \in \spaceR^L$, $V \in \spaceR^K$. Output: hankelization $\wtilde{\tX} \in \spaceR^{N}$.
    \begin{enumerate}
        \item $L^{*} \leftarrow \min(L,K)$
        \item $W \leftarrow (1,2,\ldots,L^*,\ldots,L^*,\ldots,2,1)^{\top} \in
        \spaceR^{N}$ (weights for $\bfw$~correlations)
        \item ${U'} \leftarrow  \begin{pmatrix} U \\ 0_{K-1} \end{pmatrix}$,
        ${V'} \leftarrow  \begin{pmatrix} V \\ 0_{L-1} \end{pmatrix}$
        \item $\widehat{U'} \leftarrow\fftw{N} (U')$, $\widehat{V'}\leftarrow \fftw{N} (V')$
        \item $\wtilde{\tX}' \leftarrow \fftw{N}^{-1} (\widehat{U'} \odot \widehat{V'})$
        \item $\wtilde{x}_k \leftarrow \wtilde{x}'_k/w_k$
    \end{enumerate}
\end{algorithm}

Note that in Algorithm~\ref{alg:hankelization}, we calculate the weights
explicitly, as in \citet{Korobeynikov2010}. However, writing the weights in the
expanded form (the denominator in (\ref{eq:hankeliz_def})) helps to
understand how the weights are computed for quasi-Hankelization in
Section~\ref{ssec:qhmatmul}.

%% file: implement_shaped.tex
\subsection{Fast computations with quasi-Hankel matrices}
\label{ssec:qhmatmul}
\subsubsection{Preliminaries}
For two matrices $\rmA, \rmB \in \spaceC^{\Nx \times \Ny}$ we define by
$\rmA\odot\rmB$ their elementwise product and by $\conjug{\rmA}$ the
elementwise complex conjugation. For $k = 1,\ldots,\Nx$ and $l = 1,\ldots,\Ny$,
we define elementary arrays as $\elmat_{k,l} = \unitvec{\Nx}{k}
(\unitvec{\Ny}{l})^{\top} \in \spaceR^{\Nx\times\Ny}$, where $\unitvec{N}{j} \in
\spaceR^N$ is the $j$th unit vector. For two matrices $\rmA$ and $\rmC$ their
Kronecker product is denoted as $\rmA \otimes \rmC$.

For a complex matrix $\rmX = (x_{k,l})_{k,l=1}^{\Nx,\Ny}$ its discrete Fourier
transform is defined as
\[
\fftw{\Nx,\Ny}(\rmX) = \fftwone{\Nx} \rmX \fftwone{\Ny}^{\top},
\]
where $\fftwone{N}$ is the matrix of the discrete Fourier transform
(\ref{eq:fft_def}).  By the properties of the vectorization operator, we have
that
\[
\mvec (\fftw{\Nx,\Ny}(\rmX)) = (\fftwone{\Ny} \otimes \fftwone{\Nx}) \mvec (\rmX).
\]
The \textit{TbT (Toeplitz-block-Toeplitz) circulant} constructed from the matrix $\rmX$ is, by definition,
\[
\tbtcirc{\rmX} =
\begin{pmatrix}
\bfC_{\rmT,1}     & \bfC_{\rmT,\Ny}     & \bfC_{\rmT,\Ny-1}   & \cdots & \bfC_{\rmT,3} & \bfC_{\rmT,2} \\
\bfC_{\rmT,2}     & \bfC_{\rmT,1}     & \bfC_{\rmT,\Ny}     & \cdots & \bfC_{\rmT,4} & \bfC_{\rmT,3} \\
\vdots  & \vdots  &\vdots     &   & \vdots &\vdots \\
\bfC_{\rmT,\Ny-1} & \bfC_{\rmT,\Ny-2} & \bfC_{\rmT,\Ny-3}   & \cdots & \bfC_{\rmT,1} & \bfC_{\rmT,\Ny} \\
\bfC_{\rmT,\Ny}     & \bfC_{\rmT,\Ny-1} & \bfC_{\rmT,\Ny-2}   & \cdots & \bfC_{\rmT,2} & \bfC_{\rmT,1} \\
\end{pmatrix},
\]
where $\bfC_{\rmT,j} = \tcirc{\rmX_{:,j}}$, and $\rmX_{:,j}$ denotes the
$j$th column of $\rmX$.  The \textit{HbH circulant} constructed from the matrix
$\rmX$ is, by definition,
\[
\hbhcirc{\rmX} =
\begin{pmatrix}
{\bfC_{\rmH,1}}     & {\bfC_{\rmH,2}}     & \cdots & {\bfC_{\rmH,\Ny-2}}   & {\bfC_{\rmH,\Ny-1}} & {\bfC_{\rmH,\Ny}} \\
{\bfC_{\rmH,2}}     & {\bfC_{\rmH,3}}     & \cdots & {\bfC_{\rmH,\Ny-1}}   & {\bfC_{\rmH,\Ny}}     & {\bfC_{\rmH,1}} \\
\vdots  & \vdots  &        &\vdots     & \vdots  &\vdots \\
{\bfC_{\rmH,\Ny-1}} & {\bfC_{\rmH,\Ny}}   & \cdots & {\bfC_{\rmH,\Ny-4}}   & {\bfC_{\rmH,\Ny-3}} & {\bfC_{\rmH,\Ny-2}} \\
{\bfC_{\rmH,\Ny}}     & {\bfC_{\rmH,1}}   & \cdots & {\bfC_{\rmH,\Ny-3}}   & {\bfC_{\rmH,\Ny-2}} & {\bfC_{\rmH,\Ny-1}} \\
\end{pmatrix},
\]
where $\bfC_{\rmH,j} = \hcirc{\rmX_{:,j}}$. Note that for a rank-one array $\rmX = X_x X_y^\top$ we have that
\be\label{eq:rank_one_circ}
\hbhcirc{\rmX} = \hcirc{X_y} \otimes \hcirc{X_x} \quad\mbox{and}\quad
\tbtcirc{\rmX} = \tcirc{X_y} \otimes \tcirc{X_x}.
\ee

We will also need the following relation between HbH and TbT circulants.
\bl\label{lem:hbh_tbt_elmat}
For $k = 1,\ldots,\Nx$, $l = 1,\ldots,\Ny$ and $\rmA \in \spaceR^{\Nx \times \Ny}$
\[
{\mvec(\rmX)}^{\top} \hbhcirc{\elmat_{k,l}} = \mvec(\elmat_{k,l})^{\top} \tbtcirc{\rmX}.
\]
\el
\bproof
Due to linearity of the equation, we need to prove it only for rank-one matrices
$\rmX = X_x X_y^{\top}$. Then, by (\ref{eq:rank_one_circ}) we have that
\[
\begin{split}
&{\mvec(\rmX)}^{\top} \hbhcirc{\elmat_{k,l}} =
\left(X_y \otimes X_x \right)^{\top} \left(\hcirc{\unitvec{\Ny}{l}} \otimes \hcirc{\unitvec{\Nx}{k}}\right)\\
& =
\left(X_y^{\top}\hcirc{\unitvec{\Ny}{l}} \right) \otimes \left(X_x^{\top} \hcirc{\unitvec{\Nx}{k}} \right)
 =
\left((\unitvec{\Ny}{l})^{\top}\tcirc{X_y} \right) \otimes \left((\unitvec{\Nx}{k})^{\top}\tcirc{X_x} \right)\\
& =\left(\unitvec{\Ny}{l} \otimes \unitvec{\Nx}{k} \right)^{\top} \tcirc{X_y} \otimes \tcirc{X_x} =
\mvec(\elmat_{k,l})^{\top} \tbtcirc{\rmX}.
\end{split}
\]
\eproof

\subsubsection{Multiplication by a quasi-Hankel matrix}
Analogously to (\ref{eq:tcirc_mul_dft}) and (\ref{eq:hcirc_dft}), the following equalities hold true.
\bl
For $\rmX, \rmA \in \spaceR^{\Nx \times \Ny}$ we have that
\begin{align*}
&\tbtcirc{\rmX} \mvec \left(\rmA\right) = \mvec \left(\fftw{\Nx,\Ny}^{-1}\left(\fftw{\Nx,\Ny}(\rmX) \odot \fftw{\Nx,\Ny}(\rmA)\right)\right), \\
&\hbhcirc{\rmX} \mvec \left(\rmA\right) = \mvec \left(\fftw{\Nx,\Ny}^{-1}\left(\fftw{\Nx,\Ny}(\rmX) \odot \conjug{\fftw{\Nx,\Ny}(\rmA)}\right)\right).
\end{align*}
\el
\bproof
Due to linearity of the Fourier transform and circulant matrices we can prove
the statements only for matrices $\rmX = X_x X_y^\top$ and $\rmA = A_x
A_y^\top$. In this case, by (\ref{eq:rank_one_circ}) we have that
\[
\begin{split}
&\mvec \left(\fftw{\Nx,\Ny}^{-1}\left(\fftw{\Nx,\Ny}(\rmX) \odot \fftw{\Nx,\Ny}(\rmA)\right)\right)\\
&= \mvec \left(\fftw{\Nx,\Ny}^{-1}\left((\fftwone{\Nx}X_x) (\fftwone{\Ny}X_y)^\top \odot (\fftwone{\Nx} A_x) (\fftwone{\Ny} A_y)^\top\right)\right)\\
&=\mvec \left(\fftwone{\Nx}^{-1}\left((\fftwone{\Nx}X_x)\odot (\fftwone{\Nx}A_x)\right)  \left((\fftwone{\Ny} X_y) \odot (\fftwone{\Ny} A_y)\right)^\top (\fftwone{\Ny}^{-1})^\top\right) \\
&=\mvec ((\tcirc{X_x} A_x) (\tcirc{X_y} A_y)^\top ) = \tbtcirc{\rmX} \mvec{\rmA}.
\end{split}
\]
The proof of the second statement is analogous.
\eproof

Now we consider multiplication by a quasi-Hankel matrix. Let $\Ls,\Ks,\Ns \subseteq
\{1,\ldots,\Nx\} \times\{1,\ldots,\Ny\}$ be such that $\Ls\shs\Ks= \Ns$. Then
the quasi-Hankel matrix $\trajmat{ShSSA} (\tX)$ (defined in (\ref{eq:qh_mat}))
is a submatrix of the HbH circulant $\hbhcirc{\tX}$.  Now let us write this
formally and provide an algorithm for the calculation of the matrix-vector
product.

For a set of indices $\mathfrak{A} = \{(k_1,l_1),\ldots,(k_N,l_N)\}$, which is
ordered lexicographically, we define the projection matrix
\[
\Pmat_{\mathfrak{A}} = [\mvec(\elmat_{k_1,l_1}) : \ldots : \mvec(\elmat_{k_N,l_N})].
\]
Then we have that
\[
\trajmat{ShSSA} (\tX) = \Pmat_{\Ls}^{\top} \hbhcirc{\tX} \Pmat_{\Ks}.
\]
Hence, the matrix-vector multiplication algorithm can be written as follows.

\begin{algorithm}
    Input: $V \in \spaceR^K$, $\tX \in \spaceR^{\Nx\times\Ny}$. Output: $U = \trajmat{ShSSA} (\tX) V \in \spaceR^L$.
    \begin{enumerate}
        \item $\widehat{\rmX} \leftarrow \fftw{\Nx,\Ny} (\tX)$
        \item $\rmU' \leftarrow \mvec^{-1}_{\Kx}\left(\Pmat_{\Ks} U \right)$
        \item $\widehat{\rmU'} \leftarrow \fftw{\Nx,\Ny} (\rmU')$
        \item $\rmV' \leftarrow \fftw{\Nx,\Ny}^{-1} (\widehat{\rmX} \odot \conjug{\widehat{\rmU'}})$
        \item $V \leftarrow \Pmat_{\Ls}^{\top} \mvec(\rmV')$
    \end{enumerate}
\end{algorithm}

\subsubsection{Rank-one quasi-Hankelization}
Let $\Ls,\Ks,\Ns \subseteq \{1,\ldots,\Nx\} \times\{1,\ldots,\Ny\}$ be such that
$\Ls\shs\Ks= \Ns$.  The \textit{quasi-Hankelization} operator, by definition,
computes for a given $\bfX \in \spaceR^{L \times K}$ the shaped array
$\wtilde{\tX}_{\Ns} = (\widetilde{x}_{k,l})_{(k,l)\in \Ns}$ that minimizes the
distance $\|\calT(\wtilde{\tX}_{\Ns}) - \bfX\|_\frob^2$.

As shown in Section~\ref{sec:common_alg}, quasi-Hankelization can be expressed as averaging. More
precisely,
\[
\widetilde{x}_{k,l} = \frac{\langle \bfX,
    \calT(\elmat_{k,l})\rangle_\frob}{\|\calT(\elmat_{k,l})\|^2_\frob},
\quad\mbox{for }(k,l)\in \Ns.
\]
The numerator represents summation over the set of positions in $\bfX$ that
correspond to the $(k,l)$th element of the array. The
denominator is equal to the number of such elements.

In this section, we assume that $\bfX$ is a rank-one matrix, i.e., $\bfX = U
V^{\top}$. Then
\[
\widetilde{x}_{k,l} = \frac{\langle UV^{\top}, \calT(\elmat_{k,l})\rangle_\frob}{\|\calT(\elmat_{k,l})\|^2_\frob} =
\frac{\langle UV^{\top}, \calT(\elmat_{k,l})\rangle_\frob}{\langle 1_L 1_K^{\top}, \calT(\elmat_{k,l})\rangle_\frob},
\]
where $1_Q = (1,\ldots,1)^{\top} \in \spaceR^Q$. If we define by
\[
\diagsums(U,V) = (x_{k,l})_{k,l=1}^{\Nx,\Ny}, \quad\mbox{where} \;
x_{k,l} = \langle UV^{\top}, \calT(\elmat_{k,l})\rangle_\frob,
\]
then quasi-Hankelization is given by the following algorithm.

\begin{algorithm}
    \label{alg:qhankelization}
    \quad

    Input: $U \in \spaceR^L$, $V \in \spaceR^K$, shapes $\Ls$, $\Ks$. Output: shape $\Ns' =\Ls \shs \Ks$, quasi-Hankelization $\widetilde{\tX}_{\Ns}$.
    \begin{enumerate}
        \item $\mathbb{W} \leftarrow\diagsums(1_L,1_K)$ (array of weights for w-correlations)
        \item $\Ns'\leftarrow \{(k,l)\;|\; w_{k,l} \neq 0\}$
        \item $\wtilde{\tX'} \leftarrow\diagsums(U,V)$
        \item Compute $\wtilde{x}_{k,l} =\wtilde{x}_{k,l}' / w_{k,l}$
    \end{enumerate}
\end{algorithm}
We note that the weights $\mathbb{W}$ can be precomputed, as well as the shape $\Ns'$.

The only missing part in Algorithm~\ref{alg:qhankelization} is computation of $\diagsums(U,V)$ for given $U \in \spaceR^{L}$ and $V \in \spaceR^K$. For this we note that
\[
\begin{split}
&\langle UV^{\top}, \trajmat{ShSSA}(\elmat_{k,l})\rangle_\frob =
\trace (VU^{\top}\Pmat_{\Ls}^{\top}\trajmat{ShSSA}(\elmat_{k,l})\Pmat_{\Ks}) \\
&={\mvec(\rmU')}^{\top} \hbhcirc{\elmat_{k,l}} \mvec(\rmV') = \mvec(\elmat_{k,l})^{\top} \tbtcirc{\rmU'} \mvec(\rmV'),
\end{split}
\]
where $\rmU' = \mvec^{-1}_{\Lx}(\Pmat_{\Ls} U)$, $\rmV' = \mvec^{-1}_{\Kx}(\Pmat_{\Ks} V$), and the last equality holds by Lemma~\ref{lem:hbh_tbt_elmat}. Thus, $\diagsums(U,V)$ can be computed by Algorithm~\ref{alg:diagsums}.
\begin{algorithm}\label{alg:diagsums}
    \quad

    Input: $U \in \spaceR^L$, $V \in \spaceR^K$. Output: $\tX = \diagsums(U,V) \in \spaceR^{\Nx \times \Ny}$.
    \begin{enumerate}
        \item $\rmU' \leftarrow \mvec^{-1}_{\Lx}\left(\Pmat_{\Ls} U \right)$
        \item $\rmV' \leftarrow \mvec^{-1}_{\Kx}\left(\Pmat_{\Ks} V \right)$
        \item $\widehat{\rmU'} \leftarrow \fftw{\Nx,\Ny} (\rmU')$
        \item $\widehat{\rmV'}\rm \leftarrow \fftw{\Nx,\Ny} (\rmV')$
        \item $\tX \leftarrow \fftw{\Nx,\Ny}^{-1} (\widehat{\rmU'} \odot \widehat{\rmV'})$
    \end{enumerate}
\end{algorithm}

\subsubsection{Calculation of shapes}
\label{sec:implement_shaped}
Assume that we have $\Ls \subseteq \{1,\ldots,\Lx\} \times \{1,\ldots,\Ly\}$ and
$\Ns \subseteq \{1,\ldots,\Nx\} \times \{1,\ldots,\Ny\}$. We would like to find
the maximal $\Ks$ such that $\Ls \shs \Ks \subseteq \Ns$.

Let $I^{(\Ns)} \in \cR^{\Nx \times \Ny}$ be the indicator array of $\Ns$, i.e.,
\[
i^{(\Ns)}_{k,l} =
\begin{cases}
    1, & ({k,l}) \in \Ns, \\
    0, & ({k,l}) \not\in \Ns. \\
\end{cases}
\]
Then the shape $\Ks$ is equal to the maximal $\mathfrak{A}$ such that all the elements of
the $\Ls \times \mathfrak{A}$ submatrix of $\trajmat{2DSSA} (I^{(\Ns)})$ are equal to
$1$.  Algorithm~\ref{alg:find_ks} finds such a maximal shape from the product of 
$\trajmat{2DSSA}^{\top} (I^{(\Ns)})$ and $1_L$. The elements of $\Ks$ 
correspond to the elements of the resulting vector that have the value $L$.

\begin{algorithm}\label{alg:find_ks}
    \quad

    Input: $\Ls \subseteq \{1,\ldots,\Lx\} \times \{1,\ldots,\Ly\}$, $\Ns
    \subseteq \{1,\ldots,\Nx\} \times \{1,\ldots,\Ny\}$. Output: maximal $\Ks$
    such that $\Ls \shs \Ks \subseteq \Ns$.
    \begin{enumerate}
        \item $V \leftarrow \trajmat{2DSSA}^{\top} (I^{(\Ns)}) 1_L$
        \item $\rmV' \leftarrow \mvec^{-1}_{\Kx} (V)$
        \item $\Ks \leftarrow \{(k,l) \,|\, v'_{k,l} = L \}$
    \end{enumerate}
\end{algorithm}

%% file: implement_forecast.tex
\subsection{Fast vector forecasting algorithm}
\label{sec:implement_forecast}

It is mentioned in \citet[p.~76]{Golyandina.Zhigljavsky2012} that the vector
forecasting is time-consuming, while recurrent forecasting is fast.  However,
this is so if one implements the algorithm from Section~\ref{SEC:continuation}
directly.  It appears that it is possible to considerably accelerate the vector
forecasting.  Moreover, in the current implementation in \pkg{Rssa} the vector
forecasting is slightly faster than the recurrent one.

In this section, we will use the notation from Section~\ref{SEC:continuation}.
We also denote by $\dagger$ the pseudo-inversion of a matrix.

\subsubsection{Column MSSA forecast}
As it was shown in Section~\ref{SEC:continuation}, column vector MSSA forecasting 
is reduced to performing $s$ 1D vector forecasts the same subspace $\cL^{\mathrm{col}}$.
Next, we describe the algorithm for fast vector
forecasting of 1D time series in a given subspace $\cL^{\mathrm{col}}$ 
(vector forecast in Basic SSA).

Consider the forecasting in the subspace $\cL^{\mathrm{col}}$ given by a basis
$\{P_1,\ldots,P_r\}$.  Denote $\bfP=[P_1:\ldots:P_r]$.  Each reconstructed
vector $\what{X}_k$ of the 1D time series belongs to $\cL^{\mathrm{col}}$; hence, there exist coefficients
$W_k \in \spaceR^r$ such that $\what{X}_k=\bfP W_k$.  Denote $\bfW = [W_1 :
\ldots : W_K]$. In fact, the input for the algorithm is the minimal
decomposition of $\what{\bfX}$ into the sum of elementary matrices of rank 1 in
the form $\what{\bfX}=\bfP \bfQ^\top$ and $\bfW=\bfQ^\top$.

Note that if we have a singular value decomposition of
$\what\bfX=[\what{X}_1:\ldots:\what{X}_K]$, then the left singular vectors
provide the basis of the subspace, while $W_k$ are determined by the right singular
vectors and singular values: $\bfP=[U_1:\ldots:U_r]$ and $\bfQ=[\sqrt\lambda_1
V_1:\ldots:\sqrt\lambda_r V_r]$.

In vector forecasting, we extend the reconstructed matrix as
$\bfZ = [\what\bfX:{Z}_{K+1}:\ldots:{Z}_{K+M+L-1}]$, where the vectors $Z_k$, $k\ge K+1$, are obtained as 
$Z_k = \cP_{\mathrm{Vec}}^{\mathrm{col}} Z_{k-1}$ (and $Z_{k} = \widehat{X}_k$ for $k=1,\ldots,K$). Since all $Z_k$ belong to $\cL^{\mathrm{col}}$, there exist $W_k \in \spaceR^r$, $k \ge K+1$ such that
\[
\bfZ = \bfP [W_1:\ldots:W_{K+M+L-1}].
\]
Thus, there exist a matrix $\bfD$ such that $W_k = \bfD W_{k-1}$ for $k \ge K+1$. This 
observation leads to the following algorithm for vector forecasting.

\begin{algorithm}\label{alg:fast_col_forecast}
Input: $\bfP$, $\bfW$.
Output: the forecasted values $z_{N+1},\ldots,z_{N+M}$.
\begin{enumerate}
    \item \label{item:esprit} Compute the matrix $\bfD = \last{\bfP}^\dagger
    \first{\bfP}$ using the QR-decomposition \citep{Golub.VanLoan1996}.
    \item For $k = K+1,\ldots, K+M+L-1$ compute $W_k = \bfD W_{k - 1}$.
    \item Perform the fast rank-one hankelization algorithm (Section~\ref{ssec:hmatmul}) for
    the matrix $\bfZ=\bfP [\bfW : W_{K+1} : \ldots : W_{K+M-L-1}]$ , which is
    explicitly expressed as a sum of rank-one matrices, and obtain the series
    $z_1, \ldots, z_{N + M + L - 1}$.
    \item The numbers $z_1, \ldots, z_{N + M}$ form updated reconstructed and
    forecasted series.
\end{enumerate}
\end{algorithm}

In order to prove the correctness of the algorithm, it remains to show that the formula
$\bfD= \last{\bfP}^\dagger \first{\bfP}$ is correct, which will be discussed at the end of this section.
Finally, there is also a further improvement on the computation of $\bfD$.

\begin{remark}
\label{rem:pseudoinv}
1. Item \ref{item:esprit} in Algorithm~\ref{alg:fast_col_forecast} is exactly the shift matrix from
the LS-ESPRIT method for frequency estimation \citep{Roy.Kailath1989}.\\
2. If $\{P_i\}$ is orthonormal system, then $\bfD$ can be computed without the QR decomposition as:
\begin{equation*}
    \bfD = \left(\bfI_r - \frac{1}{1 - \bfpi^\top\bfpi} \bfpi\bfpi^\top\right) \last{\bfP}^\top\first{\bfP},
\end{equation*}
where $\bfpi = \bfpi(\bfP)$, $\bfI_r$ is the $r\times r$ identity matrix.
\end{remark}

\subsubsection{Row MSSA forecast}
Row vector forecast is  slightly different from the column one, but the idea is the
same. We deal with forecasting in the row subspace
$\cL^{\mathrm{row}}=\sspan(Q_1,\ldots,Q_r)$ and continue the sequence of the row
vectors $\what{Y}_k$ of $\what{\bfX}$.  If the row vectors are equal to
$\what{Y}_k=\bfQ W_k$, $k = 1,\ldots,L$,
then $\what{\bfX}^{\top} = \bfQ \bfW$, where $\bfW = [W_1:\ldots:W_L]$.
In the following algorithm, as in column forecasting, the vectors $W_k$ are continued
instead of the vectors $Y_k$.

\begin{algorithm}\label{alg:fast_row_forecast}
Input: $\bfQ$, $\bfW$.
Output: the forecasted values $z^{(p)}_{N_p+1},\ldots,z^{(p)}_{N_p+M}$, $p=1,\ldots,s$.
\begin{enumerate}
    \item Compute the matrix $\bfD = \llast{\bfQ}^\dagger\ffirst{\bfQ}$ using
    the QR decomposition.
    \item For $k = L+1, \ldots, L+M+\max_{p=1,\ldots,s}{K_p}-1$ compute $W_k =
    \bfD W_{k - 1}$.
    \item Perform the fast rank-one hankelization algorithm (Section~\ref{ssec:hmatmul}) for each of $s$ matrices\\
    $\bfZ^{(p)}= \bfQ^{(p)}[\bfW : W_{L+1} : \ldots : W_{L+M+K_p-1}]$ for $p =
    1, \ldots, s$ and obtain $s$ series $z_1^{(p)}, \ldots z_{N_p + M + L -
        1}^{(p)}$.
    \item The numbers $z_1^{(p)}, \ldots, z_{N_p + M}^{(p)}$ form updated
    reconstructed and forecasted series.
\end{enumerate}
\end{algorithm}

\begin{remark}
\label{rem:pseudoinv_row}
If $\{Q_i\}$ is orthonormal system, then $\bfD$ from Algorithm~\ref{alg:fast_row_forecast} can be expressed as
\begin{equation*}
    \bfD = \left(\bfI_r - \bfS^\top (\bfI_s - \bfS\bfS^\top)^{-1} \bfS\right) \llast{\bfQ}^\top\ffirst{\bfQ},
\end{equation*}
where $\bfS = [\bfmu(Q_1):\ldots:\bfmu(Q_r)]$, $\bfI_r$ and $\bfI_s$ are the
$r\times r$ and $s \times s$ identity matrices.
\end{remark}

\subsubsection{Proof of the algorithms correctness}

For simplicity, we will consider the one-dimensional case, that is, SSA vector forecast,
which coincides with the column MSSA forecast for one-dimensional series.
The proof for row forecasting is analogous.

We need to prove that $Z_{k+1}=\bfP W_{k+1}$, where $W_{k+1} = \bfD W_k$,
$\bfD = \last{\bfP}^\dagger  \first{\bfP}$.
It is sufficient to prove
$\last{Z}_{k+1}=\last{\bfP} W_{k+1}$, since the last coordinate of the vector is
uniquely defined.  In the standard formulation of vector forecasting algorithm
in \citet[Section 2.3]{Golyandina.etal2001} and in Section
~\ref{SEC:continuation}, $\last{Z}_{k+1}$ is the projection on the column space
of $\last{\bfP}$, that is,
$\last{Z}_{k+1}=\last{\bfP}\last{\bfP}^\dag\first{Z}_k$. Since
$\first{\bfP} W_k$ is exactly $\first{Z}_k$ by definition of $W_k$,
we have that $\last{Z}_{k+1}=\last{\bfP}(\last{\bfP}^\dag \first{\bfP}) W_k = \last{\bfP} W_{k+1}$, and the
equivalence of the standard and fast vector forecasting algorithms is proved.

\begin{remark}
    The speedup of the algorithms implementation is explained by two reasons:
    \begin{enumerate}
    \item multiplication by matrices of small size $r\times r$ at each step
    instead of multiplication by matrices of much larger size, and
    \item the form of the matrix to be
    hankelized is suitable for application of the fast rank-one hankelization algorithm.
    \end{enumerate}
\end{remark}

%% file: app_mssa_theory.tex
\subsection{Elements of MSSA theory}
\label{sec:mssa_theory}
We put into this section several propositions related to the MSSA theory taken
from \citet{Golyandina.etal2003, Stepanov.Golyandina2005} (in Russian).

\subsubsection{Separability}
Separability is the key notion in the SSA theory, since separability of series
means the ability of the method to extract them from the given sum-total series.
Notion of separability for multidimensional time series is analogous to that for
one-dimensional series, which is briefly commented in Section~\ref{sec:general} and
is thoroughly described in \citet[Sections 1.5 and 6.1]{Golyandina.etal2001}.  There
is a weak separability, which means orthogonality of the trajectory spaces.
There is a strong separability that means empty intersection of the sets of
singular values produced by the separated series.

Generally, conditions of separability of multidimensional time series are more
restrictive than that for one-dimensional series.  The following sufficient
condition of weak separability is valid.

\begin{proposition}
\label{prop:mssa_sep}
If time series $\tF^{(1)}$ and $\tF^{(2)}$, $\tG^{(1)}$ and $\tG^{(2)}$,
$\tF^{(1)}$ and $\tG^{(2)}$, and also $\tG^{(1)}$ and $\tF^{(2)}$ are weakly
$L$-separable by SSA, then the two-dimensional time series $(\tF^{(1)},
\tF^{(2)})$ and $(\tG^{(1)}, \tG^{(2)})$ are weakly $L$-separable by MSSA and
the complex-valued time series $\tF^{(1)}+\unit\tF^{(2)}$ and
$\tG^{(1)}+\unit\tG^{(2)}$ are weakly $L$-separable by CSSA.
\end{proposition}

Proposition~\ref{prop:mssa_sep} can be extended to an analogous result for
asymptotic separability ($N_i\ra\infty$) and therefore for approximate
separability for fixed $N_i$.

\begin{example}
    Consider the example of four harmonic real-valued time series $\tF^{(1)}$,
    $\tF^{(2)}$, $\tG^{(1)}$ and $\tG^{(2)}$ of length $N$
    \bea
        f^{(1)}_{k}=A_1\cos(2\pi \omega_1 k+\varphi_1),\qquad
        f^{(2)}_{k}=B_1\cos(2\pi \omega_1 k+\varphi_2),
    \eea
    \bea
        g^{(1)}_k=A_2\cos(2\pi k \omega_2 k+\phi_1), \qquad
        g^{(2)}_k=B_2\cos(2\pi k \omega_2 k+\phi_2),
    \eea
    $\omega_1\neq \omega_2$, $k=0,\ldots,N-1$, $A_1,A_2,B_1,B_2 \neq 0$.  If $L
    \omega_i$ and $K\omega_i$, $i=1,2$, are integer, then $(\tF^{(1)},\tF^{(2)})$ and
    $(\tG^{(1)}, \tG^{(2)})$ are $L$-separable by MSSA and the
    complex-valued time series $\tF^{(1)}+\unit\tF^{(2)}$ and
    $\tG^{(1)}+\unit\tG^{(2)}$ are weakly $L$-separable by CSSA.
\end{example}

Weak separability is not enough for extraction of time series
components. Therefore, let us look at strong separability related to eigenvalues
produced by time series components. It appears that each time series
$(\tF^{(i)}, \tG^{(i)})$ can produce different eigenvalues in SSA, MSSA and
CSSA. Therefore, the application of an appropriate multidimensional modification
of SSA can improve the strong separability.

Note that if $L \omega_i$ or $K\omega_i$ is not integer, then the series become
approximately separable.

\begin{example}
\label{ex:mssa_ev}
Let
\bea
     f_k^{(1)}=A\cos(2\pi \omega k+\varphi_1),\qquad
     f_k^{(2)}=B\cos(2\pi k \omega k+\varphi_2).
\eea
If $L\omega$ and $K\omega$ are integer, then $(\tF^{(1)}, \tF^{(2)})$ produces two equal
eigenvalues in MSSA: $\lambda_1=\lm_2=(A^{2}+B^{2})LK/4$, while
$\tF^{(1)}+\unit\tF^{(1)}$ produces two nonequal eigenvalues in CSSA:
\bea
    \lambda_{1}&=&(A^2+B^2+2AB\sin \varphi)LK/4,\\
    \lambda_{2}&=&(A^2+B^2-2AB\sin \varphi)LK/4,
\eea
where $\varphi=\varphi_1-\varphi_2$ and $|\varphi|\neq\pi/2 \mmod{\pi}$
(otherwise, the series produces one non-zero eigenvalues equal to
$(A^2+B^2)LK/2$).  Note that the series $\tF^{(1)}$ itself produces two eigenvalues
equal to $A^{2}LK/4$.
\end{example}

\subsubsection{Multi-dimensional time series and LRRs}
Consider a system of infinite time series $\tX^{(1)}, \tX^{(2)}, \ldots,
\tX^{(s)}$, choose the window length $L$ and denote $\cX^{(1)}, \ldots,
\cX^{(s)}$ the column trajectory spaces of the series 
(subspaces spanned by the $L$-lagged vectors of the series).  Let
$\cX=\sspan(\cX^{(1)}, \ldots, \cX^{(s)})$ be the column trajectory space of the
collection of time series $(\tX^{(1)}, \tX^{(2)}, \ldots, \tX^{(s)})$.  As well
as for one-dimensional time series, we call the dimension of the trajectory
space (equal to the rank of the trajectory matrix of the series collection) the
rank of the series collection, see Section~\ref{sec:general} for short
description of general notions.

Denote the ranks of $\tX^{(l)}$ by $r_l=\dim \cX^{(l)}\leq L$, $l=1,\ldots,s$.
For each time series $\tX^{(l)}$ we can write out the minimal LRR governing the
time series:
\be
\label{EQ:Def:LRF12}
\begin{array}{l}
    x_{j+r_l}^{(l)}=\sum\limits_{k=1}^{r_l}a^{(l)}_k x_{j+r_l-k}^{(l)},
    \quad \mbox{where} \quad a^{(l)}_{r_l}\neq 0, \;\; l=1,\ldots,s.
\end{array}
\ee
The corresponding characteristic polynomials of the LRR (\ref{EQ:Def:LRF12}) are
\be
\label{EQ:Def:ChPol12}
\begin{array}{l}
    P_{r_l}^{(l)}(\mu)=\mu^{r_l}-\sum\limits_{k=1}^{r_l}a^{(l)}_k\mu^{r_l-k},
    \quad l=1,\ldots,s.
\end{array}
\ee
The roots of the characteristic polynomial of the minimal LRR governing the
series called \emph{characteristic roots}.

Let
\bea
\begin{array}{lcl}
    p^{(l)}          &\mbox{be}& \mbox{number of different roots of the polynomial}\quad P_{r_l}^{(l)}(\lm),\\
    \mu_{m}^{(l)}&\mbox{be}& m\mbox{-th root of the polynomial}\quad P_{r_l}^{(l)}(\lm),\\
    k_{m}^{(l)}      &\mbox{be}& \mbox{the multiplicity of the root}\quad \mu_{m}^{(l)}.
\end{array}
\eea
Then, from standard theory of time series of finite rank \citep{Golyandina.etal2001}, we have that 
\bea
    k_{1}^{(l)}+\ldots+k_{p^{(l)}}^{(l)}=r_l, \ \  l=1,\ldots,s.
\eea
 The characteristic roots determine the series behavior. For example, if $k_{m}^{(l)}=1$, then
 the time series has the form
 \bea
 x_n^{(l)}=\sum_{j=1}^{r_l} C_j^{(l)} \left(\mu_j^{(l)}\right)^n.
 \eea

 Also let
\bea
\begin{array}{lcl}
    \mu_{1},\ldots,\mu_{p} &\mbox{be}& \mbox{the pooled set of roots of all the polynomials}
                                             \;\; P_{r_1}^{(1)}, \ldots, P_{r_s}^{(s)},\\
    k_{1},      \ldots, k_{p}      &\mbox{be}& \mbox{the multiplicities of the roots}\; \mu_{1},\ldots,\mu_{p},
\end{array}
\eea
where multiplicity of a root in the pooled set is equal to the maximal 
multiplicity of the corresponding root across all the polynomials.

Since the roots are determined by the structure of the trajectory space, the
following proposition can be proved.

\begin{proposition}
\label{PR:LL:MSSA}
Rank of the infinite multi-dimensional time series $(\tX^{(1)}, \tX^{(2)},
\ldots, \tX^{(s)})$ is equal to $r=\sum_{i=1}^p k_i$, for $L>r$.
\end{proposition}

Consider a simple example.
\begin{example}
Let  $\tF=(f_1,\ldots,f_{N})$ and
    $\tG=(g_1,\ldots,g_{N})$ with
\be
\label{eq:2harms}
  f_k^{(1)}=A\cos(2\pi\omega_1 k+\varphi_{1}), \quad
  f_k^{(2)}=B\cos(2\pi\omega_2 k+\varphi_{2}),
\ee
where $0<\omega < 1/2$, $0\leq \vphi_1,\vphi_2<2\pi$ and $A,B\neq 0$.  Let us
fix the window length $L$ and find the SSA-rank of the time series
$\tF$, the MSSA-rank of $(\tF^{(1)},\tF^{(2)})$ and the CSSA-rank of $\tF^{(1)}+\unit\tF^{(2)}$:
\label{pr:spaces}
\begin{enumerate}
\item
For $\omega_1=\omega_2$ the SSA and the MSSA ranks of the sinusoid
(\ref{eq:2harms}) is equal to 2. The CSSA rank is equal to 1 if $A=B$ and
$|\vphi_1-\vphi_2|=\pi/2 \mmod{\pi}$
and is equal to 2 otherwise.
\item
For $\omega_1 \neq \omega_2$ the SSA, MSSA and CSSA ranks equal 4.
\end{enumerate}
\end{example}

%% file: app_2dssa_theory.tex
\subsection{Elements of 2D-SSA theory}
\label{sec:app_2dssa_theory}
\subsubsection{Arrays of finite rank}
The theory of arrays of finite rank is mainly contained in
\citet{Golyandina.Usevich2009}.  We present here a short summary.

In this section we consider a class of infinite arrays $\tX =
(x_{m,n})_{m,n=0}^{\infty, \infty}$ given in a parametric form:
\begin{equation}
x_{m,n} = \suml_{k=1}^{r} c_k \lambda^{m}_k \mu^{n}_k,
\label{eq:sods}
\end{equation}
where $(\lambda_k,\mu_k) \in \spaceC^2$ are distinct pairs of complex numbers.
It can be shown that for large enough $\Nx, \Ny, \Lx, \Ly$, the rank of the
trajectory matrix $\bfX$ is equal to $r$ (or equivalently, the arrays of the
form (\ref{eq:sods}) are \textit{arrays of finite rank}. Note that the exponentials
can be also represented in the form
\begin{align*}
\lambda_k = \rho_{x,k} \exp(2\pi \unit \omega_{x,k} + \phi_{x,k}), \quad
\mu_k =  \rho_{y,k} \exp(2\pi \unit \omega_{y,k} + \phi_{y,k}).
\end{align*}

We should note that the class of arrays of finite rank also contains bivariate
polynomials and their products with the functions from (\ref{eq:sods}). An
algebraic characterization of the arrays of finite rank can be found in
\citet{Golyandina.Usevich2009}.

\subsubsection{Real arrays}
Now let us summarize what happens in the case of real arrays $\tX$. If
(\ref{eq:sods}) is real then for any pair $(\lambda,\mu) \in \spaceC^2$ its complex
conjugate $(\overline{\lambda}, \overline{\mu})$ also should be present in
(\ref{eq:sods}). We order the roots such that $(\lambda_k,\mu_k)$ are real for
$1 \le k \le d$, and other $2s$ roots are at least partially complex (there are
complex conjugate pairs), and are arranged as $(\lambda_k, \mu_k) =
(\conjug{\lambda_{k-s}},\conjug{\mu_{k-s}})$ for $d+s+1 \le k \le d+2s$.  Then
the roots have a representation
\[
(\lambda_k, \mu_k) =
\begin{cases}
    (\rho_{x,k}, \rho_{y,k} ) = \left(\rho_{y,k} \exp(2\pi \unit \omega_{x,k}), \rho_{y,k} \exp(2\pi \unit \omega_{y,k})\right), & 1 \le k \le d, \\
    \left(\rho_{x,k} \exp(2\pi \unit \omega_{x,k}), \rho_{y,k} \exp(2\pi \unit \omega_{y,k})\right), & d+1 \le k \le d + s, \\
    \left(\rho_{x,k-s} \exp(-2\pi \unit \omega_{x,k-s}), \rho_{y,k} \exp(-2\pi \unit
        \omega_{y,k-s})\right), & d +s+1 \le k \le r,
\end{cases}
\]
where $(\rho_{x,k}, \rho_{y,k}, \omega_{x,k}, \omega_{y,k}) \in \spaceR^2 \times
[0; 1/2)^{2}$ are distinct $4$-tuples of real numbers, such that $\omega_{x,k} =
\omega_{y,k} = 0$ for $1 \le k \le d$.

Then the representation (\ref{eq:sods}) becomes a sum of $d+s$ planar modulated
sinewaves:
\[
x_{m,n} = \sum\limits_{k=1}^{d+s} b_k
\rho_{x,k}^{m} \rho_{y,k}^{n} \cos \left(2 \pi (\omega_{x,k} m+ \omega_{y,k} n) + \phi_k\right),
\]
where $b_k$ and $\phi_k \in [0;2\pi)$ are unique real coefficients obtained from
$c_k$.

\begin{example}
    The product of sines can be uniquely represented as a sum of two planar
    sines.
    \[
    2 \cos (2 \pi \omega_{x} m+   \phi_1) \cos (2 \pi  \omega_{y} n + \phi_2) =
    \cos(2 \pi (\omega_{x} m+ \omega_{y} n) + \phi_1 + \phi_2)+
    \cos(2 \pi (\omega_{x} m- \omega_{y} n) + \phi_1 - \phi_2)
    \]
\end{example}